\documentclass[longauth]{aa}
\usepackage{natbib}
\usepackage[varg]{txfonts}
\usepackage{graphicx}
\usepackage{wasysym}
\usepackage{color}
\usepackage[draft]{hyperref}
\usepackage{url}
\bibpunct{(}{)}{;}{a}{}{,}

\begin{document}

\title{Stellar masses, sizes, and radial profiles for 465 nearby
  early-type galaxies: an extension to the \emph{Spitzer}
  Survey of Stellar Structure in Galaxies (S$^{4}$G)}

\author{A. E. Watkins\inst{1,2,3}
  \and H.~Salo\inst{1}
  \and E.~Laurikainen\inst{1}
  \and S.~D\'{i}az-Garc\'{i}a\inst{6,7,8}
  \and S.~Comer\'{o}n\inst{6,7}
  \and J.~Janz\inst{1,4,5}
  \and A.~H.~Su\inst{1}
  \and R.~Buta\inst{9}
  \and E.~Athanassoula\inst{10}
  \and A.~Bosma\inst{10}
  \and L.~C.~Ho\inst{11,12}
  \and B.~W.~Holwerda\inst{13}
  \and T.~Kim\inst{14}
  \and J.~H.~Knapen\inst{6,7}
  \and S.~Laine\inst{15}
  \and K.~Men\'{e}ndez-Delmestre\inst{16}
  \and R.~F.~Peletier\inst{17}
  \and K.~Sheth\inst{18}
  \and D.~Zaritsky\inst{19}}

\institute{Space Physics and Astronomy Research Unit, University of Oulu,
  FI-90014, Finland
 \and Centre for Astrophysics Research, School of Physics, Astronomy
 and Mathematics, University of Hertfordshire, Hatfield AL10 9AB, UK
 \and Astrophysics Research Institute, Liverpool John Moores
  University, IC2, Liverpool Science Park, 146 Brownlow Hill,
  Liverpool L3 5RF, UK
 \and Finnish Centre of Astronomy with ESO (FINCA), Vesilinnantie 5,
 FI-20014 University of Turku, Finland
 \and Specim, Spectral Imaging Ltd., Elektroniikkatie 13, FI-90590 Oulu, Finland
 \and Departamento de Astrof\'{i}sica, Universidad de La Laguna,
 E-38200, La Laguna, Tenerife, Spain
 \and Instituto de Astrof\'{i}sica de Canarias E-38205, La Laguna,
 Tenerife, Spain
 \and Department for Physics, Engineering Physics and Astrophysics,
 Queen's University, Kingston, ON K7L 3N6, Canada
 \and  Department of Physics and Astronomy, University of Alabama, Box 870324, Tuscaloosa, AL 35487, USA
 \and Aix Marseille Univ, CNRS, CNES, LAM, Marseille, France
 \and Kavli Institute for Astronomy and Astrophysics, Peking University, Beijing 100871, China
 \and Department of Astronomy, School of Physics, Peking University, Beijing 100871, China
 \and University of Louisville, Department of Physics and Astronomy, 102 Natural Science Building, 40292 KY Louisville, USA
 \and Department of Astronomy and Atmospheric Sciences, Kyungpook National University, Daegu, 41566, Korea
 \and IPAC, Mail Code 314-6, Caltech, 1200 E. California Blvd., Pasadena, CA 91125, USA
 \and Valongo Observatory, Federal University of Rio de Janeiro, Ladeira Pedro Ant\^{o}nio, 43, Sa\'{u}de CEP 20080-090 Rio de Janeiro, RJ, Brazil
 \and Kapteyn Astronomical Institute, University of Groningen, PO Box 800, NL-9700 AV Groningen, the Netherlands
 \and NASA Headquarters Mary W. Jackson Building, 300 E Street SW, Washington DC 20546
 \and Steward Observatory and Department of Astronomy, University of Arizona, 933 N. Cherry Ave., Tucson, AZ 85721, USA}

\abstract{The \emph{Spitzer} Survey of Stellar Structure in Galaxies
  (S$^{4}$G) is a detailed study of over 2300 nearby galaxies in
  the near-infrared (NIR), which has been critical to our understanding of
  the detailed structures of nearby galaxies.  Because the sample galaxies were
  selected only using radio-derived velocities, however, the survey
  favored late-type disk galaxies over lenticulars and ellipticals.}
  {A follow-up \emph{Spitzer} survey was
  conducted to rectify this bias, adding 465 early-type galaxies (ETGs) to
  the original sample, to be analyzed in a manner consistent with the initial
  survey.  We present the data release of this ETG
  extension, up to the third data processing pipeline (P3): surface
  photometry.}
   {We produce curves of growth and radial surface brightness profiles
    (with and without inclination corrections) using reduced and
    masked \emph{Spitzer} IRAC 3.6\,$\mu$m and 4.5\,$\mu$m images produced through Pipelines 1 and 2, respectively.
    From these profiles, we derive the following integrated
    quantities: total magnitudes, stellar masses, concentration
    parameters, and galaxy size metrics.  We showcase NIR scaling relations
    for ETGs among these quantities.}
   {We examine general trends across the whole S$^{4}$G and ETG
     extension among our derived parameters, highlighting differences
     between ETGs and late-type galaxies (LTGs).  ETGs are, on average, more massive and
     more concentrated than LTGs, and also show subtle distinctions among ETG
     morphological sub-types.  We also derive the following
     scaling relations and compare with previous results in visible
     light: mass--size (both half-light and isophotal),
     mass--concentration, mass--surface brightness (central, effective,
     and within 1~kpc), and mass--color.}
   {We find good
     agreement with previous works, though some relations (e.g.,
     mass--central surface brightness) will require more careful
     multi-component decompositions to be fully understood.  The
     relations between mass and isophotal radius and between mass and
     surface brightness within 1~kpc, in particular, show notably
     small scatter.  The former provides important constraints on the limits of size growth in galaxies, possibly related to star formation thresholds, while the latter---particularly when paired with the similarly tight relation for LTGs---showcases the striking self-similarity of galaxy cores, suggesting these evolve little over cosmic time.  All of the profiles and parameters described in this paper will be
     provided to the community via the NASA/IPAC database on a
     dedicated website.}

\keywords{galaxies: evolution - galaxies: photometry - galaxies:
  spiral - galaxies: structure - galaxies: statistics - galaxies:
  elliptical and lenticular, cD}
\titlerunning{S${4}$G+ETG}
\maketitle

\section{Introduction}\label{sec:intro}

Over cosmic time, galaxies acquire and process baryons, turning
into stars and redistributing those stars through interactions with
neighboring systems or through their own internal
gravitational processes.  Under $\Lambda$CDM cosmology \citep{white78,
  frenk85}, galaxies form via the hierarchical merging of dark matter
halos, which in the early Universe served as the potential wells into
which baryons initially settled.  Given the hierarchical nature of the
merging, the centers of galaxies likely formed first, and subsequent
gas or satellite accretions \citep[e.g.,][]{kauffmann93, bendo00,
  aguerri01, balcells03, bournaud07, athanassoula16} and secular evolution
\citep[e.g.,][]{combes81, kormendy82, fathi03, kormendy04,
  athanassoula05} then conspired to modify both these centers and the
structures built atop them in complex collisional and dissipational
interactions.  Nearby galaxies display the present end-results
of these interactions; therefore, detailed study of the stellar
structure of nearby galaxies is critical to understanding
galaxy evolution.

It was to this end that the \emph{Spitzer} Survey of Stellar Structure
in Galaxies \citep[S$^{4}$G;][]{sheth10} was carried out, providing
excellent quality deep imaging of over 2300 galaxies within $\sim$40
Mpc, in a wavelength regime (near-infrared, NIR) that suffers little
from dust extinction \citep{draine84, calzetti01} and contains few
emission or absorption features \citep[e.g.,][]{willner77, tokunaga91,
  kaneda07}.  Stellar mass-to-light ratios are also nearly constant
across stellar populations in this wavelength regime, making this
imaging a nearly direct tracer of stellar mass
\citep[e.g.,][]{pahre04, jun08, peletier12, meidt14, driver16}.  The survey has been
indispensible to the study of galaxy evolution, having resulted in
\textrm{over one hundred} published papers \citep[e.g.,][]{buta10, comeron11,
  martinnavarro12, zaritsky13, laine14, herrera15, kim16, laine16,
  laurikainen17, bouquin18, diazgarcia19, ikiz20}.  Yet, until now, it
has suffered from a serious bias: the velocity cut in the original survey
used only radio measurements, thereby excluding gas-poor---mostly
early-type---galaxies.

Early-type galaxies (ETGs), being typically gas- and dust-poor and
dominated by evolved stellar populations, are often referred to as
``red and dead.''  This is an unfortunate term, connoting a lack of
interesting properties or phenomena.  The term ETG---referring to the classification scheme of increasing morphological complexity proposed by
\citet{hubble26}---covers both the disk-dominated and
rotation-supported lenticular (S0) galaxies, as well as the more
pressure-supported elliptical (E) galaxies.  What the exact
relationship is linking these two classes with each other---and with
their more gas-rich, late-type galaxy (LTG) counterparts---is a
long-standing question.

For example, due to their many similarities with late-type spirals
\citep[including bulge-to-total mass or light ratios, range of bulge luminosities,
  bar properties, and kinematics; e.g.,][]{laurikainen11,
  cappellari11, kormendy12}, S0 galaxies may have formed from LTGs
through the removal or rapid consumption of gas and dust.  S0s also form a
parallel morphological sequence with spirals \citep{spitzer51,
  vandenbergh76, laurikainen11, cappellari11, kormendy12}, lending
further credence to this idea.  Detailed photometric analysis has also
provided evidence linking the two disk galaxy classes;
\citet{laurikainen10} found that the sizes and masses of bulges in
both early- and late-type disk galaxies correlate strongly with those
of the disk, suggesting a similar kind of concurrent evolution
\citep[see also:][]{kormendy04, aguerri05}, and disk scaling relations
between S0s and LTGs also appear similar \citep[e.g.,][]{eliche15}.

Even if the two are evolutionarily linked, however, the precise
mechanisms behind the transformation from LTG to ETG are still open to debate.
Because of the relative abundance of S0s in galaxy clusters, one
long-favored such mechanism is infall into dense environments
\citep[e.g.,][]{gunn72, dressler80, moore96, bekki02}.  This seems to happen quite early; detailed study of nearby cluster S0 stellar populations frequently shows quenching epochs over 10~Gyr ago \citep[e.g.,][]{silchenko13, comeron16, katkov19}.  However,
isolated field S0s, though rare, do exist \citep{karachentsev10},
ruling out cluster or group infall as the sole formation pathway.  \citet{falcon19} also argued that simple quenching and
fading of stellar populations is unlikely to be the dominant
spiral--S0 transformation mechanism given differences in angular
momentum between the classes \citep[with simulations suggesting major
  mergers may instead explain these S0 kinematics;][]{querejeta15},
though such quenching may still occur in low-density environments
\citep[e.g.,][]{rizzo18}.  Environmental influence may not be necessary, either; \citet{saha18} showed that galaxies with properties very similar to S0s can form in isolation, via instabilities in gaseous disks.  \citet{burstein05} also argued that the relative \emph{K}-band luminosities of S0s and spirals demonstrate that direct quenching, while feasible in individual cases, may not be the dominant transformation mechanism either.  Indeed, detailed comparisons between
environments suggest there may be multiple quenching pathways behind
the formation of S0 galaxies \citep[e.g.,][]{wilman12}. \citet{aguerri12} provides a thorough review of different proposed S0 formation pathways.

The origins of elliptical galaxies, despite their relatively simple
appearances, are likely equally complex.  Early theories of elliptical galaxy
formation relied on a monolithic collapse \citep{eggen62}, either via
the low-rotation, dissipational collapse of gas clouds
\citep[e.g.,][]{larson74a, larson74b, larson75}, or the dissipationless
collapse of stars formed prior to the gas collapse time
\citep[e.g.,][]{gott73, thuan75, gott77}.  As $\Lambda$CDM cosmology
came into favor, hierarchical models of elliptical galaxy formation did
alongside \citep[see the review by][]{dezeeuw91}, and the discussion
became more granular.  The origin of elliptical galaxy rotation (or lack
thereof), for example, received a lot of focus \citep{bertola75,
  illingworth77, binney78, davies83, wyse84, barnes88} given its
strong implications for elliptical galaxy merger histories; indeed, rotation may be critical to understanding ETG formation pathways as a whole \citep{cappellari16}.

From simulations, it became clear that major mergers of gas-poor disk galaxies can result in
remnants that look very much like elliptical galaxies, albeit with nearly zero
net rotation \citep[e.g.,][]{toomre77, white83, barnes88, barnes90,
  hernquist92}.  The prevalence of elliptical galaxies in dense environments
\citep[e.g.][]{oemler74, dressler80, cappellari11b, calvi12} also
points toward mergers or interactions as key formation mechanisms.  Elliptical
galaxies also show a diversity in their core profiles that might
result from varying merger histories, with some showing continuous
S\'{e}rsic profiles, while the most luminous ellipticals show central
light deficits \citep[e.g.,][and many others]{king66, gudehus73,
  davies83b, kormendy96, faber97, ravindranath01, emsellem07,
  kormendy09, dullo12, krajnovic15, krajnovic20}.  Others still show
central light enhancements \citep[e.g.,][]{kormendy99, cote06,
  kormendy09}.  Core light deficits may result from scouring by
supermassive black hole (SMBH) pairs \citep[e.g.,][]{begelman80,
  ebisuzaki91, milosavljevic01, thomas14} donated by merged galaxies (implying at least one major merger event), while core
enhancements might occur through central starbursts induced by
dissipative mergers \citep[e.g.,][]{mihos94, kormendy99, kormendy09}. Mergers may also be necessary to explain a seeming coupling of baryonic and dark matter kinematics in some ETGs \citep[if one assumes a constant stellar initial mass function, IMF, and with some dependence on local environment;][]{wegner12, corsini17}; otherwise, a non-universal IMF is required.

Mergers thus appear to be important for elliptical galaxy formation for most of cosmic history, but the
dominant type of merger history remains ambiguous.  Contrary to expectations from a
major-merger formation scenario \citep{nipoti03, hopkins08,
  bezanson09, naab09}, mass evolution of ellipticals has been
slow relative to size evolution \citep[e.g.,][]{trujillo07,
  buitrago08, vandokkum08, szomoru12}, and metallicity gradients in
elliptical outskirts (particularly in low-luminosity ellipticals) are
quite flat \citep[e.g.,][]{peletier90, labarbera04, denbrok11}.  Both
indicate that, for most of cosmic history, elliptical galaxies may
have grown via a complex series of minor mergers
\citep[e.g.,][]{bournaud07, naab09}. \citet{kim12} and \citet{ramos15} also showed that tidal debris in their sample of ellipticals, when present, makes up only a small fraction ($\sim10$\%) of the hosts' total mass, again implying a contribution primarily from minor mergers.  Indeed, even the most
well-behaved classical ellipticals show noticeable deviations from
simple S\'{e}rsic profiles \citep[e.g.,][]{schombert86}, suggesting
perhaps a more piecemeal evolutionary history \citep{schombert15}.  The most massive ellipticals, however, typically show less rotational support \citep{cappellari07} and more frequently have cored light deficits compared to low-mass ellipticals \citep{lauer07, kormendy09}, implying that they resulted from larger mass-ratio mergers \citep[though in the case of disk mergers, dissipational processes may be necessary to produce realistic kinematics in the resulting remnants; e.g.,][]{thomas09}.  \citet{davari17} also showed that high-redshift progenitors of massive ellipticals (red nuggets) often have surrounding disks that are mostly destroyed by $z=0.5$, which should only occur under strong gravitational perturbations.  Elliptical galaxy formation pathways thus seem to have a mass dependence.
That said, one commonality all of these varied scenarios must
reproduce is that they must somehow conspire to yield galaxies that
fall on the tight scaling relation linking their structure and
kinematics known as the fundamental plane \citep{djorgovski87,
  dressler87}.

Given the unique contribution of ETGs to the overall picture of galaxy
evolution, the S$^{4}$G Team carried out an additional \emph{Spitzer} survey to fill
in the initial S$^{4}$G's morphological gap \citep[S$^{4}$G$+$, \emph{Spitzer} project ID 10043;][]{sheth13}.  To avoid confusion, as these new observations are an extension of the S$^{4}$G, we hereafter refer to the full survey to date (the original S$^{4}$G and the new ETG extension) as S$^{4}$G+ETG.  This new survey echoed the initial
survey's methodology, now targeting an additional 465
ETGs with the \emph{Spitzer} Space Telescope's \citep{werner04}
Infrared Array Camera \citep[IRAC;][]{fazio04} in the 3.6\,$\mu$m and
4.5\,$\mu$m bands.  This new sample selection used the same criteria as
the S$^{4}$G, but rectified one of its blind spots: distance estimates
based only on radio-derived redshifts, which excluded many ETGs.

With these new data in hand, we continue the S$^{4}$G data processing
by conducting the same photometric analyses applied to the previous
2352 S$^{4}$G galaxies as part of Pipeline 3 \citep[hereafter, P3;
  see][hereafter MM2015]{munoz15}.  This includes: intensity and
surface brightness profiles, position angle and ellipticity profiles,
total magnitudes, stellar masses, measures of central concentration,
and measures of galaxy size.  MM2015 provides in their Introduction a
long list of the applications of this type of structural analysis, all
of which remain applicable to this ETG extension sample (which
includes still both disk and elliptical galaxies).  This will lead
into future analyses of this sample as well, including a detailed
study of morphology (morphological classifications already having been
completed by R. Buta, which are used throughout this paper) and
multi-component decompositions.  In an era of large-scale deep
surveys, most of which are primarily focused on probing large cosmic
volumes \citep[e.g., the Hyper-Suprime Cam Subaru Strategic Program,
  the Vera C. Rubin Observatory's Legacy Survey of Space and Time, the Roman Space Telescope, and the Euclid
  Mission;][]{aihara18, ivezic19, spergel15, laureijs10}, it is critical to have
a complete census of massive galaxies at redshift $\textrm{z}=0$ for a
current evolutionary end-state comparison.  This is ultimately what
the S$^{4}$G hopes to provide.

This paper is organized as follows.  In Section~\ref{sec:sample}, we
discuss the S$^{4}$G+ETG sample selection, observing strategy, and
give a brief overview of the data reduction steps.  In Section
\ref{sec:p3}, we describe our photometric analyses.  Section
\ref{sec:histograms} provides an overview of the photometric parameter
distributions for the entire sample, isolating ETGs for comparison.
In Section~\ref{sec:scaling}, we showcase our results via a number of
different scaling relations derived using the aforementioned
photometric parameters.  We discuss what we supply to the community
and where to find it in Section~\ref{sec:access}, and finally provide
a summary of our results in Section~\ref{sec:summary}.

This paper presents an introduction to the
S$^{4}$G+ETG sample and the quantities derived from P3.  Given the
extraordinarily broad scope of the topic, more detailed investigations
into the relations demonstrated in this paper will be published in
future papers.

\section{S$^{4}$G ETG extension sample, observations, and data reduction}\label{sec:sample}

\begin{figure*}
  \centering
  \includegraphics[scale=1.0]{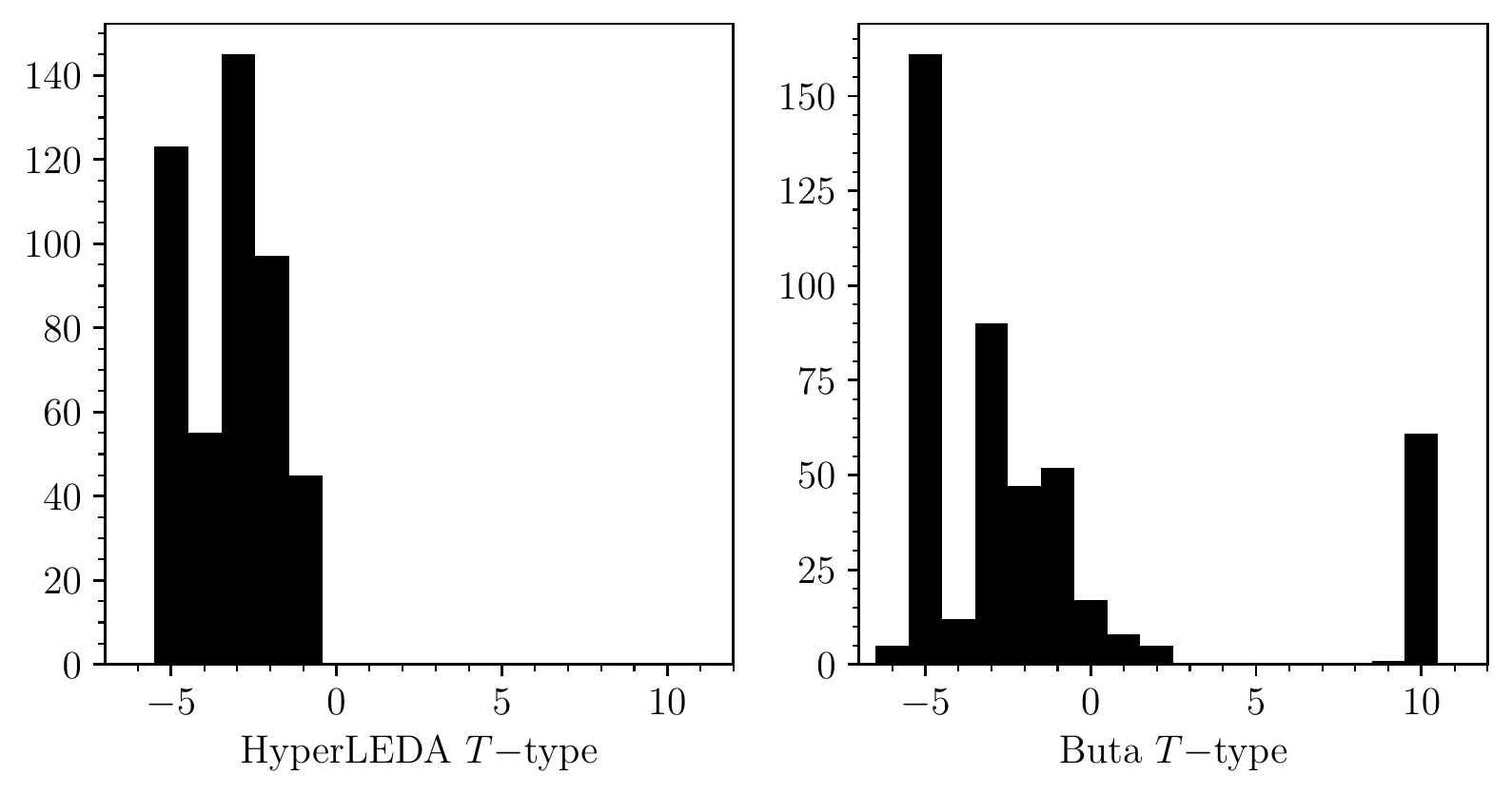}
  \caption{HyperLEDA $T-$type distribution (\emph{left} panel) and
    $T-$type distribution from 3.6 $\mu$m classifications by Ron Buta
    (\emph{right} panel) for the 465 newly-observed ETGs in the S$^{4}$G+ETG sample.  We use the Buta $T-$types for the
    remainder of this paper, given their improved
    accuracy. \label{fig:ttypes}}
\end{figure*}

Developed by the original S$^{4}$G Team, who also carried out the observations, the S$^{4}$G ETG extension sample (initially denoted S$^{4}$G$+$) followed the S$^{4}$G selection
criteria.  This includes all galaxies in the HyperLEDA database
\citep{paturel03} with radial velocities $v < 3000$ km s$^{-1}$, total
extinction-corrected blue magnitudes $m_{\rm B,corr}<15.5$, blue
isophotal diameters $D_{25}>$~1\farcm0, and Galactic latitude $\lvert
b \rvert \geq30$\degr.  Previously, only radio-derived velocities were
used \citep{sheth10}, biasing the sample toward gas-rich LTGs.  To
resolve this, the ETG extension included visual band spectroscopic redshifts as well for
galaxies with morphological type $T\leq0$.

We note, however, that this $T-$type criterion induced another
previously unnoticed bias, as it excludes an additional 425 LTGs
lacking radio velocity measurements.  We are currently addressing this
bias using ground-based $i$-band observations, which, due to the
vastly different observing strategy and data reduction procedure, will
be the subject of a future paper.

In total, S$^{4}$G+ETG contains 2817 galaxies.  Among these are 690
galaxies with HyperLEDA Hubble $T<=0$, 225 of which were already
present in the initial 2352 galaxy S$^{4}$G sample \citep{munoz15,
  salo15}.  This left an additional 465 ETGs to be subsequently
observed as part of \emph{Spitzer} proposal ID 10043 \citep{sheth13}.
Among these are the following Milky Way satellites: ESO~351-30
(Sculptor Dwarf), ESO~356-4 (Fornax Dwarf), PGC~88608 (Sextans Dwarf),
UGC~5470 (Leo I Dwarf), UGC~6253 (Leo II Dwarf), and UGC~10822 (Draco
Dwarf).  These galaxies are fully resolved into individual stars with the \emph{Spitzer} IRAC resolution,
hence cannot be analyzed using the surface photometry methods we
describe here.  Therefore, we applied the P3 procedure we describe
below to 459 ETGs total, of which $\sim$67\% and $\sim$33\% are
classified in HyperLEDA as S0s and Es, respectively.  For our results
and discussion, we focus on ETGs as a population, and therefore
include all of the ETGs in the full S$^{4}$G+ETG sample that are not MW
satellites (684 in total).

We show the distribution of the HyperLEDA $T-$types used to select this sample in
the left panel of Fig.~\ref{fig:ttypes}, demonstrating two peaks at $T
= -3$ (S0$^{-}$) and $T = -5$ (E).  Morphological classifications in
HyperLEDA come from a compilation of sources, including indirect
estimates derived from concentration indices, hence can be somewhat
unreliable.  We thus derived our own morphological classifications based on the 3.6\,$\mu$m images as
in \citet{buta15}, the distribution of which is shown in the rightmost
panel in Fig.~\ref{fig:ttypes} for comparison.  On closer visual
inspection, not all of the ETG extension galaxies are best classified as
ETGs ($T>0$ for 75 galaxies in our new classifications), so, given this discrepancy, we opt to use our own
classifications for the remainder of this paper in lieu of those
found in HyperLEDA.  We will make these new classifications available alongside all other quantities we have derived from this sample; we show a sub-sample of the full morphology table in the Appendix.

The new observations mirrored the original 2352-galaxy S$^{4}$G sample
observing strategy.  In summary, we observed all 465 new galaxies with
the \emph{Spitzer} IRAC camera in both 3.6\,$\mu$m and 4.5\,$\mu$m during
the post-cryogenic mission, with total on-source exposure times of
$8\times23.6$~s~px$^{-1}$ for channel 1 (3.6\,$\mu$m) and $8\times26.8$~s~px$^{-1}$ for channel 2 (4.5\,$\mu$m).  For galaxies with diameters $<$~3\farcm3, we split
exposures into two visits separated by at least 30 days to allow the
telescope to rotate to a new orientation, both as an aid in correcting
large-scale defects (asteroid or satellite streaks, scattered light)
and to allow for subpixel sampling.  For galaxies with diameters
$\geq$~3\farcm3, we used an 8-position dither (also split between two
visits separated by $\geq$~30 days), with 146\farcs6 offsets applied
between the maps and frame times of 30~s at each position.  For a
more detailed description of the observation strategy, see
\citet{sheth10}.

The initial data processing procedure is described in detail in
MM2015, but we provide a brief summary here.  Pipeline 1 (P1)
creates science-ready mosaics by matching the background levels of
individual exposures, then combines the images using standard
dithering and drizzle procedures \citep{fruchter02}.  The final
mosaics have a pixel scale of 0\farcs75 and units of MJy sr$^{-1}$.
The S$^{4}$G uses the AB magnitude system \citep{oke74}, hence the
magnitude zeropoint for both 3.6\,$\mu$m and 4.5\,$\mu$m is 21.0967 mag,
translating to a surface brightness zeropoint of 20.4720 mag
arcsec$^{-2}$.

Pipeline 2 (P2) produces masks of contaminating sources, including
foreground stars, background galaxies, or scattered light artifacts.
Proper masking is crucial for measuring accurate surface photometry of
extended sources, particularly at low surface brightness \citep[see,
  e.g., the Appendix of][]{watkins19}.  For the S$^{4}$G, we first
automatically generate three different masks using SExtractor
\citep{bertin96} on the 3.6 $\mu$m mosaic images, with high, medium,
and low detection thresholds in order to identify sources both far
from the target galaxy and overlapping in projection with the
galaxy.  The best of these---for example, high-threshold masks are necessary for objects embedded within galaxies to avoid masking galaxy light, while low-threshold masks help reduce the impact of low surface brightness scattered light artifacts---is then visually identified and
subsequently edited by hand \citep[using a custom IDL editing code;
  see][]{salo15} to improve masking of diffuse wings of stellar point
spread functions (PSFs) or to remove masks targeting bright parts of
the galaxy such as bar ansae or HII regions.  Once completed, we
transfer these masks to the 4.5\,$\mu$m images.  The
masks typically suffice for both detectors without changes;
nonetheless, we visually inspect the 4.5\,$\mu$m masks to ensure no such
changes are required.  These masked mosaic images are then ready
for the third pipeline, described below.

The 2D structural multi-component decomposition analysis (Pipeline 4, P4) for the
S$^{4}$G+ETG galaxies will be presented in a forthcoming article (Comer\'{o}n et al.,
in preparation). Nevertheless, we make use here of their single-component
GALFIT \citep{peng02, peng10} models where the galaxy light distribution is
fitted with a single S\'{e}rsic law. In these decompositions, the same
PSF-function (covering the inner $30$\arcsec$ \times 30$\arcsec) and the same method for
constructing the \emph{sigma-images} are used as in \citet{salo15} for the
original S$^{4}$G sample. The galaxy centers, and the isophotal ellipticity
and position angles, were fixed to values determined before
the decomposition analysis, leaving the total magnitude, effective radius
and S\'{e}rsic $n$ parameter as free parameters of the fit.  In Sect.~\ref{sec:concentration}
below we will compare the concentration derived from the $n$ parameter and the total S\'{e}rsic-fit magnitudes with those obtained directly from light profiles. We also utilize the S\'{e}rsic fits in an approximate deconvolution of the observed light profiles (Sect.~\ref{sec:radprofs}).

\section{P3: surface photometry}\label{sec:p3}

Pipeline 3 derives galaxy centroids, sky backgrounds, asymptotic
magnitudes, and isophotal properties such as surface brightness
profiles, isophotal radii, and concentration indices.  To maintain
consistency across the whole S$^{4}$G, we adhere closely to the
methodology of MM2015, which we briefly summarize here.  To improve
the accuracy of our results, however, and to account for necessary
changes in procedure given the steepness of ETG light profiles, we
have made some modifications to this methodology, which we describe in
detail below.

\subsection{Sky Measurement and Uncertainty}\label{sec:skysigma}

Accurate sky subtraction is critical for correctly measuring both
surface brightness profiles (hence isophotal radii) and asymptotic
magnitudes and any derived properties.  While the random noise of the
S$^{4}$G imaging is quite low \citep{sheth10}, the image backgrounds
themselves are often dominated by large-scale systematic sources of
uncertainty, such as background gradients or scattered light.  Because
it affords more control, for the ETG sample we chose to use the sky
subtraction and uncertainty estimation used in the S$^{4}$G Pipeline 4
analysis \citep{salo15} in lieu of that described by MM2015.

Briefly, we measure the sky in each image using 10--20
40~px~$\times$~40~px (30\arcsec$\times$30\arcsec) boxes, placed far from
the galaxy by hand to avoid scattered light, image artifacts, and
heavily masked areas.  We take the median of the median values of
unmasked pixels within each box as the image sky value
\citep[hereafter, SKY, to use the nomenclature from][]{salo15} and the
standard deviation of these medians as its associated uncertainty
(hereafter, DSKY).  Finally, we measure the sky root mean square
(hereafter, RMS) as the median of the standard deviations within each
box.  Although this method is an improvement due
to its enhanced control over sky box placement, the derived sky values deviate only slightly (showing a median difference in sky background of 0.003 MJy sr$^{-1}$ between the P3 and P4 methods) from those of MM2015
\citep[see Fig.~6 of][]{salo15}.

\subsection{Radial Profiles}\label{sec:radprofs}

\begin{figure*}
  \centering
  \includegraphics[scale=1.0]{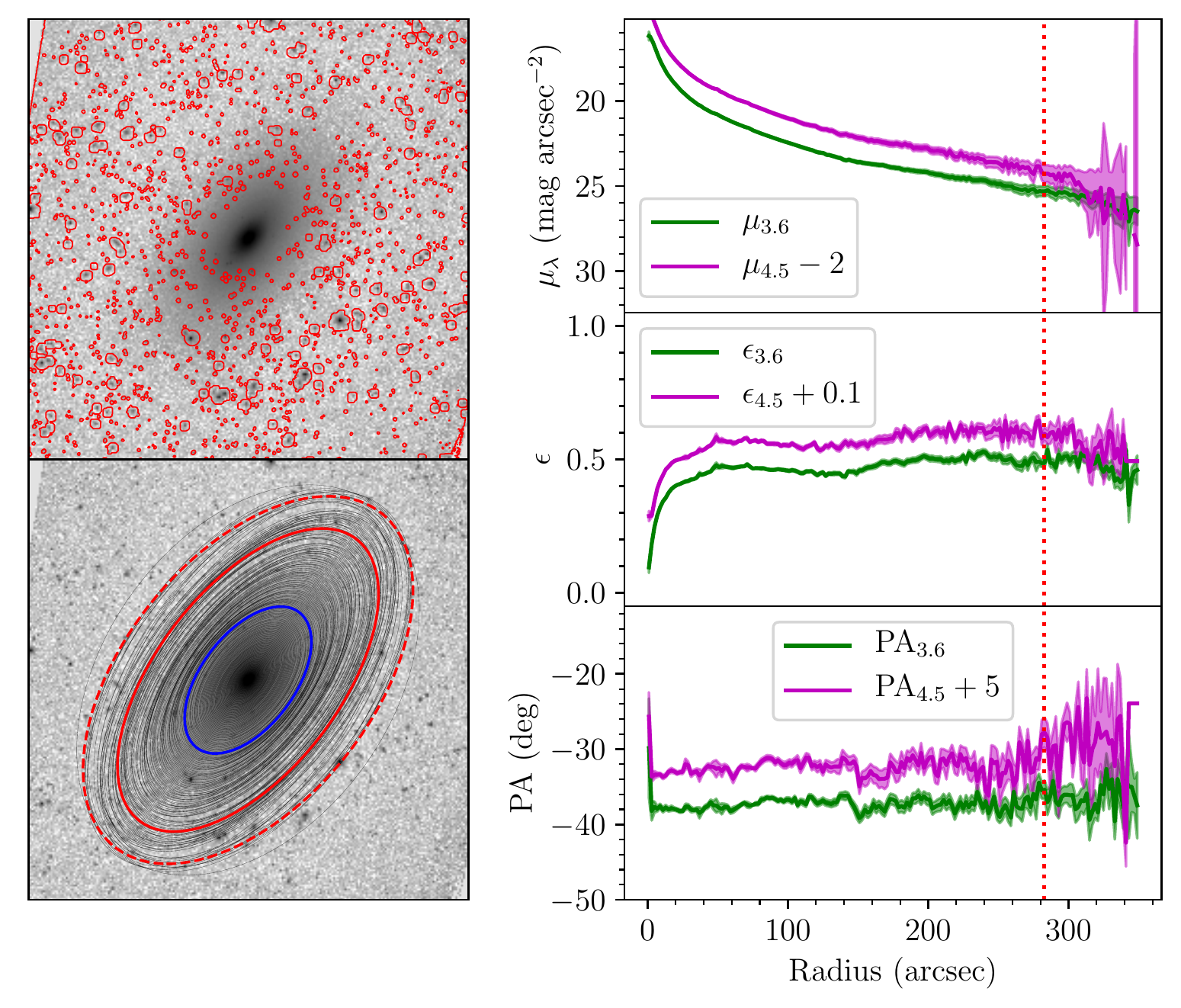}
  \caption{Showcasing examples of radial profiles, as described in
    Section~\ref{sec:radprofs}, using NGC~720.  The \emph{top left}
    panel shows the 3.6\,$\mu$m image, with masks shown via the red
    contours.  The \emph{bottom left} panel shows the same image, with
    2\arcsec \ radial bin width, variable PA and $\epsilon$ isophotal
    ellipses overlaid in black.  The blue ellipse shows \emph{B}-band
    $R_{25}$ from HyperLEDA, while the two red ellipses show our
    values of $R_{25.5, 3.6}$ and $R_{26.5, 3.6}$ (solid and dashed,
    respectively).  The \emph{rightmost} panels show radial profiles
    of, from top to bottom: surface brightness, ellipticity, and
    position angle.  The 3.6\,$\mu$m and 4.5\,$\mu$m profiles are offset
    from each other by the amounts shown in the legends to avoid
    significant overlap.  Profiles shown here are truncated in radius
    compared to those available online, to avoid clutter from
    poor-quality fits in the far outskirts. Vertical dotted red lines show $R_{25.5, 3.6}$.\label{fig:profdemo}}
\end{figure*}

The P3 analysis relies on radial surface brightness profiles and
curves of growth.  Again, we generally follow MM2015 in
constructing these radial profiles, with some deviations.  We first
use a custom IDL routine to locate the centroids of the galaxies (which we use to fix the galaxy's center coordinates), then
construct profiles using the IRAF \emph{ellipse} task \citep{jedrzejewski87,
  busko96}, varying or holding fixed the ellipticities
($\epsilon$) and position angles (PA).  We initially create three
different kinds of profiles, as MM2015:

\begin{itemize}

  \item {\it Fixed center, free $\epsilon$ and PA, radial bin width
    $\Delta \textrm{r} =6$\arcsec}: these profiles use wide radial bins for
    high S/N.  We thus use them to robustly measure the outermost
    structures of each galaxy.  From these profiles, we provide values
    of $\epsilon$ and PA at both the $\mu_{\lambda}=25.5$ and 26.5 mag
    arcsec$^{-2}$ isophotes.  If one or both of those isophotes was
    not found (generally due to high background noise), we record
    their values as -999.

  \item {\it Fixed center, free $\epsilon$ and PA, radial bin width
    $\Delta \textrm{r}=2$\arcsec}: these profiles have lower S/N and are less
    reliable in the galaxies' outskirts, but provide more detailed
    information on isophote shapes in the galaxies' inner regions.
    The value 2\arcsec \ is chosen to match the IRAC PSF at 3.6\,$\mu$m
    and 4.5\,$\mu$m, hence reflects one resolution element.  These
    profiles are most useful for assessing the properties of bars and
    other internal structures.  We show an example of these profiles
    in Fig.~\ref{fig:profdemo}.

  \item {\it Fixed center, fixed $\epsilon$ and PA, radial bin width
    $\Delta \textrm{r}=2$\arcsec}: these profiles have parameters fixed to
    those of the $\mu_{\lambda}=25.5$ mag arcsec$^{-2}$ isophotes
    measured using the $\Delta r=6$\arcsec profiles described above.
    With fixed isophote shapes, these profiles are most useful for
    measuring disk scalelengths, break radii, and for performing 1D
    bulge-disk decompositions.  With these profiles we measure the
    curves of growth, from which we derive asymptotic magnitudes (Sect.~\ref{sec:p3mags}) and
    derivative values thereof.  We note that the use of these profiles relies on the assumption that the $\mu_{\lambda}=25.5$ mag arcsec$^{-2}$ isophote shapes accurately reflect those of the disk's underlying structures, such as spiral arms; more detailed analysis may require a more careful outer isophote selection to ensure this is true.

\end{itemize}

\begin{figure}
  \centering
  \includegraphics[scale=1.0]{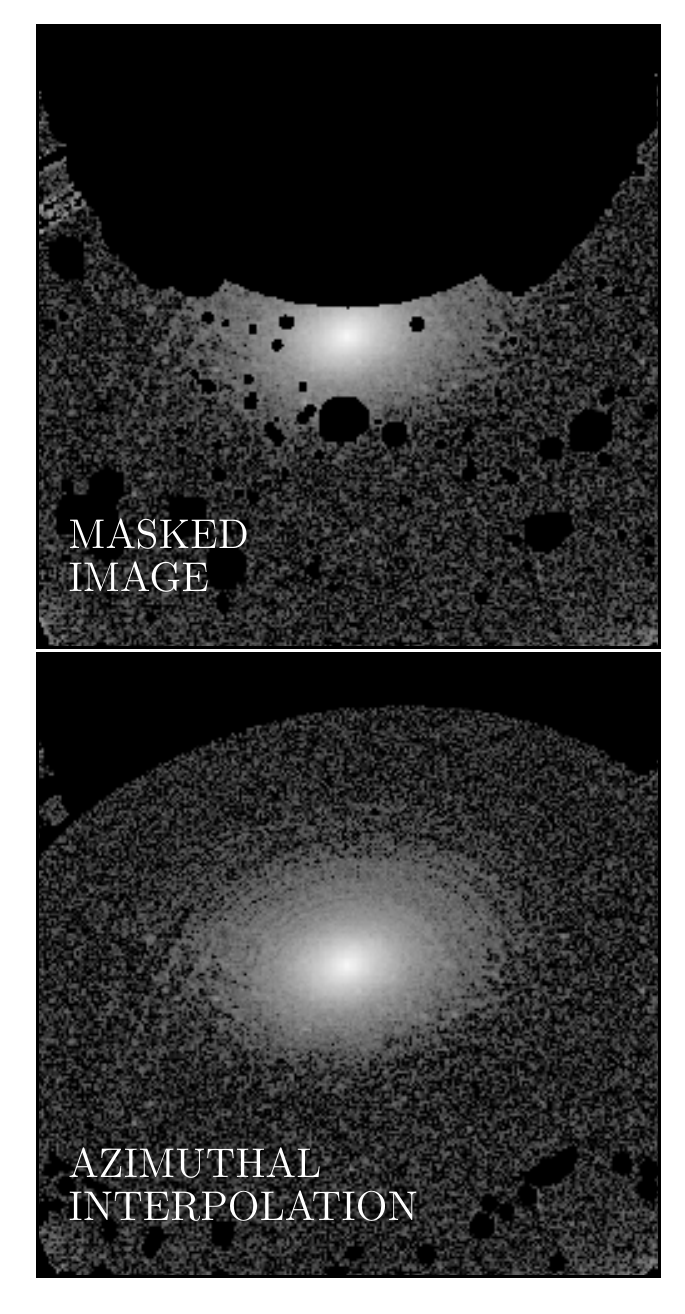}
  \caption{Demonstrating interpolation across masks for a heavily masked
    galaxy (IC~3381).  The top panel shows the masked image, while the bottom panel shows the same image but with the galaxy's light interpolated azimuthally across the masked pixels (see Section~\ref{sec:radprofs}).  \label{fig:interp}}
\end{figure}

Generating these profiles required multiple steps.  First, IRAF's
\emph{ellipse} task simply ignores masked pixels, biasing the curves of
growth.  While there are options to estimate masked flux within IRAF, because many of the sample galaxies have spiral arms, bars, and other such symmetrical structures, we opt instead to estimate masked flux through our own interpolation algorithm, demonstrated in Fig.~\ref{fig:interp}.  For this ETG sample only, we interpolate across masks
using the galaxy's azimuthal light profile, measured in one-pixel-wide
radial bins from the center outward.

We demonstrate this using the galaxy IC~3381, which is contaminated by a bright nearby star and so has been heavily masked.  First, in each radial bin we measured the mean and standard deviations
within 10-degree-wide azimuthal bins.  As a first pass, we linearly
interpolated this azimuthal profile across the masked pixels and
recorded the median standard deviation among azimuthal bins
(hereafter, $\sigma_{\rm rad}$).  In order to preserve any periodicity
induced by, e.g., bars, we then decomposed the resulting interpolated
azimuthal profiles into Fourier modes and used the first $m=20$
Fourier modes to more accurately reconstruct the flux across masked
pixels.  We then added Gaussian noise to the interpolated values,
distributed as $\sim\mathcal{N}(0,\sigma_{\rm rad})$.  We show an example
of this interpolation for the same example galaxy, IC~3381, in the
bottom panel of Fig.~\ref{fig:interp}.

With the missing flux now preserved, we constructed radial profiles by
running \emph{ellipse} twice: first, using the software described by
\citet{salo15} to select sensible starting values for semi-major axis
length (SMA), PA, and $\epsilon$, then automatically running
unconstrained \emph{ellipse} fits, in the three manners described
above, using these starting parameters.  We chose not to constrain the
automated fits in order to measure profiles to as low a surface
brightness level as possible, but the outskirts of these unconstrained
profiles evidently showed large swings in $\epsilon$ and PA where the
fits were dominated by noise.  Therefore, we derived maximum radii for
the fits using the point at which the values of $\epsilon$ or PA in
the $\Delta r=6$\arcsec \ profiles began to show regular
3$\sigma$ divergences from their mean values as measured beyond the
$\mu_{3.6} = 25.5$ mag arcsec$^{-2}$ isophote (or, failing this, where the
\emph{ellipse} uncertainty in these values, beyond $\mu_{3.6} = 25.5$ mag arcsec$^{-2}$, first
was recorded as INDEF).  We chose as the maximum fitting radius
1.2 times this value, which ensures in most cases that profiles
reach at least the $\mu_{3.6} = 25.5$ mag arcsec$^{-2}$ isophote if the image's noise
limit is below this value, even if the isophotal fits there are poor.
We then re-fit all profiles out to these maximum radii for each
galaxy, checking the robustness of each by overplotting the fitted
isophotes on the 3.6\,$\mu$m images of each galaxy.  Where the fitting
failed (evident when the isophotal shapes were constant even at small
radii), we simply re-ran the fits with different starting SMA until they
succeeded.  Due to the ETG sample galaxies' often simple morphology,
we needed only re-run in this manner one or two times.

\begin{figure*}
  \centering
  \includegraphics[scale=1.0]{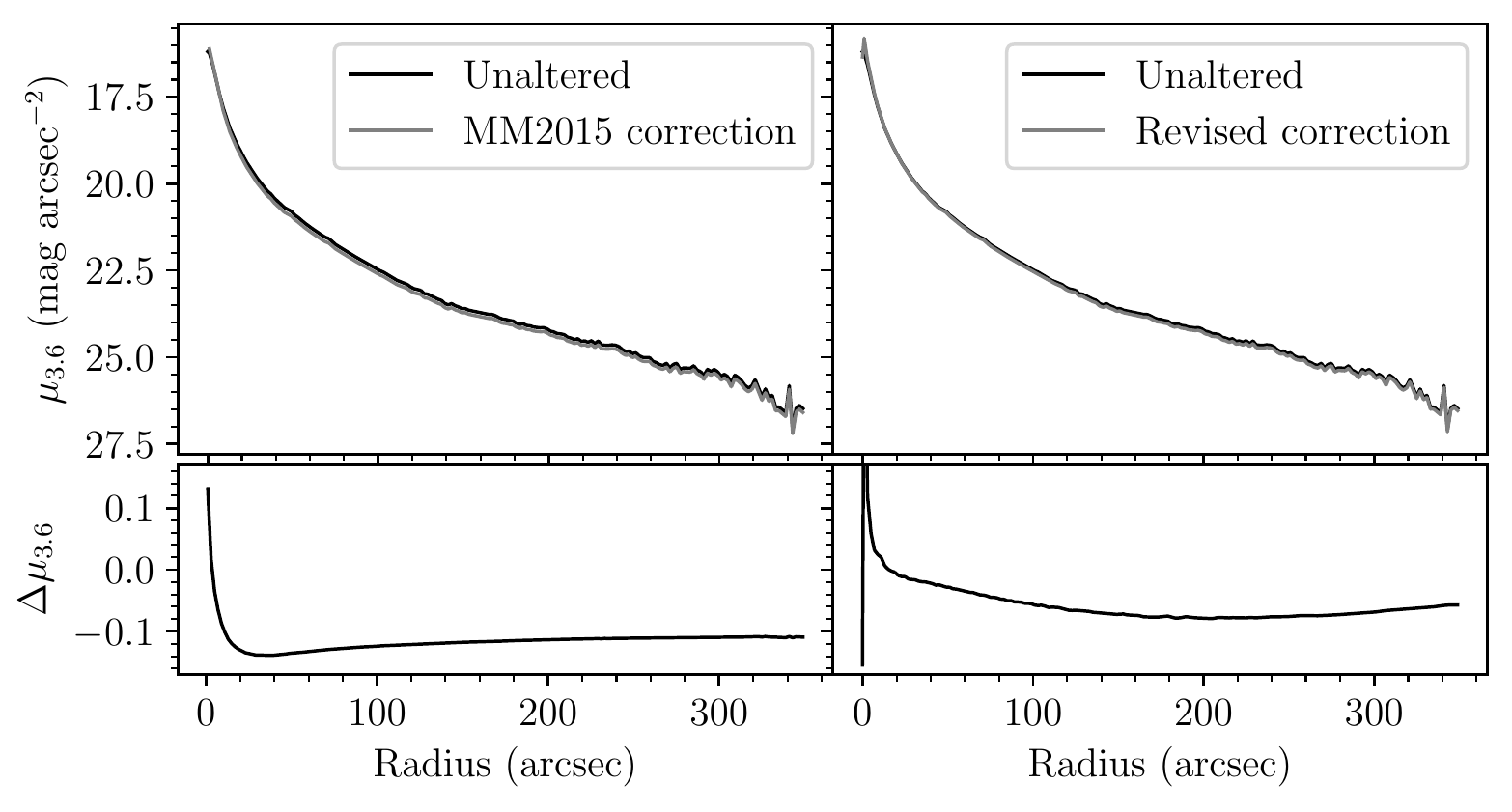}
  \caption{\emph{Left} panels show the effect of the aperture
    correction described by MM2015.  Black shows the unaltered surface
    brightness profile for NGC~720, while the gray line shows the same
    profile after the aperture correction is applied.  The
    \emph{bottom left} panel shows the difference between the two
    (unaltered$-$corrected).  \emph{Right} panels are the same as the
    left panels, but show our more robust aperture correction using
    GALFIT models (see Section~\ref{sec:radprofs}).  We derived all
    magnitudes and related quantities using the MM2015 correction, to
    maintain consistency with that study, however for any future
    analysis using our datasets we recommend instead using the revised
    correction.  \label{fig:corrcomp}}
\end{figure*}

As in MM2015, we provide these fits as ASCII tables (Sect.~\ref{sec:access}).  We provide the same parameters as in MM2015, including harmonic deviations from
perfect ellipses and uncertainty estimates.  We also include photometric
profiles corrected for PSF scattering in the same manner as in MM2015,
who used correction factors taken from the IRAC Instrument
Handbook\footnote{\url{https://irsa.ipac.caltech.edu/data/SPITZER/docs/irac/iracinstrumenthandbook/}}.
Briefly, the light from any extended source (e.g., galaxies) in IRAC
is a convolution of the astrophysical light distribution and the
extended IRAC PSF.  This means that, in a given photometric aperture,
light from the astrophysical source is both scattered out of the
aperture into the PSF wings, and scattered into the aperture from
astrophysical sources outside, including other regions of the same
galaxy.

Given the complexity of the IRAC PSFs, modeling this effect requires
using model galaxies convolved with the measured PSF for each IRAC
band.  From the IRAC Instrument Handbook, PSF correction factors were measured
using single-S\'{e}rsic profile model ETGs with sizes ranging up to
200\arcsec.  For the total flux $F$ within an aperture, these are
provided in the following form:
\begin{equation} \label{eq:appcorr}
  F_{\rm corr}(r_{\rm eq}) = F_{\rm obs}(r_{\rm eq}) \times
  (Ae^{-r_{\rm eq}^{B}} + C)
\end{equation}
where $r_{\rm eq} = \sqrt{ab}$ is the equivalent elliptical aperture
radius in arcseconds, and the constants $A$, $B$, and $C$ are equal to
0.82, 0.37, and 0.91 for 3.6\,$\mu$m and 1.16, 0.433, and 0.94 for
4.5\,$\mu$m.  Likewise, as shown by MM2015, the correction for the
surface brightness within an elliptical aperture $I_{\rm obs}$ can be
made via a series expansion of Eq.~\ref{eq:appcorr}:

\begin{equation} \label{eq:sbcorr}
  I_{\rm corr}(r_{\rm eq}) = I_{\rm obs}(r_{\rm eq}) \times
  (Ae^{-r_{\rm eq}^{B}} + C) - ABr_{\rm eq}^{B-2}e^{-r_{\rm eq}^{B}}F_{\rm obs}(r_{\rm
    eq})/(2\pi)
\end{equation}

These correction factors are nominally accurate to within $\sim$10\%,
however they may also contain systematic errors given the limited
kinds of models used to derive them.  We therefore now also include a
second correction for each galaxy in our tables.  Using the GALFIT S\'{e}rsic fit parameters from the fortcoming P4 analysis, we construct both the theoretical non-convolved and the
PSF-convolved 2D model images, MODEL and PSF~$\otimes$~MODEL,
respectively. The convolution is made with the same symmetrized PSF as
was used in \citet{comeron18} obtained from the data in \citet{hora12}. To make sure the PSF has sufficient coverage, we continued the PSF tails so that the final coverage was $450$\arcsec$\times450$\arcsec.  We then performed IRAF \emph{ellipse} fits to both the
non-convolved and convolved model images, using the same ellipticity
and PA as for the observed profile. We denote these profiles as
$I_{\rm non-convolved}$ and $I_{\rm convolved}$, giving an estimate 
\begin{equation}\label{eq:newcorr}
    I_{\rm corr}(r_{\rm eq}) = f I_{\rm obs}(r_{\rm eq}) \times (I_{\rm
    non-convolved}(r_{\rm eq})/I_{\rm convolved}(r_{\rm eq}))
\end{equation}
Here $f$ is the infinite aperture
correction used to correct flux of extended sources \citep[$f=0.944$
  for 3.6\,$\mu$m and 0.937 for 4.5\,$\mu$m;][]{reach05}.  The correction
for the total flux $F$ has the same form.

Note that in this analysis we ignore the small difference in the
orientation parameters used in P4 \citep{salo15} compared to our current
ones: they were also estimated from IRAF \emph{ellipse} profiles, but instead
of using a fixed isophotal level, correspond to the visually selected
outer isophote.

Fig.~\ref{fig:corrcomp} shows example profiles using each kind of
aperture correction, for comparison.  We note here that, for consistency across the full S$^{4}$G+ETG sample, we derive
all P3 values using the aperture correction from MM2015; however,
because it is tailored to specific galaxies, we recommend any future
analyses using these profiles are done with the revised corrections.
The choice has no noticeable impacts on the results presented in this
paper; it primarily results in very slight deviations in our derived
magnitudes, well below the photometric uncertainty.  It does not
impact any of our broader scientific conclusions. We make these new corrections available along with our data release.

Finally, we correct all magnitudes and surface brightnesses for
Galactic extinction using the extinction tables from
\citet{schlafly11}.  These corrections are typically extremely small ($<
0.01$ mag), but we include them in the interest of accuracy.

\subsection{Asymptotic Magnitudes}\label{sec:p3mags}

\begin{figure*}
  \centering
  \includegraphics[scale=1.0]{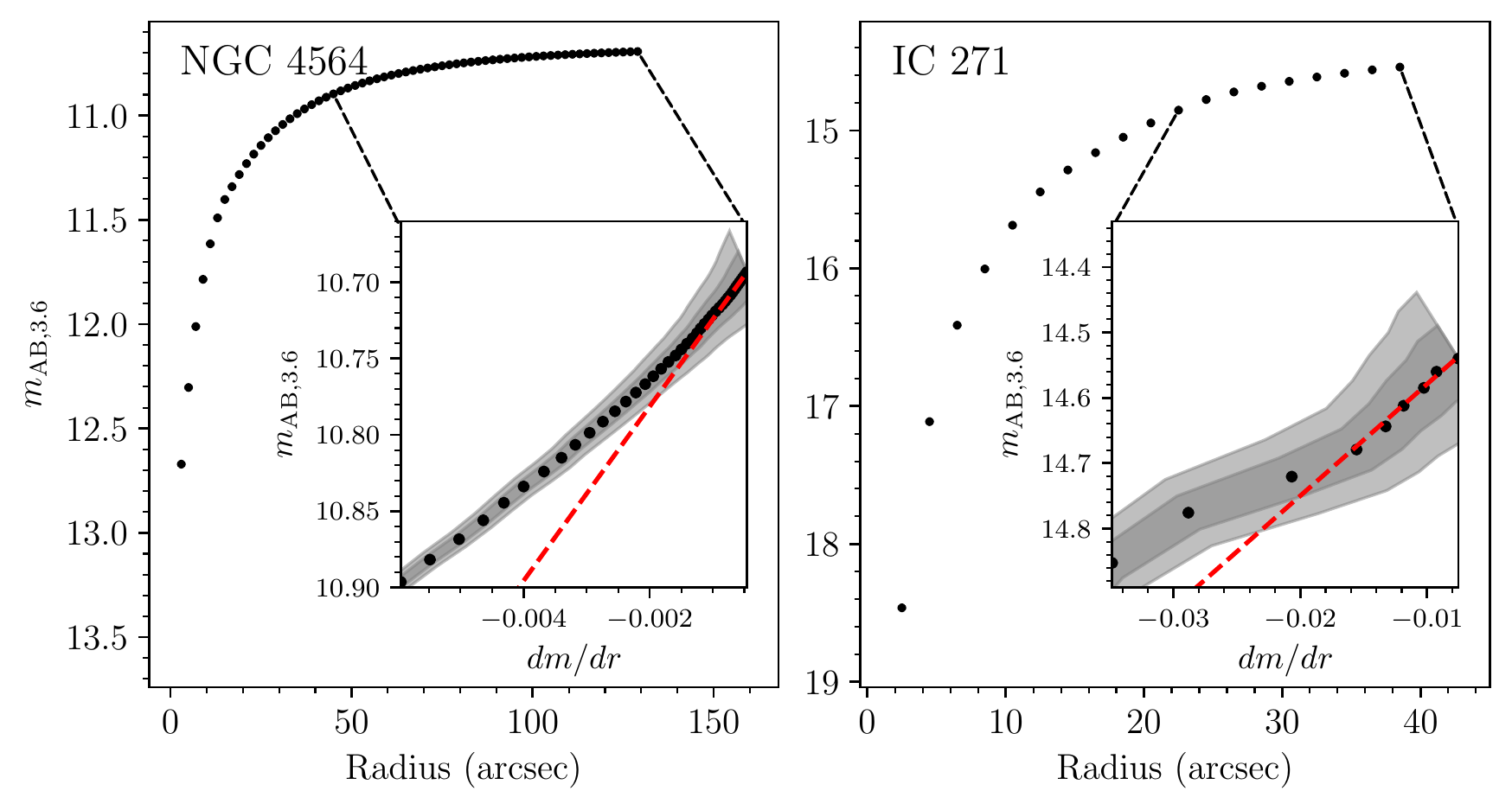}
  \caption{Example curves of growth for two galaxies, one bright with
    a large angular size (NGC~4564, \emph{left}), the other faint with
    a small angular size (IC~271, \emph{right}).  Black points in the
    main panels show the total magnitudes enclosed within each radius
    (bin width 2\arcsec).  The inset panels show the local gradients
    in the outskirts of the curves of growth.  Red dashed lines show
    fits to the linear parts of these gradients chosen by eye; the
    $y$-intercepts of these red lines are the asymptotic magnitudes.
    Gray shaded regions show the $1\sigma$ (dark gray) and $2\sigma$
    (light gray) perturbations to the curves of growth from adjusting
    the sky subtraction (see Section
    \ref{sec:p3mags}).  \label{fig:cog}}
\end{figure*}

To derive asymptotic magnitudes, we follow MM2015, using each galaxy's
curve of growth (hereafter, c.o.g.).  We derive a given galaxy's
c.o.g. as the cumulative sum of the galaxy's flux within elliptical
apertures of constant $\epsilon$ and PA, grown in semi-major axis in
steps of 2\arcsec \ and corrected for PSF effects using
Eq.~\ref{eq:appcorr}.  Given the depth of the S$^{4}$G+ETG, for most
galaxies in our sample the relationship between the local gradient and
the magnitude enclosed within each elliptical aperture becomes roughly
linear near the outskirts of the c.o.g.  We therefore fit a line
between the local gradient and the magnitude in this linear regime;
the $y$-intercept of this line is, by definition, the asymptotic
magnitude.

Fig.~\ref{fig:cog} shows examples of the c.o.g (top panels) and linear
fits (inset panels) for two galaxies with different radial extents and
luminosities.  Particularly for galaxies with small angular sizes
(hence few points in their surface brightness profiles) and low
luminosities (low S/N at all radii), it is not always clear whether we
are indeed reaching the flat part of the c.o.g.  Additionally, because
many galaxies in our sample are ellipticals with steep luminosity
profiles (S\'{e}rsic $n>2$), the slopes become approximately
linear only at very large radii.  For each galaxy, we therefore
hand-picked the limits within which to fit these lines.

This contributes some uncertainty into the final derived magnitudes,
but this uncertainty is quite small compared to the systematic
uncertainty induced by the sky subtraction. Because the c.o.g. relies on summing flux in
concentric apertures, a slight under- (over-)subtraction
of the sky will result in a positive (negative) gradient
in the outermost areas of the galaxy disk, where the sky
level becomes comparable to the galaxy disk brightness,
thus potentially altering the $y$-intercept used to derive
the magnitude.  To derive the uncertainties associated with the sky
subtraction, we subtract from each c.o.g. 1000 different sky levels
randomly sampled from a normal distribution $\sim\mathcal{N}($SKY,
DSKY$)$ (see Section~\ref{sec:skysigma}).  We take the $1\sigma$
amplitude of the variation in the final derived asymptotic magnitudes
as this sky uncertainty, using our initial hand-picked fitting limits
to derive $y$-intercepts for every perturbation.  The gray shaded
regions in the sub-panels of Fig.~\ref{fig:cog} show the effects of
these perturbations, with the dark and light gray regions showing the
$1\sigma$ and $2\sigma$ variations on the c.o.g., respectively.  We
provide these systematic uncertainties in addition to the photometric
uncertainties.

To measure absolute magnitudes, we derived distances (again following
MM2015) by either taking the average of all redshift-independent
distance estimates for each galaxy from the NASA Extragalactic
Database (NED)\footnote{The NASA/IPAC Extragalactic Database (NED) is
  funded by the National Aeronautics and Space Administration and
  operated by the California Institute of Technology.}, if available
(60.4\% of the new sample), or else by taking the average of all redshift
distances available from HyperLEDA \citep{paturel03}.

We derive stellar masses using these distances and asymptotic
magnitudes following the dust-corrected \citep{meidt12} expression given by Eq.~8 of
\citet{querejeta15}:
\begin{equation}\label{eq:mass}
  \mathcal{M}_{*} =
  10^{8.35} \times F_{3.6}^{1.85} \times F_{4.5}^{-0.85} \times D^{2}
\end{equation}
where $\mathcal{M}_{*}$ is in solar masses, $F_{\lambda}$ are the
total fluxes in Jy, and $D$ is the distance to the galaxy in Mpc.
This expression was derived using the independent component analysis technique
demonstrated by \citet{meidt12}.  This differs from MM2015, who used
the value $\mathcal{M}_{*}/L_{3.6} = 0.56$ from \citet{eskew12}.  We
found the impact of this choice to be almost unnoticeable.  One might
expect some noticeable differences given that \citet{querejeta15} had
few ETGs available for their calibration, but ETGs are also relatively
dust-poor compared to LTGs \citep[e.g.,][]{skibba11}, hence any dust
corrections should be, and seemingly are, slight.

In the following cases, the target galaxy overlaps strongly in
projection with another, much brighter object: ESO~400-30, NGC~3636,
NGC~4269, NGC~5846A, and PGC~93119.  The c.o.g. method is incapable of
correctly measuring asymptotic magnitudes for such galaxies, as their
flux cannot be disentangled from that of their brighter neighbors.  We
will therefore obtain more accurate magnitudes for these galaxies at a
future date through 2D decompositions (Pipeline 4), taking into
account both the galaxy and its neighbor simultaneously.  For now, we
urge caution regarding the magnitudes we provide for these particular
galaxies.

\subsection{Concentration Indices}\label{sec:concinds}

The radial distribution of light---the light concentration---is easily
measurable and, when used in conjunction with galaxy asymmetry and
clumpiness \citep{conselice03}, can serve as a quantitative
alternative to visual morphological class \citep[e.g.,][]{hubble26,
  morgan57, morgan58, okamura84, kent85, abraham94, bershady00,
  conselice03, conselice06, taylor07, holwerda14}.  In elliptical
galaxies, the light concentration---being explicitly related to the
S\'{e}rsic index $n$ \citep[e.g.,][]{bershady00,
  trujillo01}---correlates with galaxy luminosity
\citep[e.g.,][]{caon93, conselice03, ferrarese06, mahajan15, graham19}
and thereby with stellar mass, suggesting that the light profile is
somehow indicative of an elliptical galaxy's evolutionary history.  It
is thus a rather useful first measure of a given galaxy's intrinsic
properties.

Again following MM2015, we measured the following two concentration
parameters: $C_{31}$ \citep{devau77} and $C_{82}$ \citep{kent85},
defined as: 
\begin{equation}\label{eq:c31}
  C_{31} = R_{75}/R_{25}
\end{equation}
\begin{equation}\label{eq:c82}
  C_{82} = 5\log\,(R_{80}/R_{20})
\end{equation}
where $R_{x}$ is the radius containing $x\%$ of the total luminosity of the
galaxy \citep[see also][]{fraser72}.  As in MM2015 and
\citet{bershady00}, to avoid assumptions about the shapes of the light
profiles, we measure $R_{x}$ from the c.o.g.  Also, in constrast with,
e.g., \citet{bershady00}, we extrapolate the galaxies' total
luminosities out to infinity rather than measuring them within set
apertures.

\subsection{Galaxy Sizes}

ETG mass--size relations are of considerable interest,
particularly given their connection to ETG formation and evolution
mechanisms \citep[e.g.,][]{daddi05, trujillo07, vandokkum08,
  vanderwel14}.  Recently, the mass--size relation for galaxies
generally has come under additional scrutiny due to the recent
rediscovery of sometimes abundant populations of extremely extended
but low-mass galaxies in clusters, originally identified by \citet{binggeli85} in the Virgo Cluster, now broadly referred to as ultra-diffuse galaxies \citep[e.g.,][]{vandokkum15, koda15, mihos15, iodice20}.  Depending on the size metric used,
such galaxies either appear as extreme outliers in the mass--size
relation, or show little difference from other dwarf galaxies
\citep{chamba20}, leading to the argument that a more physically
motivated definition of galaxy size (based on, e.g., star formation
thresholds) is called for \citep[and references therein]{trujillo20}.

For our ETGs, we provide 3.6 $\mu$m and 4.5\,$\mu$m half-light (or
effective) radii, as well as two isophotal radii: $R_{25.5}$ and
$R_{26.5}$, denoting the radii at which the galaxy's surface
brightness profiles reach 25.5 and 26.5 AB mag arcsec$^{-2}$,
respectively.  As NIR imaging is such a close tracer of stellar mass,
the effective radius is nearly equivalent to the half-mass radius,
while $R_{25.5, 3.6}$ and $R_{26.5, 3.6}$ denote the $\sim 3.6\,
{\mathcal M}_{\odot}\,{\rm pc}^{-2}$ and $\sim 1.5\,\mathcal{M}_{\odot}\,{\rm pc}^{-2}$ surface
mass density contours, respectively \citep[using the $\mathcal{M}/L$ calibration
  from][]{eskew12}.  We measure all radii along the galaxies' major
axes, i.e., all radii are deprojected.  We also derive isophotal radii
both from the raw and inclination-corrected surface brightness profiles.

We measure the half-light radii in the same manner as the
concentration parameters described above, taking them as the radii at
which the c.o.g. reaches 50\% of the galaxies' total flux (converted
from the asymptotic magnitudes).  We measure the isophotal radii
directly from the $\Delta r=6$\arcsec surface brightness
profiles described in Section~\ref{sec:radprofs} (i.e., using elliptical apertures), to take advantage of
the improved S/N.  Given these profiles' rather coarse sampling, we
interpolate the exact radii between the two bins that enclose the
desired isophote.  We record as well the full isophotal parameters at
each radius (PA and $\epsilon$), interpolated in the same manner.  If
the background noise precluded our ability to find either a 25.5 or
26.5 mag arcsec$^{-2}$ isophote for a given galaxy, we recorded these
values as -999 (for 3.6\,$\mu$m, 20 cases and 177 cases, respectively;
for 4.5\,$\mu$m, 178 and 204 cases, respectively).  For the remainder of
this paper we focus on the $\mu_{3.6} = 25.5$ isophotal radii, given
their higher S/N.

All of these values---magnitudes, concentrations, sizes, and
associated uncertainties (including SKY, DSKY, and RMS)---will be made
available in table form on the NASA/IPAC Infrared Science Archive (IRSA).  We now showcase them,
starting with a summary of observed parameters across the S$^{4}$G+ETG
population, then moving to a variety of NIR scaling relations.

\section{Overview: general trends}\label{sec:histograms}

\begin{figure*}
  \centering
  \includegraphics[scale=1.0]{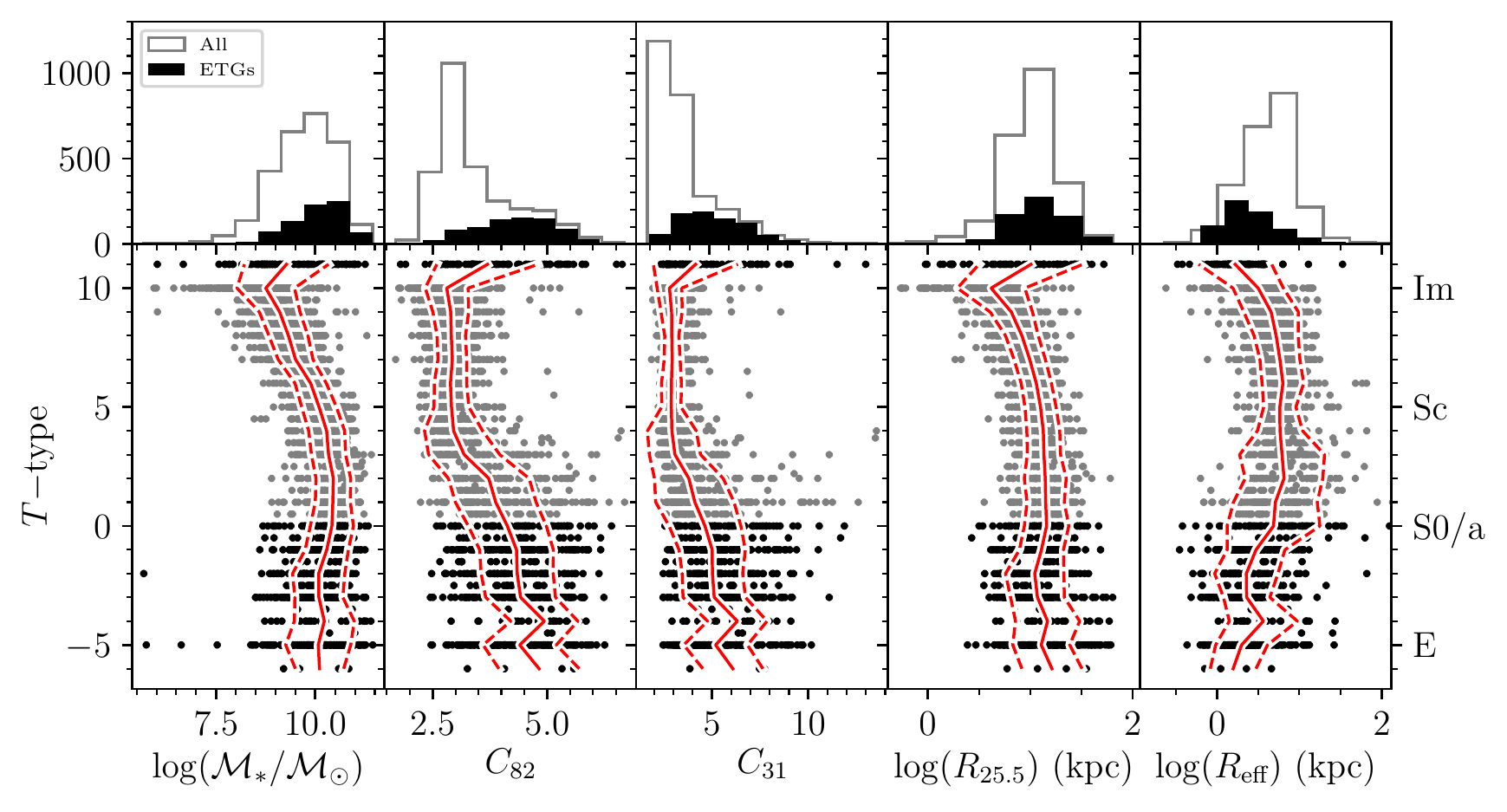}
  \caption{Distributions of the photometric parameters discussed in
    Section~\ref{sec:p3} (\emph{top}), and parameter values plotted
    against morphological $T-$type (\emph{bottom}).  From left to right,
    the parameters shown are: decimal logarithm of stellar mass (in
    solar units); concentration parameter $C_{82}$ (equation
    \ref{eq:c82}); concentration parameter $C_{31}$ (equation
    \ref{eq:c31}); logarithm of the $\mu_{3.6, \textrm{AB}}=25.5$\,mag\,
    arcsec$^{-2}$ isophotal radius in kiloparsecs; and logarithm of
    the half-light ($\sim$half-mass) radius in kiloparsecs.  Solid red
    lines show the median value of each parameter measured in bins of
    $\Delta T=1$, while dashed red lines show the associated first
    and third quartiles.  Unfilled histograms and gray points show
    distributions for all the galaxies, while filled histograms and black
    points show distributions for galaxies with $T \leq 0$ and $T = 11$
    (ETGs).  \label{fig:hists}}
\end{figure*}

\begin{table}
  \caption{Numerical $T-$types summary}
  \label{tab:ttypes}
  \centering
  \begin{tabular}{lc}
    \hline\hline
    Stage & Numerical index ($T-$type) \\
    \hline
    cE & -6 \\
    E & -5 \\
    E$^{+}$ & -4 \\
    S0$^{-}$ & -3 \\
    S0$^{0}$ & -2 \\
    S0$^{+}$ & -1 \\
    S0/a & 0 \\
    Sa & 1 \\
    Sab & 2 \\
    Sb & 3 \\
    Sbc & 4 \\
    Sc & 5 \\
    Scd & 6 \\
    Sd & 7 \\
    Sdm & 8 \\
    Sm & 9 \\
    Im & 10 \\
    dE, dS0, dSph & 11 \\
  \hline
  \end{tabular}
  \tablefoot{References: \citet{devau91}, \citet{buta15}}
\end{table}

The top panels in Fig.~\ref{fig:hists} show the distributions of
stellar mass (luminosity), concentration, and size for the entire
S$^{4}$G+ETG sample, with ETGs ($T\leq 0$ and $T=11$; filled histograms) also displayed separately from the whole sample (unfilled histograms). The bottom panels show
individual galaxies' parameter values as a function of morphological
$T-$type.  ETGs are plotted as black points, with the full population
plotted underneath in gray.  Red solid lines show the median trends,
with red dashed lines showing the first and third quartiles.

We use $T-$types from \citet{buta15} for the original S$^{4}$G sample,
and new classifications by R. Buta for the ETG extension galaxies.
These classifications follow the Comprehensive de Vaucouleurs revised
Hubble-Sandage \citep[CVHRS;][]{buta07} system, which follows the de
Vaucouleurs revised Hubble-Sandage system \citep{devau59} but is
modified to include many additional morphological features.  For quick
reference, we reproduce the integer $T-$types associated with each
broad morphological classification from \citet{buta15} in Table
\ref{tab:ttypes}.  A sample of the full classification table can be found in Table \ref{tab:app_ttypes}.

We focus here on the ETG population, in its context within the general
population of nearby galaxies.  We replicate a few previous
observations.  For example, the high-mass end of the galaxy stellar mass function tends to
be dominated by ETGs at low redshift \citep[e.g.][]{baldry04,
  pozzetti10, calvi12}, and in our sample the bootstrap median mass (measured as the median of the medians of 10000 randomly resampled mass distributions, with replacement) of
ETGs is indeed higher than that of either LTGs or the total galaxy population (ETGs and LTGs combined):
$\log\,(\mathcal{M}_{*} / \mathcal{M}_{\odot}) = 10.15 \pm 0.03$ for ETGs only, vs. $9.63 \pm 0.03$ for LTGs only, and $9.78 \pm 0.02$ for all the galaxies, where uncertainties are the $1\sigma$
bootstrap uncertainty (measured as the standard deviation of the bootstrapped median values described above) on the median (we use the bootstrap medians and uncertainties
in all following comparisons).  Bright ETGs also have the highest concentration ($C_{82} = 4.40 \pm 0.04$ vs. $3.12
\pm 0.02$ for the whole sample, and $C_{31} = 5.13 \pm 0.09$ vs. $3.13 \pm
0.02$).  Nearby ETGs are also larger in isophotal size
compared to either LTGs or the whole galaxy population (median $R_{25.5} = 12.74
\pm 0.36$ kpc, vs. $R_{25.5} = 9.91 \pm 0.16$ kpc for LTGs, and $R_{25.5} = 10.47 \pm 0.15$ kpc for all the galaxies);
combined with the previous point, this results in them having smaller
effective radii on average \citep[median $R_{\rm eff} = 2.30 \pm 0.11$\,kpc
  vs. $R_{\rm eff} = 4.47 \pm 0.08$\,kpc for all the galaxies; a similar trend
  was also noted by][]{jarrett03}.

More granular trends are also visible.  S0s ($-3 \leq T \leq 0$) are
undermassive compared to early-type spirals (Sa, Sab, Sb), with median
$\log\,(\mathcal{M}_{*} / \mathcal{M}_{\odot}) = 10.23 \pm 0.03$
compared to $\log\,(\mathcal{M}_{*} / \mathcal{M}_{\odot}) = 10.39 \pm
0.01$.  S0s are closer in mass to Sc galaxies ($\log\,(\mathcal{M}_{*} /
\mathcal{M}_{\odot}) = 10.10 \pm 0.01$).  This too was shown
previously, by \citet{vandenbergh98}.  Most of this low-mass tendency
is driven by early-type S0s (S0$^{-}$ and S0$^{0}$), which have median
mass $\log\,(\mathcal{M}_{*} / \mathcal{M}_{\odot}) = 10.09 \pm 0.05$,
compared to $\log\,(\mathcal{M}_{*} / \mathcal{M}_{\odot}) = 10.36 \pm
0.01$ for S0$^{+}$ and S0a galaxies, making Sc and early-type S0s the
most similar in mass.  S0s are also smaller on average in isophotal
radius ($R_{25.5} = 12.65 \pm 0.38$\,kpc) than all but Sc ($R_{25.5} =
12.62 \pm 0.35$\,kpc) and dwarf galaxies ($T\geq 8$, with median
$R_{25.5} = 6.49 \pm 0.16$\,kpc), although S0s are more concentrated
(hence have smaller $R_{\rm eff}$) than such dwarf disks.  Again, this
tendency toward small size among S0s is driven by the early-type S0s
($R_{25.5} = 11.30 \pm 0.72$\,kpc, compared to $R_{25.5} = 13.36 \pm
0.63$\,kpc for late-type S0s).

Many ETGs classified as $T=11$, which are morphologically dwarf galaxies, have masses of $\log\,(\mathcal{M}_{*} / \mathcal{M}_{\odot}) > 10.0$, suggesting that they are not best classified as dwarfs in terms of their masses.  The reason for this is unclear; either their morphological classifications do not match their high stellar masses, or the distances we use to derive their masses are incorrect.  The median distance to these $T=11$ galaxies is 17.6~Mpc; at this distance, a dwarf galaxy with a diameter $D_{25} < 1$~kpc would have a angular size of $< 12$\arcsec, hence should have been excluded from the sample via the criterion of $D_{25} > 1^{\prime}$ (Sec. \ref{sec:sample}), which biases the sample against dwarf galaxies even within the selected distance constraints ($D \lesssim 40$~Mpc).  However, because these $T=11$ galaxies tend to follow the rest of the ETGs in each scaling relation we examine, we will defer a detailed discussion of them for future work.

On average, LTGs and ETGs show clear disparities in concentration,
as noted already by \citet{devau48}.  Specifically, this
disparity sets in for galaxies with $T\leq 3$ (Sb), with both
concentration parameters rising monotonically with decreasing T below
this value \citep[this trend is visible already with the smaller ETG
  sample from][e.g., their Fig.~9]{munoz15} save for a slight dip
among early-type S0s ($-3 \leq T \leq -2$).  \citet{diazgarcia16}
found similar morphological separation using stacked profiles, which
was echoed in the central stellar contribution to the circular
velocity curves. This likely reflects the similar steady increase of
the bulge-to-total luminosity ratio (B$/$T) with decreasing
morphological type observed by, e.g., \citet{laurikainen10}.

\section{Scaling relations}\label{sec:scaling}

Here we showcase several ETG scaling relations derived from the P3
isophotal parameters.  We begin with concentration, then move to size,
surface brightness, and finally discuss integrated $m_{3.6}-m_{4.5}$
colors.  At the end of each sub-section, we briefly summarize our results.

\subsection{Concentration}\label{sec:concentration}

\begin{figure}
  \centering
  \includegraphics[scale=1.0]{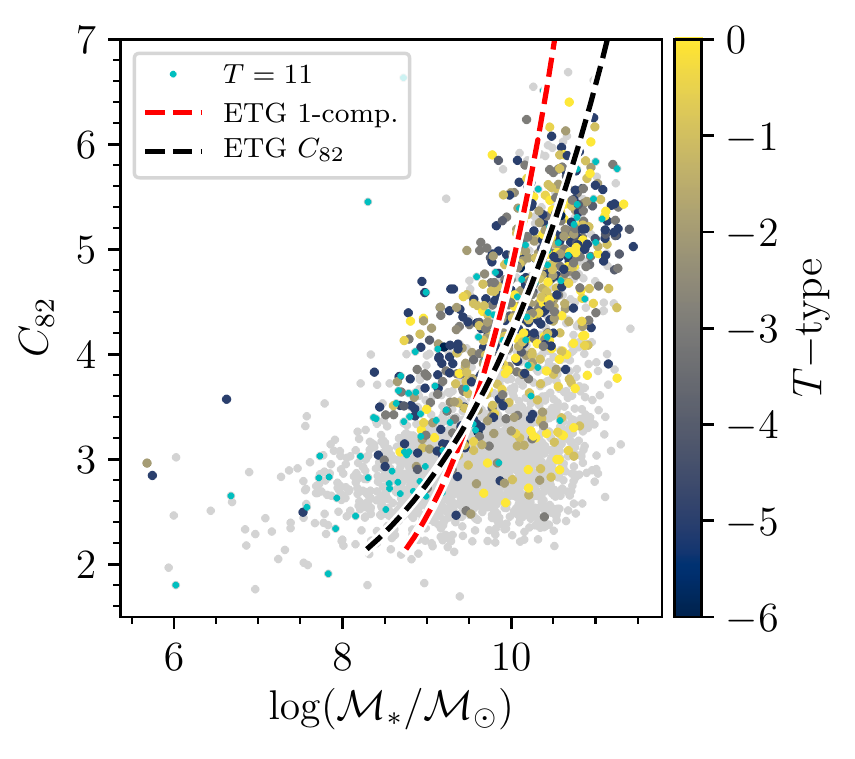}
  \caption{Correlation between concentration and stellar mass.  Here
    and in all subsequent figures with this color scheme, ETGs ($T
    \leq 0$) are color-coded by $T-$type, while all S$^{4}$G galaxies
    are shown in grey.  Dwarf ETGs ($T=11$) are shown in cyan.  The dashed black line shows a linear fit
    between $\log\,(\mathcal{M}_{*}/\mathcal{M}_{\odot})$ and $\log\,(n)$,
    converted from $C_{82}$, for ETGs (Eq.~\ref{eq:nfitc82}).  The
    red dashed line shows a similar fit, but here $n$ was estimated
    for each ETG via single-component 2D decompositions
    (Eq.~\ref{eq:nfitdecomp}).
    \label{fig:c82mass}}
\end{figure}

\begin{figure*}
  \centering
  \includegraphics[scale=1.0]{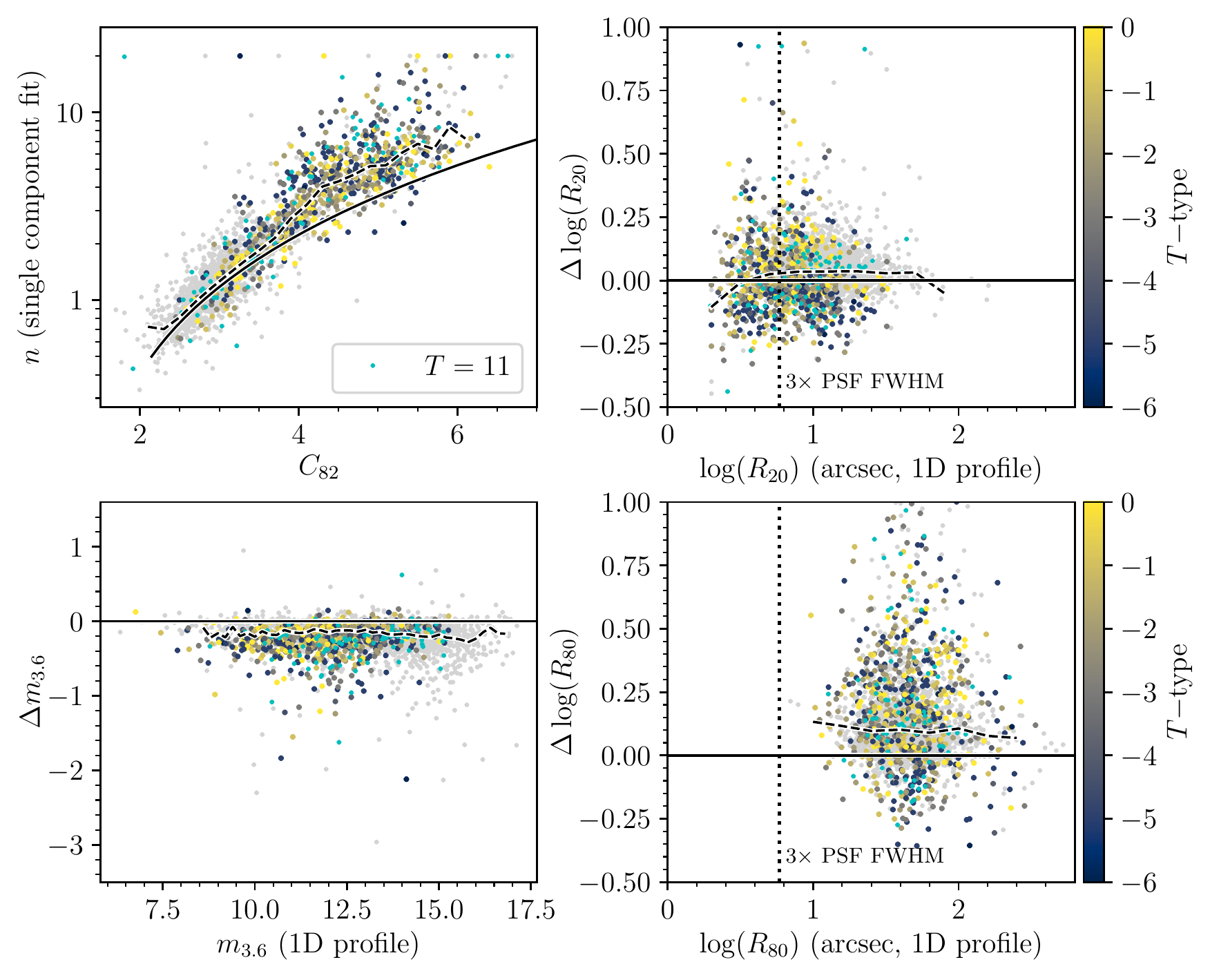}
  \caption{Investigating the divergence between the S\'{e}rsic $n$ derived
    from $C_{82}$ and that measured from single-component 2D
    decompositions.  The \emph{top-left} panel shows concentration
    parameter vs. S\'{e}rsic index, the latter derived from
    single-component 2D decompositions for all S$^{4}$G galaxies.  The
    solid black line shows the theoretical behavior of $C_{82}$ for
    S\'{e}rsic profiles of varying $n$.  Here and in subsequent panels,
    the dashed black line shows the median of points in equal-width
    bins (bins with fewer than five points are ignored).  In the
    \emph{bottom-left} panel, we show the difference between our
    asymptotic magnitudes and those derived from decompositions
    (negative numbers mean the asymptotic magnitudes are fainter).
    The solid black line here and in the following panels shows a
    value of 0 (no difference).  The \emph{top-right} panel shows the
    difference between $R_{20}$ derived from single-component
    decompositions and those derived from our curves of growth for
    3.6\,$\mu$m imaging.  The vertical dotted line shows the size of
    three IRAC 3.6\,$\mu$m resolution elements.  The \emph{bottom-right}
    panel is the same as the \emph{top-right} panel, but shows
    the behavior of $R_{80}$.  ETG data points are multi-colored (with dwarfs in cyan), while LTGs are shown in gray.
    \label{fig:sersicfits}}
\end{figure*}

\begin{figure*}
    \centering
    \includegraphics[scale=1.0]{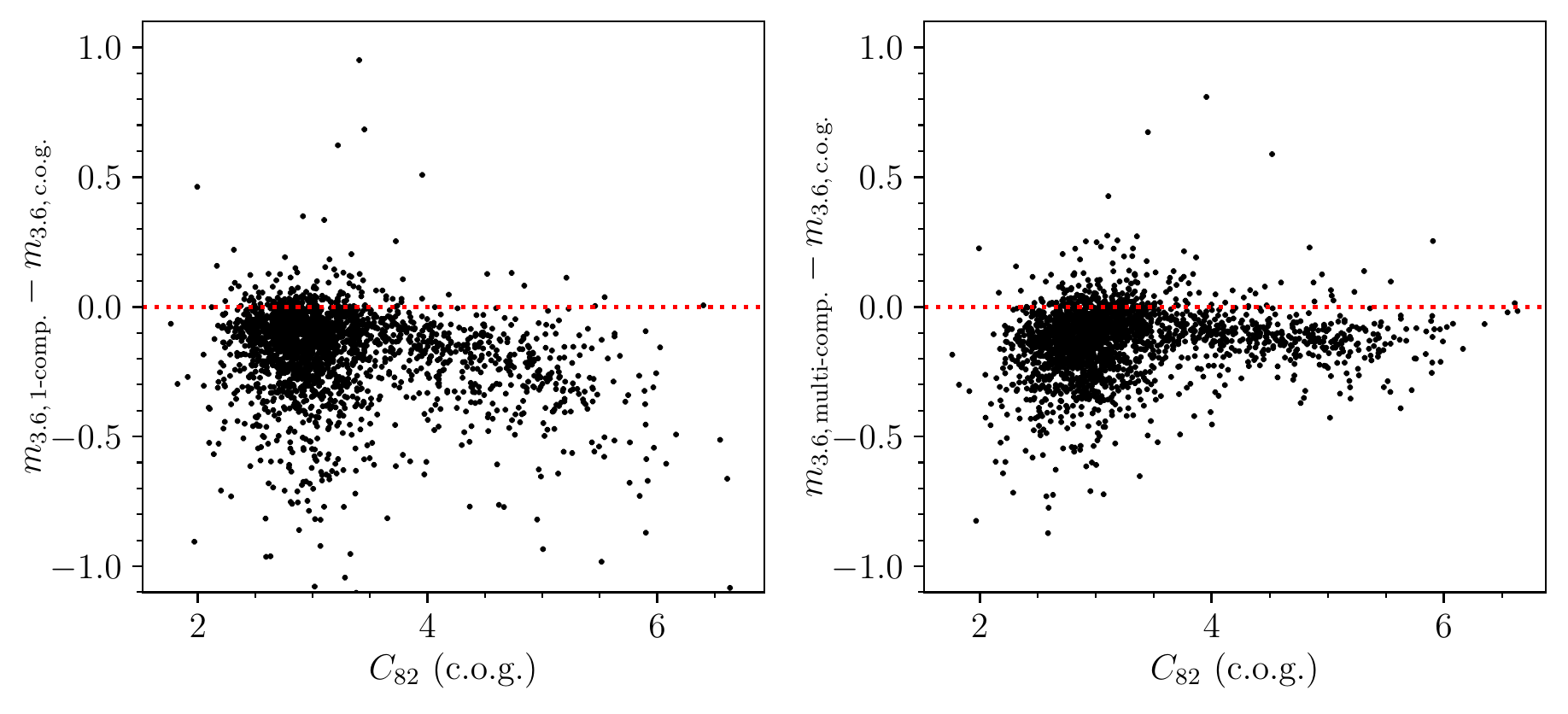}
    \caption{\textbf{Left} panel: difference between single-component decomposition total magnitudes and asymptotic magnitudes (P3; MM2015), as a function of curve of growth concentration parameters (P3) for the original S$^{4}$G sample.  \textbf{Right} panel: as the left panel, but showing the difference between multi-component decomposition total magnitudes \citep[P4;][]{salo15} and asymptotic magnitudes.  While both decomposition types show slight offsets from the asymptotic magnitudes on average, as well as downward trends (brighter magnitudes) with increasing concentration, the downward trend in the single-component decompositions is much stronger than that in the multi-component decompositions.  Also, at least $\sim 0.06$ magnitudes of offset in the multi-component magnitudes arise because we applied no infinite aperture correction to the total flux \citep[see Eq. \ref{eq:newcorr}, and][]{reach05}.
    \label{fig:multicomp}}
\end{figure*}

Particularly for high-redshift galaxies, where individual
morphological features are unresolvable, light concentration is useful
for separating ETGs and LTGs \citep[e.g.,][]{morgan58, devau73,
  okamura84, kent85, conselice03}.  Concentration also scales with
luminosity, showing an almost log-linear relationship for ETGs
\citep[e.g.,][]{schombert86, binggeli91, graham01b, conselice03,
  kauffmann04}, which we demonstrate as well in
Fig.~\ref{fig:c82mass}.  Structural parameters such as concentration in
turn show little dependence on local environment, but environment
correlates well with star formation rate \citep[e.g.,][]{kauffmann04,
  vanderwel08}, suggesting that the structure of the bright, visible
parts of galaxies was in place early on in most galaxies
\citep{kauffmann04}.  Still, secular evolution and subsequent star
formation episodes can modify galaxy structure over time, so even
under this early formation scenario one might expect that galaxies
with bars, spiral arms, and other dynamical structures capable of
moving mass \citep[e.g.,][]{sellwood02}, as well as those with
ongoing, wide-spread star formation, might show the largest scatter in
the concentration--stellar mass relation.  Indeed,
\citet{mendezabreu21} found that disk galaxy bulges evolve little over
cosmic time, while their disks can change dramatically
\citep[depending on the central concentration of the galaxy, e.g.,][]{athanassoula05b}.  Additionally, \citet{sheth08} showed that the distribution of bars among massive S0s at $z=0.84$ was quite similar to that of massive S0s at $z=0$, suggesting that the disk structures were in place early on in these galaxies; lower-mass LTGs, by contrast, appear to have formed their bars much more gradually during this time, but evidently continued forming stars in their disks much longer than their S0 counterparts.  We see evidence for this in
Fig.~\ref{fig:c82mass}.  Ellipticals (navy blue points) show a fairly tight
correlation; S0s (yellow and gold points) of the same stellar mass scatter toward
low $C_{82}$ values, with the tightest scatter found at the high-mass end; and the scatter and distribution of $C_{82}$ values among LTGs is quite similar to that of S0s, with the highest scatter at the lowest stellar masses.

Though this correlation appears roughly log-linear, it does show some
mild curvature, suggesting the true log-linear correlation is between
stellar mass and S\'{e}rsic index \citep[e.g.,][]{caon93, young94,
  graham01a, ferrarese06, savorgnan13, graham19}.  We demonstrate this
via the black dashed curve in Fig.~\ref{fig:c82mass}, which we derived
by first converting our measured values of $C_{82}$ for ETGs (colored
points) to their corresponding values of $n$ \citep[see,
  e.g.,][]{graham05, janz14}, then fitting the resulting
$\log\,(\mathcal{M}_{*}/\mathcal{M}_{\odot})$--$n$ relationship as a
line:
\begin{equation}\label{eq:nfitc82}
  \log\,(\mathcal{M}_{*}/\mathcal{M}_{\odot}) = 2.47\log\,(n) + 9.03
\end{equation}
We then converted this linear fit back to a relation between
$\log\,(\mathcal{M}_{*}/\mathcal{M}_{\odot})$ and $C_{82}$ using the
relation between $n$ and $C_{82}$, which produces the aforementioned
black dashed curve.  Using the single-component decompositions
described in Sect.~\ref{sec:sample}, we also derived empirical values of $n$
for all ETGs for comparison.  These too showed a linear relationship
with stellar mass, which we overplot in Fig.~\ref{fig:c82mass} via
the red dashed line (again, converted to $C_{82}$).  This has the
form:
\begin{equation}\label{eq:nfitdecomp}
  \log\,(\mathcal{M}_{*}/\mathcal{M}_{\odot}) = 1.52\log\,(n) + 9.21
\end{equation}
which has a similar zeropoint but a slope different by nearly a factor
of $\sim1.5$.  Even though we are using the PSF-corrected radial profiles to
derive $C_{82}$, the values of $n$ implied by our concentration
parameters thus agree poorly with those derived via single-component
2D fits.

We expand on this in the top-left panel of Fig.~\ref{fig:sersicfits},
which shows our $C_{82}$ values plotted against the values of $n$ for all our ETGs
derived from single-component GALFIT decompositions.  The solid black line
here shows the theoretical relationship between the $C_{82}$ and $n$,
while the dashed line (here and in all subsequent panels) shows the
median trend for all the galaxies (again shown in grey) as measured within
consecutive equal-width $C_{82}$ bins.  Clearly, at the high concentration
(high mass) end, the measured values of $C_{82}$ are too low compared
to the theoretical expectation from the measured $n$.

\citet{venhola18} found superficially similar behavior for galaxies in the Fornax
Cluster region using Fornax Deep Survey \citep{peletier20} $u^{\prime}$-, $g^{\prime}$-, $r^{\prime}$-, and $i^{\prime}$-band imaging (see their Fig.~13).  Like us, \citet{venhola18} observed that the S\'{e}rsic $n$ values from single
component GALFIT decompositions were systematically higher than the
theoretical values of $n$ estimated from $C_{82}$.  In their case, the aperture
magnitudes and the corresponding $C_{82}$ values were calculated using a
region within one Petrosian radius, which was also taken into account in
their theoretical $n$ vs. $C_{82}$ relation (the $n$ corresponding to a given $C_{82}$
is larger than if infinite extent is assumed).  They concluded that the offset, seen for all values of $n$,
was due to resolution effects: the GALFIT decompositions included
PSF-convolution, whereas the aperture photometry did not\footnote{Note that the galaxies shown in Fig.~13 in \citet{venhola18} are
mainly background galaxies at distances much larger than the 20~Mpc
distance of the Fornax Cluster, making the resolution a more pronounced problem
than for our sample, even if the FWHM of the FDS observations was
smaller (about 1\arcsec \ instead our \emph{Spitzer} 2\arcsec).}.

We investigate this explicitly in the right two panels of
Fig.~\ref{fig:sersicfits}.  In these two panels, we show the values
$\Delta\log\,(R_{x}) = \log\,(R_{x,~1~\textrm{comp.}}) -
\log\,(R_{x,~\textrm{c.o.g.}})$, where $x$ is the fraction of light
contained within the radius, ``1 comp.'' represents values derived
from single-component decompositions, and ``c.o.g.'' represents values
we derive from the radial curves of growth.  We show both components
used for $C_{82}$, the 20\% (top-right panel) and 80\% (bottom-right
panel) flux radii.  Vertical dotted lines in each panel show three times
the 3.6\,$\mu$m FWHM; if resolution was purely at fault for the
diverging $C_{82}$--$n$ profiles, because $R_{20}$ often falls within
this limit, it might show the strongest method-to-method difference.
However, while $\Delta\log\,(R_{20})$ is systematically slightly
positive (albeit with large scatter), the offset is small compared to
that of $\Delta\log\,(R_{80})$, for which resolution's impact should
be mild.  Resolution does not thus appear to be the driving factor
here.

Typically, we find $\Delta\log\,(R_{80}) > 0$, indicating that
single-component-fit $R_{80}$ are systematically $\sim 1.3$ times larger on average (albeit with large scatter) than those
derived from the curves of growth.  In the bottom-left panel of
Fig.~\ref{fig:sersicfits}, we also show the difference between the
magnitudes estimated from single-component S\'{e}rsic fits and our
asymptotic magnitudes.  These tend to be negative, meaning
single-component fit magnitudes are systematically $\sim 0.2$ mag brighter than
asymptotic magnitudes.  This, combined with the larger values of
$R_{80}$, suggests that single-component fitting often overestimates
the flux in the galaxies' outskirts.  This is expected for disk
galaxies, which often show down-bending breaks in their surface
brightness profiles \citep[e.g.,][]{pohlen06, erwin08}.  However, from
the color distribution of the points, it is clear that the magnitude
of the difference is similar for all ETG morphological types.  Various
studies have shown that elliptical galaxies, much like disks, are poorly
characterized by single S\'{e}rsic component models
\citep[e.g.,][]{schombert86, hopkins09}, which may result in the
behavior we see here.

We provide further evidence for this in Fig.~\ref{fig:multicomp}.  In each panel, we show the difference between decomposition magnitudes---both single- and multi-component---and asymptotic, c.o.g-derived magnitudes, as a function of $C_{82}$, for the original S$^{4}$G sample.  Asymptotic magnitudes and $C_{82}$ here come from MM2015, while multi-component decomposition magnitudes come from \citet{salo15}.  Both decomposition magnitudes show slight systematic offsets from the asymptotic magnitudes, as well as trends with $C_{82}$: the higher the concentration, the brighter the estimated decomposition magnitude relative to the asymptotic value.  However, this trend is much stronger for the single-component decompositions than for the multi-component decompositions.  Additionally, $\sim 0.06$ magnitudes of the offset in the multi-component magnitudes arises because these magnitudes were derived directly from the images, without applying the $3.6\mu$m infinite aperture correction of $0.944$ \citep[see Eq. \ref{eq:newcorr}, and][]{reach05}.  This confirms our suspicion that single-component fits are not as reliable as either multi-component fits or the curve of growth method for estimating total galaxy magnitudes.  Combined with Fig.~\ref{fig:sersicfits}, it is clear that most of the systematic errors in single-component fits occur in the galaxy outskirts.

In summary, we reproduce well the known trend between $C_{82}$ and stellar mass for ETGs, thereby reproducing the trend between $n$ and stellar mass (Fig. \ref{fig:c82mass}).  In comparing our values of $C_{82}$ derived from radial profiles to those derived from single-component GALFIT decompositions, we find that such decompositions tend to overestimate the concentration and the total flux of these galaxies (Fig. \ref{fig:sersicfits}).  Most of this over-estimate of flux occurs in the galaxy outskirts, in a manner that correlates with the galaxies' concentration.  Multi-component decompositions fare much better---total magnitude estimates from these agree more closely with radial profile--derived asymptotic magnitudes, with only small systematic differences due to galaxy concentration (Fig. \ref{fig:multicomp}).

\subsection{Mass--size}\label{sec:masssize}

\begin{figure*}
  \centering
  \includegraphics[scale=1.0]{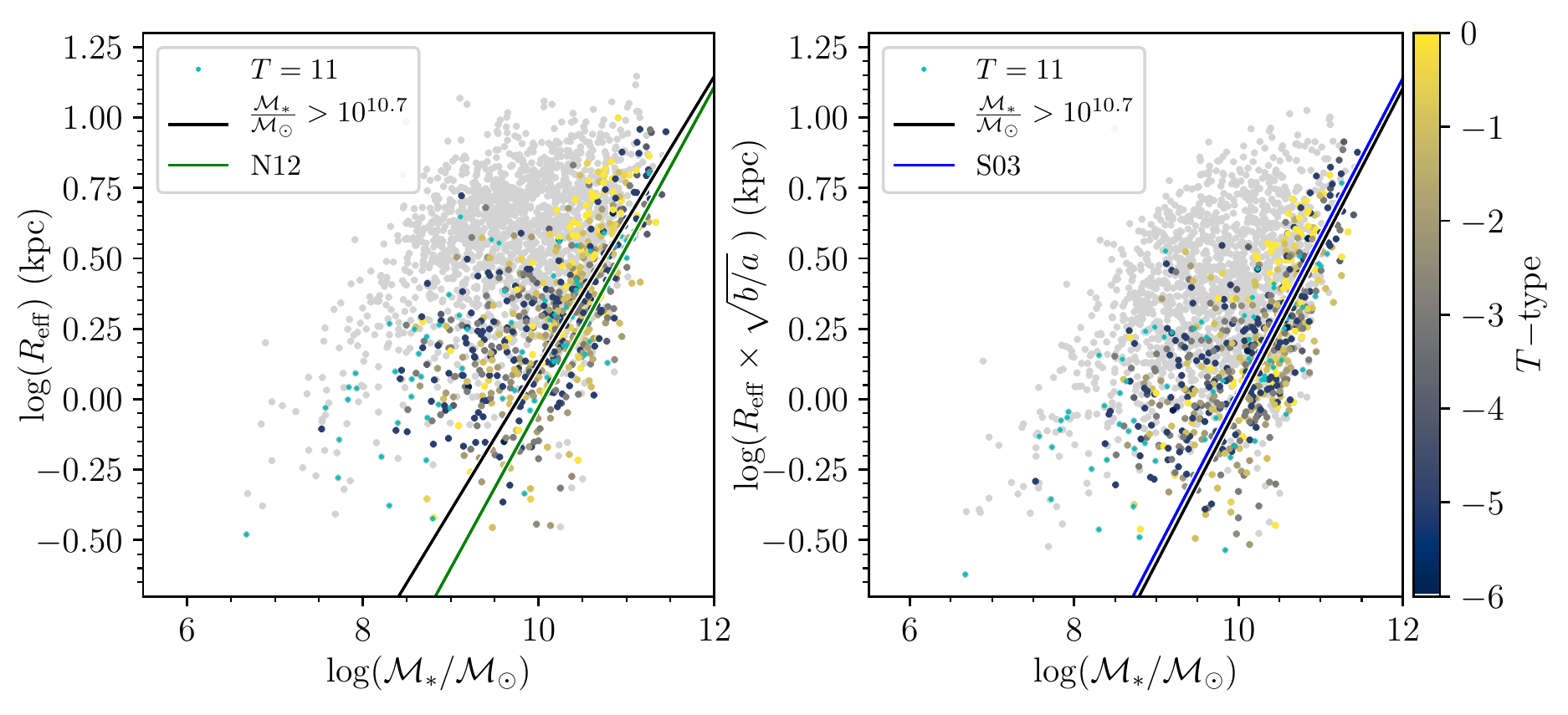}
  \caption{Effective radius mass--size relation.  The \emph{left} panel
    shows our effective radius values (derived from 3.6\,$\mu$m imaging)
    plotted against stellar mass, with a linear fit to ETGs with
    masses $>10^{10.7} \mathcal{M}_{\odot}$ shown in black and the
    $z=0.25$ ETG mass--size relation from \citet{newman12} shown in
    green (labeled N12).  The \emph{right} panel is the same as the left, but
    here $R_{\rm eff}$ has been adjusted by a factor of $\sqrt{b/a}$
    to adjust the values to those measured using circular apertures,
    following \citet{shen03}, whose fit is overplotted in blue (labeled S03).  ETG data points are multi-colored (with dwarfs in cyan), while LTGs are shown in gray.
    \label{fig:reff}}
\end{figure*}

\begin{figure*}
  \centering
  \includegraphics[scale=1.0]{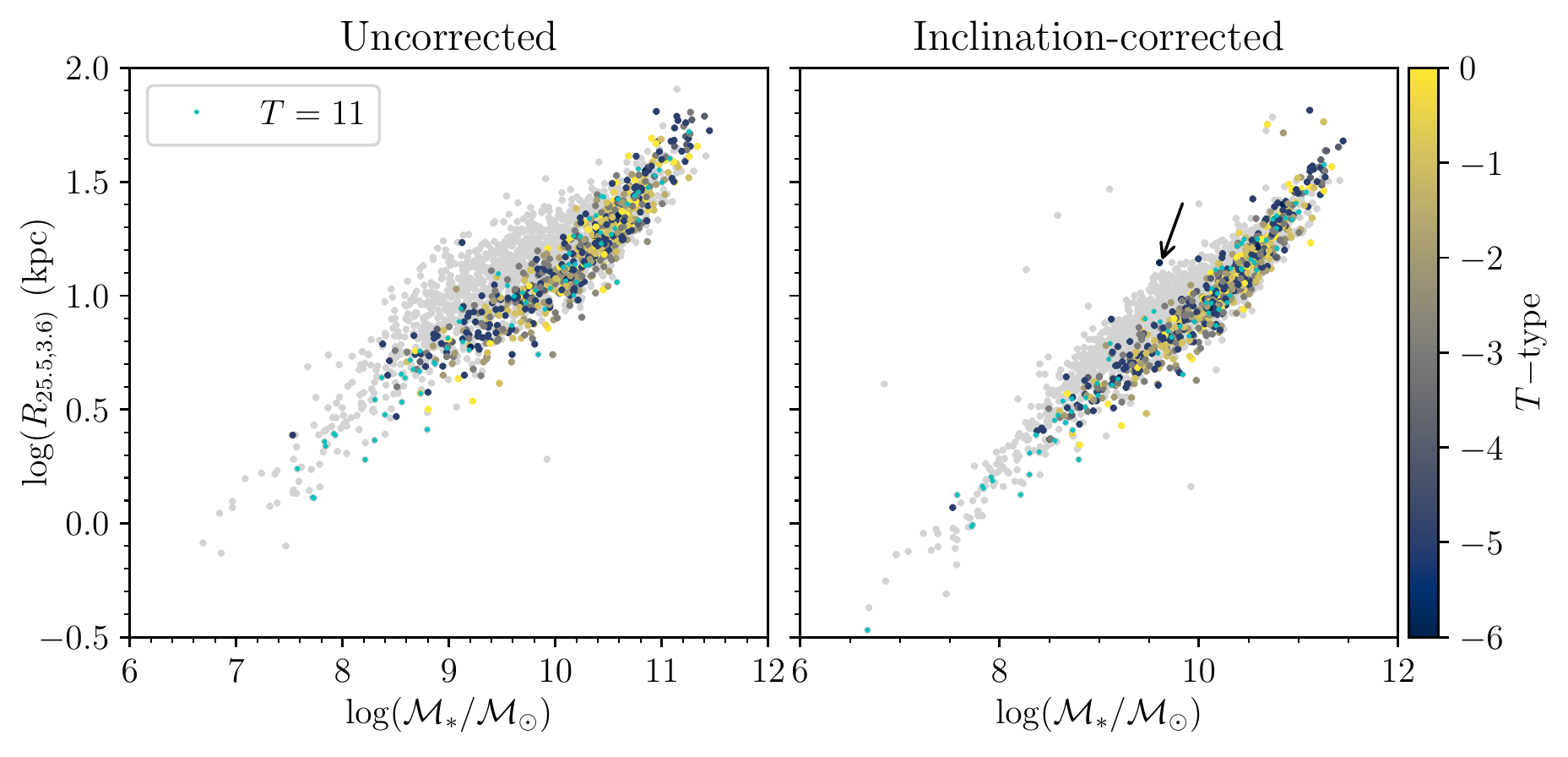}
  \caption{Isophotal mass--size relation.  The \emph{left} panel shows
    the relation between $R_{25.5, 3.6}$ and stellar mass, in which
    $R_{25.5, 3.6}$ was derived from the raw surface brightness profiles.
    The \emph{right} panel shows the same but with $R_{25.5, 3.6}$
    corrected for inclination, measured from the outermost isophote
    shapes' axial ratios.  ETG data points are multi-colored (with dwarfs in cyan), while LTGs are shown in gray. The arrow indicates the galaxy IC~3413 (see text).
    \label{fig:r255}}
\end{figure*}

\begin{figure}
  \centering
  \includegraphics[scale=1.0]{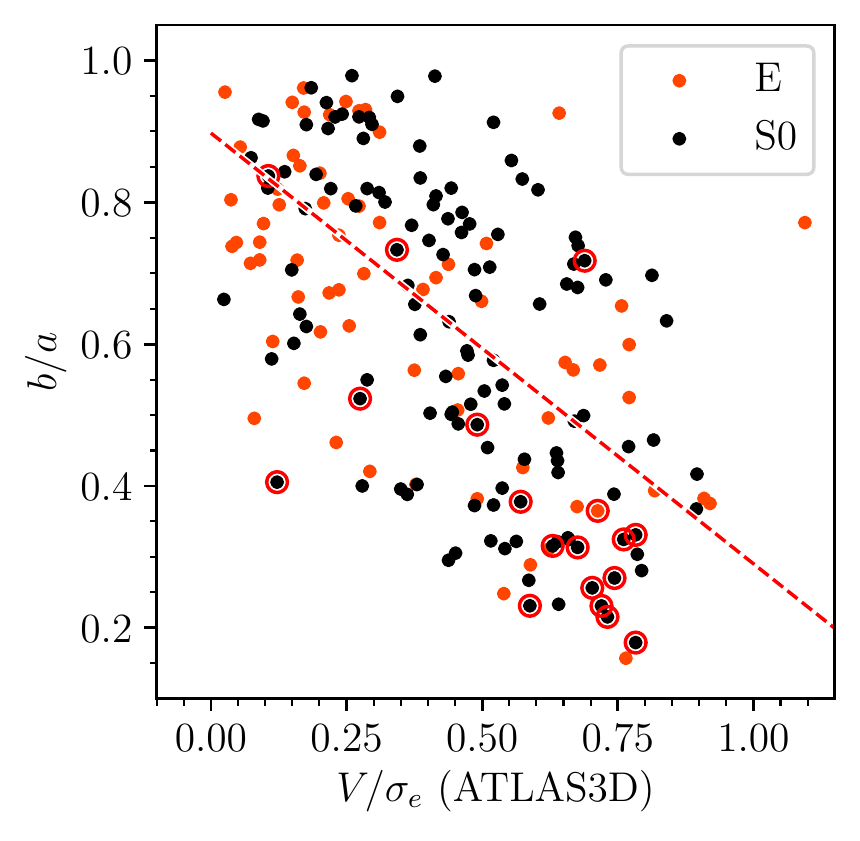}
  \caption{Comparing $V/\sigma_{e}$ values (denoting those measured at
    1\,$R_{\rm eff}$) from the ATLAS$^{\rm 3D}$ survey to isophotal
    axial ratios measured at $R_{25.5}$ for galaxies matching between
    the S$^{4}$G+ETG and ATLAS$^{\rm 3D}$ samples.  Orange point denote elliptical galaxies, while black points denote S0s.  The red dashed
    line shows a linear fit to all points.  Points circled in red are
    edge-on disks \citep[denoted `spindle' by][]{buta15}, with
    unreliable axial ratios.
    \label{fig:atlas}}
\end{figure}

The mass--size relation for ETGs has been crucial to understanding the
growth of ETGs throughout cosmic history \citep[e.g.,][]{daddi05,
  longhetti07, toft07, trujillo07, hopkins08, vandokkum08, hopkins09,
  vanderwel09, szomoru12, andreon16}.  A series of gas-poor, low
mass-ratio mergers can explain elliptical galaxies' rapid growth in size
compared to stellar mass \citep[e.g.,][]{nipoti03, naab09, szomoru12},
though dissipational processes seem necessary to explain specific
aspects of their evolution, e.g., the tilt in the fundamental plane
\citep{hopkins08}. Typically, size is measured via the effective
radius, given the utility of S\'{e}rsic functions in describing their
light profiles (though see Sect.~\ref{sec:concentration}), but it has long been known that all galaxies also show correlations between
isophotal size and the absolute magnitude \citep[or, modulo the
  mass-to-light ratio, stellar mass;][]{schombert86, binggeli91}.
These correlations are always positive, suggesting that whatever
drives the mass evolution of most galaxies drives the size evolution
in a similar way.

MM2015 showed the stellar mass--$R_{\rm eff}$ relation for the initial
2352-galaxy S$^{4}$G sample (their Fig.~15).  They demonstrated a clear
morphological segregation, with ETGs occupying the lower-right (low-luminosity, compact) corner
of the point distribution and the latest type galaxies (Sm, Irr)
occupying the lower-left (low luminosity, extended).  We broadly reproduce this trend in
Fig.~\ref{fig:reff}.  Gray points here are LTGs, while ETGs are
color-coded by morphological type.  Among ETGs, ellipticals and S0s
are mostly intermingled, though for stellar masses $\gtrsim 10^{10}
\mathcal{M}_{\odot}$, S0s show a mild tendency toward higher $R_{\rm
  eff}$ than ellipticals at equal stellar mass.  ETGs as a population
form a visibly tighter stellar mass--$R_{\rm eff}$ relation than LTGs,
even including the increase in dispersion emerging below stellar
masses $\lesssim 10^{10.5} \mathcal{M}_{\odot}$.

Only the high-mass end of the relation is roughly linear: curvature in the mass--$R_{\rm eff}$ relation is known to result from the dependence of light profile shapes on stellar mass (Sect.~\ref{sec:concentration}), alongside structural differences between dwarf and massive ETGs \citep[e.g.,][]{graham03, boselli08, janz08, kormendy09}. Massive galaxies are also not well-described by single S\'{e}rsic profiles (Figs. \ref{fig:sersicfits} and \ref{fig:multicomp}), likely leading to divergences even from the theoretical curved relationship described by \citet{graham03} and others.  For comparison with past studies, however, in Fig.~\ref{fig:reff} we show a linear fit to the S$^{4}$G mass--size relation via the black lines, including
only those ETGs with stellar masses $>10^{10.7} \mathcal{M}_{\odot}$
(126 galaxies total).  This has the following form:
\begin{equation}\label{eq:reff_massive}
  \log\,(R_{\rm eff}/\rm{kpc}) =
  0.51\log\,(\mathcal{M}_{*}/\mathcal{M}_{\odot}) - 5.02
\end{equation}
We overplot similar fits from two previous studies of nearby galaxies, derived above similar stellar mass thresholds: in
the left panel, we show the fit for $z=0.06$ ETGs from Table 1 of
\citet{newman12}; and in the right panel, we show the ETG sample fit from
Eq.~17 of \citet{shen03}.  These have the following forms:
\begin{equation}\label{eq:reff_nwm}
  \log\,(R_{\rm eff}/\rm{kpc}) =
  0.57\log\,(\mathcal{M}_{*}/\mathcal{M}_{\odot}) - 5.73
\end{equation}
for the \citet{newman12} relation (labeled N12 in Fig. \ref{fig:reff}), and:
\begin{equation}\label{eq:reff_shen}
  \log\,(R_{\rm eff}/\rm{kpc}) =
  0.56\log\,(\mathcal{M}_{*}/\mathcal{M}_{\odot}) - 4.46
\end{equation}
for the \citet{shen03} relation (labeled S03 in Fig. \ref{fig:reff}) .  We have corrected our rightmost panel
effective radii by a factor of $\sqrt{b/a}$ (with axial ratios
measured at $R_{25.5, 3.6}$) for this comparison, as \citet{shen03}
used circular apertures for their photometry (see MM2015).

Our fit is very close to those of both \citet{newman12} and
\citet{shen03}.  Our derived slope is slightly shallower than those of
either study, and our intercept is very close to that of
\citet{shen03} but noticeably smaller than that of \citet{newman12}.
While this may have physical significance---for example, the change in
intercept compared with that of \citet{newman12} could imply a size
evolution since $z=0.06$---we note that both \citet{newman12} and
\citet{shen03} used S\'{e}rsic fits (via GALFIT and from the radial
profiles, respectively) to derive their masses and radii, so the
differences could also be explained by differences in methodology (see
Sect.~\ref{sec:concentration}). For example, \citet{vanderwel14} found
a substantially different slope of 0.75, which they attribute to
differences in sample selection and their use of Petrosian
\citep{petrosian76} half-light radii.  Additionally, the aforementioned studies used visible light photometry to estimate stellar mass for their $z=0$ samples, which can be more subject to, e.g., metallicity effects \citep[e.g.,][]{meidt14}.  Within these ambiguities,
therefore, we find our fits broadly consistent with those of previous
studies, confirming the $z=0$ calibration for the observed tendency
for the slope in this relation to remain stable since as early as $z=3$
\citep[e.g.,][]{vanderwel14}.

We show the stellar mass--isophotal size relation for S$^{4}$G+ETG
galaxies in Fig.~\ref{fig:r255}, using $R_{25.5,3.6}$ for the
isophotal radii.  This relation looks similar when using $R_{26.5,
  3.6}$ or the isophotal radii measured in 4.5\,$\mu$m, but these have higher
scatter due to the lower S/N in the latter.  The left panel shows the relation
using unmodified surface brightness profiles, while the right panel
shows the relation after correcting the surface brightness profiles
to their face-on equivalents. We do this by multiplying the flux in each
  radial bin by the axial ratio, following \citet{kent85}; we note that, while such a correction may be an appropriate inclination correction for thin disks, for ETGs, many of which have spheroidal or triaxial shapes, this correction should be considered an isophotal circularization rather than a true inclination correction.  The scatter
decreases substantially once this correction is applied.  Extreme
outlier points in the rightmost panel (mostly LTGs) are all nearly
edge-on galaxies with unreliable axial ratios.  Outliers among the
ETGs have a muddier origin; for example, the point at
($\log\,(\mathcal{M}_{*}/\mathcal{M}_{\odot}) = 9.657$, $\log\,(R_{25.5,
  3.6}) = 1.129$, shown by the arrow) is IC~3413, a seemingly undisturbed, mildly
inclined, and mostly featureless galaxy, though it is a Virgo Cluster
member.  Deeper imaging of some of these outliers could reveal
evidence of tidal interactions, which may move them away from their
expected positions in this relation.

MM2015 show the stellar mass--size relation for the initial S$^{4}$G sample in their
Fig.~14, again demonstrating clear morphological segregation, which
accounts for much of the scatter in the left panel of
Fig.~\ref{fig:r255}.  Once an inclination correction is
applied\footnote{While MM2015 did apply some manner of inclination
  correction to their isophotal radii, the scatter in their relation
  much more closely resembles our uncorrected relation's scatter.  Our
  axial ratio correction reduces the scatter significantly, so it would
  appear to be the more appropriate choice.}, this morphological
segregation reduces, as the wider array of axial ratios displayed by
thin disk galaxies results in stronger corrections on average compared
to the vertically thicker ellipticals.  Yet, with only an axial ratio
correction applied, the scatter reduces just as much for elliptical galaxies as
for S0s, surprising given that ellipticals are usually oblate systems.  Possibly this correction is as effective for elliptical galaxies due
to their significant rotational support \citep[e.g.,][]{cappellari07},
which implies that their axial ratios are in some ways reflective of
their instrinsic alignments.  We show this explicitly in
Fig.~\ref{fig:atlas}, in which we plot the axial ratios of galaxies
overlapping the S$^{4}$G+ETG and the ATLAS$^{\rm 3D}$
\citep{cappellari11, cappellari11c} surveys against their rotational
velocity--to--velocity dispersion ratios \citep[$V/\sigma_{\rm e}$,
  measured at one effective radius;][]{emsellem11}.  Even with the
ambiguity presented by nearly edge-on galaxies (marked with red
circles), there is a clear downward trend, as expected if more
rotationally supported galaxies are also intrinsically vertically
thinner.  We show ellipticals as orange points and S0s as black points---the downward trend is visible using both classes of galaxies.

The tightness in the stellar mass--isophotal radius relation is
noteworthy, and has been mentioned before in studies utilizing
different isophotes and photometric bands \citep[e.g.,][]{schombert86,
  binggeli91, saintonge11, hall12}.  Indeed, several papers have
recently begun to explore it in the context of star formation
thresholds, assessing the shape and scatter of the relation for the
radius at which the stellar mass surface density reaches 1
$\mathcal{M}_{\odot}$\,pc$^{-2}$ \citep{chamba20, trujillo20,
  sanchezalmeida20}, which is comparable to previous estimates of star-formation thresholds using gas surface density \citep[3--10 $\mathcal{M}_{\odot}$\,pc$^{-2}$,
  e.g.,][]{schaye04}.  The scatter in the right panel of
Fig.~\ref{fig:r255} is roughly equivalent to that found by
\citet{trujillo20}, who used $g$- and $r$-band imaging from the IAC
Stripe82 Legacy Project \citep{fliri16, roman18} to derive stellar
masses.  Comparing to their Fig.~4, the curvature we find in our
relation is even quite similar, as is the morphological segregation.
This is not particularly surprising, as 3.6\,$\mu$m imaging is a very
close tracer of stellar mass---the 1 $\mathcal{M}_{\odot}$\,pc$^{-2}$
radius explored by these papers corresponds roughly to $R_{27, 3.6}$.
\citet{sanchezalmeida20} demonstrated that the likely origin of this
low scatter lies in the anticorrelation between $R_{\rm eff}$ and $n$,
and found that it minimizes when using the 2.4 $\mathcal{M}_{\odot}$\,
pc$^{-2}$ mass surface density radius (very close to our $R_{25.5,
  3.6} \sim 3.5 \mathcal{M}_{\odot}$\,pc$^{-2}$).  Still, the physical
origin behind such an optimal value is unknown.  More careful analysis
of this relation is required \citep[akin to that done by,
  e.g.,][]{ouellette17}, which will be the subject of a separate
paper.

In summary, we find good agreement with previous studies in the stellar-mass--$R_{\rm eff}$ relation for high-mass ETGs, modulo methodological differences.  We also find that the stellar-mass--isophotal radius relation, once the surface brightnesses are corrected for inclination, has very little scatter even among elliptical galaxies, which serves as a secondhand demonstration of their rotational support.

\subsection{Surface Brightness}\label{sec:surfbri}

\begin{figure*}
  \centering
  \includegraphics[scale=1.0]{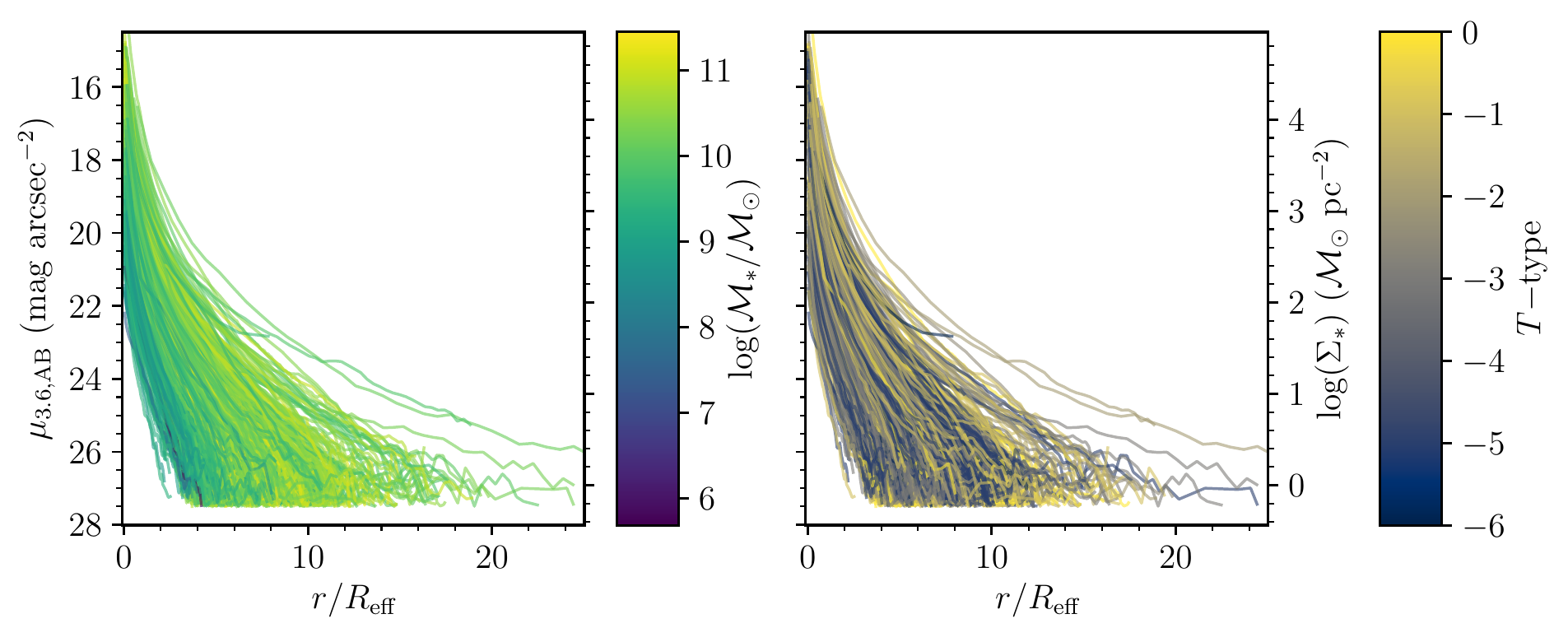}
  \caption{All ETG 3.6\,$\mu$m surface brightness profiles (fixed
    isophotal parameters, $\Delta r =2$\arcsec; see Section
    \ref{sec:radprofs}), normalized by effective radius.  The {\it
      left panel} shows profiles color-coded by total stellar mass, while
    the {\it right panel} shows profiles color-coded by morphological
    $T-$type.  Surface brightness is converted to mass surface density
    in the right panel as well, for convenience.  Data quality at very
    low surface brightness is poor, hence all profiles are truncated
    at $\mu_{3.6, \textrm{AB}} = 27.5$.
    \label{fig:allprofs}}
\end{figure*}

\begin{figure*}
  \centering
  \includegraphics[scale=1.0]{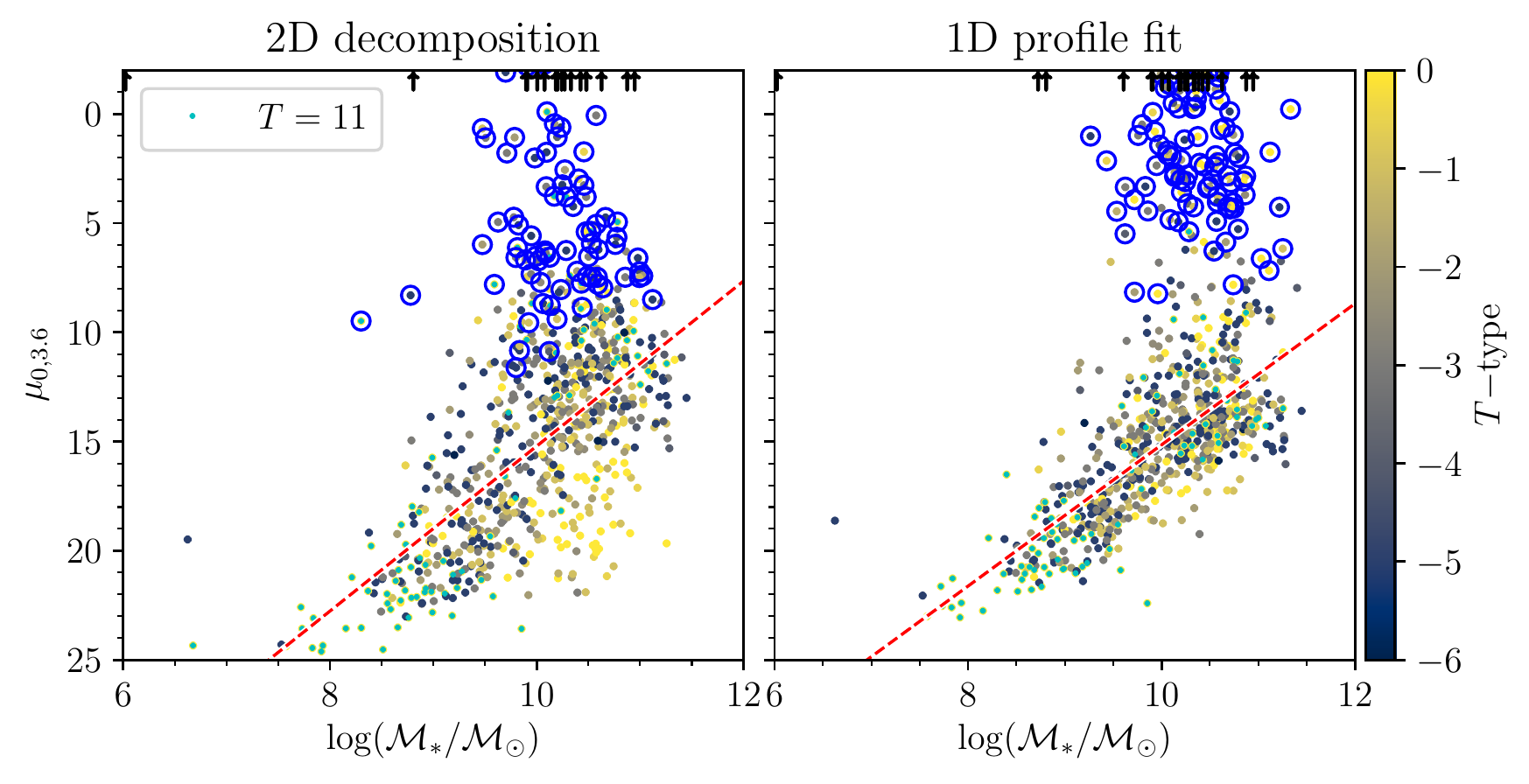}
  \caption{3.6 $\mu$m central surface brightnesses ($\mu_{0,3.6}$)
    vs. stellar mass.  The \emph{left panel} shows $\mu_{0,3.6}$
    derived from single-component 2D decompositions, while the
    \emph{right panel} shows $\mu_{0,3.6}$ derived by fitting
    generalized single S\'{e}rsic functions to each galaxy's radial
    surface brightness profile (fixed isophotal parameter, $\Delta
    r =2$\arcsec, with aperture correction applied; see
    Section~\ref{sec:radprofs}).  Galaxies with fitted $n > 7$ are
    circled in blue.  Linear fits to the two distributions are shown
    via the red dashed lines, excluding galaxies with $\mathcal{M}_{*}/\mathcal{M}_{\odot} < 8$ and fitted $n > 7$).  Points lying above the axis limits are
    shown as up-pointing arrows; each of these had fitted $n >
    7$.  ETG data points are multi-colored (with dwarfs in cyan), while LTGs are shown in gray.
    \label{fig:mu0}}
\end{figure*}

While concentration parameters provide an easy-to-measure
approximation of galaxy structure, surface brightness profile shapes
contain much more detailed information, including the true central
surface brightness, $\mu_{0}$.  In a typical hierarchical accretion
scenario, a galaxy's core will form early, during a cosmological era
in which the horizon was small and therefore mergers were more
frequent, whereas the outskirts form later from less frequent mergers
\citep{frenk85} and from reorganization of matter due to secular
processes \citep[e.g.,][]{kormendy04}. The former process builds the
very dense innermost structures that $\mu_{0}$ traces \citep[classical
  bulges;][]{sandage61}.  Subsequent secular processes can lead to
mass growth in galaxy cores as well, if the bar induces angular
momentum loss in the gas and dust \citep[e.g.,][]{athanassoula92,
  maciejewski02, carles16}, but the resulting structures are often
less dense and more disklike \citep[pseudobulges or disky bulges, in
  the nomenclature of][respectively]{kormendy04, athanassoula05} than
classical bulges.  In the absence of star formation, the
reorganization of mass in the disk then depends on the growth of
bars, whose pattern speeds set the locations of resonances.  This in turn depends on a large number of properties of the galaxy \citep{athanassoula03}, including the mass and size of the central mass concentration \citep[e.g.,][]{sellwood80, norman96, athanassoula05b, debattista06, kataria20}, which is
often small or negligible in S0s \citep{laurikainen07, laurikainen10}.
Elliptical galaxies lack large-scale bars and the disks necessary to form them; they also have the densest cores, which show little evolution over cosmic time \citep[e.g.,][]{szomoru12}, implying a two-phase evolutionary process, with early violent growth in the core followed by a slower, merger-fueled growth in the outskirts that had little impact on subsequent core evolution.  In the highest-mass ellipticals, however,
subsequent core evolution can occur through scouring effects of SMBH
mergers and the subsequent reorganization of matter while the systems
relax \citep[e.g.,][]{milosavljevic01}.  Modulo resolution effects (for example, the resolution of our profiles is likely too coarse to accurately probe the SMBH's sphere of influence), surface brightness profile
shapes can reflect all of these complex histories.

For an overall picture, Fig.~\ref{fig:allprofs} shows the radial
surface brightness profiles of every ETG ($T \leq 0$) in the
S$^{4}$G+ETG sample, with radii normalized by $R_{\rm eff, 3.6}$.  We
show here the fixed isophotal parameter, 2\arcsec \ radial bin width
curves described in Section~\ref{sec:radprofs}.  In the left panel,
each curve is color-coded by the galaxy's total stellar mass, while in the
right panel the curves are color-coded by $T-$type.  Additionally, to
reduce noise in the plot, we truncate each curve at $\mu_{3.6} =
27.5$.

Mass segregation is clear in the left panel, with high-mass galaxies
typically showing higher surface brightnesses than low-mass galaxies
at all radii.  High-mass galaxies are also larger than low-mass
galaxies relative to their effective radii (see also Section
\ref{sec:masssize}).  This, combined with the tendency for high-mass
ETGs to have more concentrated light profiles
(Fig.~\ref{fig:c82mass}), leads to high-mass ETGs showing longer
profiles in this space.  By contrast, the profiles are well-mixed by
morphological $T-$type, suggesting that stellar mass is the primary
driver behind ETG surface brightnesses.  A more detailed examination
of individual profiles is evidently necessary to disentangle the
various contributions to the shapes of these profiles, including bars,
rings, disk breaks \citep[e.g.,][]{pohlen06}, and cored vs. S\'{e}rsic
ellipticals \citep[e.g.,][]{kormendy96, faber97}.

We thus expand on the stellar-mass--surface brightness relation in
Fig.~\ref{fig:mu0}, in which we plot only the central surface
brightnesses $\mu_{0, 3.6}$ of each galaxy against their stellar
masses.  As in Fig.~\ref{fig:sersicfits}, we show $\mu_{0, 3.6}$ derived
both from single-component 2D decompositions, as well as those derived
through single-component least-squares fitting to the curves shown in
Fig.~\ref{fig:allprofs}.  We correct for PSF effects in the 1D
curve fits by using the aperture-corrected profiles described in
Section~\ref{sec:radprofs}.

Either method demonstrates clear stellar mass--$\mu_{0}$ trends, such
that higher mass galaxies have higher central surface brightnesses, in
agreement with, e.g., \citet[][and references therein]{graham19}.
Linear fits to these relations result in the following (excluding galaxies with $\mathcal{M}_{*}/\mathcal{M}_{\odot} < 8$ and fitted $n > 7$):
\begin{equation}\label{eq:mu0fit2d}
  \mu_{0,3.6, \textrm{2D}} = -3.78\log\,(\mathcal{M}_{*}/\mathcal{M}_{\odot}) + 52.99
\end{equation}
for the 2D decompositions, and
\begin{equation}\label{eq:mu0fit1d}
  \mu_{0,3.6, \textrm{1D}} = -3.23\log\,(\mathcal{M}_{*}/\mathcal{M}_{\odot}) + 47.49
\end{equation}
for the values derived from curve-fitting. The two fits are
somewhat similar despite the differing methods used to estimate $\mu_{0,3.6}$, though the profile-fit slope is noticeably shallower than the decomposition slope.

In both cases, many galaxies with masses between $\sim 10^{9.5}
\mathcal{M}_{\odot}$ and $\sim 10^{11} \mathcal{M}_{\odot}$ lie far
above the relation, meaning they have large estimated central light
enhancements.  Most of these have very high fitted S\'{e}rsic indices
($n > 7$, which we have marked using blue circles).  Examination of
individual profiles shows that these galaxies typically have strong
central peaks indicative of large, bright bulges (relative to their disks), which thus dominate the
profiles' curvature.  The distribution of these galaxies' $T-$types is
also heavily weighted toward disks, with the strongest peak at $T=0$.  Due to the complex structure of disk galaxies (which can host bulges, bars, disks, lenses, rings, nuclear bars, nuclear rings, and even active galactic nuclei), the very high fitted values of $n$ and $\mu_{0,3.6}$ we find for these galaxies may mostly result from poor fits to the galaxies' total light profiles.

Among ellipticals, we see some hint of a light deficit for many
galaxies with masses $> 10^{10.5} \mathcal{M}_{\odot}$.  The distribution of a fair number of points falls well below the fitted line in this mass regime, seemingly leveling out
at a central surface brightness of $\sim 15$ (this is clearer in the right panel of Fig.~\ref{fig:mu0}).  These may be examples
of cored ellipticals, which do tend to be high-mass galaxies \citep[e.g.,][]{kormendy96, emsellem07, kormendy09, dullo12}, though we note that many ellipticals in this mass range lie
roughly on or above the best-fit relation as well.  Whether the massive galaxies that fall on the relation are truly uncored or not is unclear given the large uncertainties inherent to our estimates of $\mu_{0}$, as evidenced by the many poor, $n>7$ fits, described above; higher resolution profiles for the cores, in conjunction with more detailed fitting procedures, are needed to verify this.

Thus, while we broadly
reproduce the known relation between stellar mass and $\mu_{0}$, evidently great care is
required when estimating the central surface brightnesses of galaxies.
As we demonstrated with the concentration index in Sect.~\ref{sec:concentration},
single-S\'{e}rsic component fits can be inaccurate and misleading.

\begin{figure*}
\includegraphics[scale=1.0]{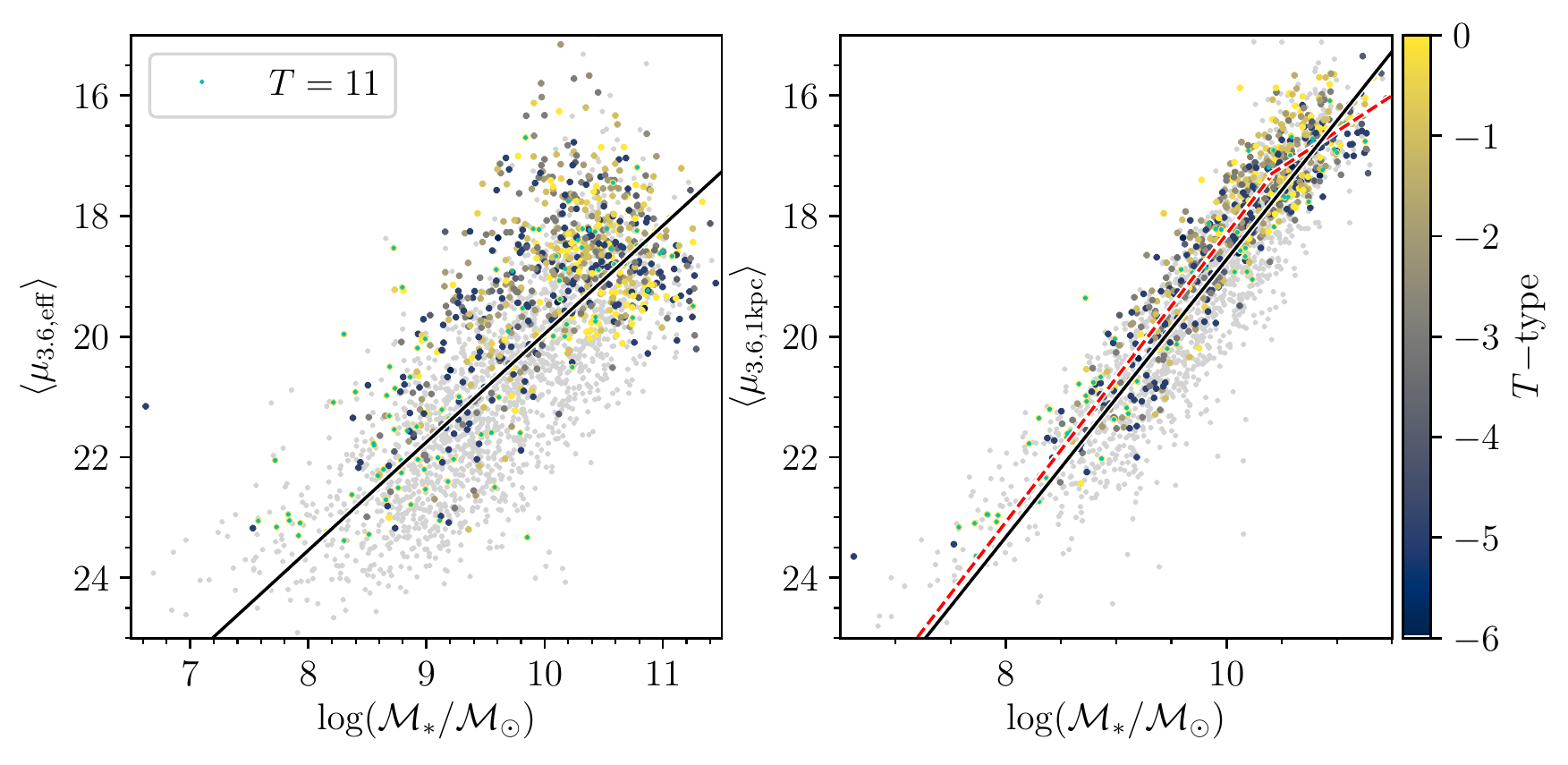}
\caption{Mean surface brightness within the effective radius
  (\emph{left panel}) and within 1~kpc (\emph{right panel})
  vs. stellar mass.  The solid black lines in both panels show linear
  fits to all S$^{4}$G+ETG galaxies.  The dashed red lines in the right
  panel show fits to only ETGs, separated at
  $\log\,(\mathcal{M}_{*}/\mathcal{M}_{\odot}) = 10.4$ (see text).  ETG data points are multi-colored, while LTGs are shown in gray.
    \label{fig:mu_relations}}
\end{figure*}

Given this, we also investigate scaling relations using the mean
surface brightnesses within fixed radii.  We show two scaling
relations of this nature in Fig.~\ref{fig:mu_relations}: the mean
surface brightness within $R_{\rm eff}$, $\langle \mu_{3.6, \rm{eff}}
\rangle$ (left panel); and the mean surface brightness within 1~kpc
radius, $\langle \mu_{3.6, 1\rm{kpc}} \rangle$ (right panel), both
plotted against stellar mass.  The former relation has been
demonstrated since photometry of low-mass, low surface brightness
galaxies became possible, and was always known to have a fairly high
scatter \citep[e.g.,][]{impey96}.  A more recent demonstration of this
relation came from \citet{sedgwick19}, who found a correlation between
effective $r$-band surface brightness and stellar mass for
core-collapse supernova host galaxies with a slope of $\sim -1.35$.
Our regression line is shown in Fig.~\ref{fig:mu_relations}, and has
the following form:
\begin{equation}\label{eq:mueff_fit}
  \mu_{3.6, \rm{eff}} = -1.79\log\,(\mathcal{M}_{*}/\mathcal{M}_{\odot})
  + 37.90
\end{equation}
where the steeper slope than that of \citet{sedgwick19} likely arises
from the difference in photometric band and galaxy sample
(core-collapse SNe hosts typically being star-forming galaxies).  In
NIR, ETGs and LTGs both follow similar relations, although below
$\log\,(\mathcal{M}_{*}/\mathcal{M}_{\odot}) \sim 10$ ETGs trend
slightly higher in effective surface brightness than LTGs of the same stellar mass.  This is to be expected, however, as ETGs in this mass range
also have smaller $R_{\rm eff}$ on average than LTGs
(Fig.~\ref{fig:reff}) and higher $C_{82}$ (Fig.~\ref{fig:c82mass}).
Much of the scatter in this relation thus might arise from the large
range of relative physical radii probed by $R_{\rm eff}$ for galaxies
with a variety of photometric profiles.

A more consistent choice of physical radius yields a much tighter
correlation, as we demonstrate in the rightmost panel of
Fig.~\ref{fig:mu_relations} via the $\langle \mu_{3.6, 1\rm{kpc}}
\rangle$--$\log\,(\mathcal{M}_{*}/\mathcal{M}_{\odot})$ relation.  This,
too, is a known feature of ETGs; \citet{saracco17}, for example, found
a tight relation between stellar mass and the mass surface density
within 1~kpc ($\Sigma^{*}_{1\rm{kpc}}$) among ETGs at a wide range of
redshifts \citep[see also][]{saracco12, tacchella16, arora21}, indicating that
the average core densities of these galaxies have not evolved much
during the past $\sim$10~Gyr \citep[echoing, in a different manner, the
  results found by][and references therein]{szomoru12}.

The linear fits shown by the red dashed lines mirror those done by
\citet{saracco17}, who found a kink in the
$\Sigma^{*}_{1\rm{kpc}}$--$\log\,(\mathcal{M}_{*}/\mathcal{M}_{\odot})$
relation at $\log\,(\mathcal{M}_{*}/\mathcal{M}_{\odot}) \sim 10.4$.
Here we fit only ETGs, with $T \leq 0$, for better comparison
with that study.  \citet{tacchella16} found a similar bend using
galaxy models from a hydrodynamical cosmological zoom simulation,
although they placed its location at
$\log\,(\mathcal{M}_{*}/\mathcal{M}_{\odot}) \sim 10.2$.  Both find a
low-mass-end slope in this relation of nearly 1 and a high-mass-end
slope of $\sim0.6$.  If we split our ETG population as
\citet{saracco17}, we find the following linear fits for the low- and
high-mass end, respectively:
\begin{equation}\label{eq:mu1kpc_low_mu}
  \mu_{3.6,~\log\,(\mathcal{M}_{*}/\mathcal{M}_{\odot}) < 10.4} =
  - 2.39\log\,(\mathcal{M}_{*}/\mathcal{M}_{\odot}) + 42.19
\end{equation}
\begin{equation}\label{eq:mu1kpc_high_mu}
  \mu_{3.6,~\log\,(\mathcal{M}_{*}/\mathcal{M}_{\odot}) > 10.4} =
  - 1.19\log\,(\mathcal{M}_{*}/\mathcal{M}_{\odot}) + 29.65
\end{equation}
For easier comparison, we convert our surface brightnesses to stellar
mass surface densities using the relation given by Eq.~A5 from \citet{munoz13},
which yields:
\begin{equation}\label{eq:mu1kpc_low_mstar}
  \log(\Sigma^{*}_{1\rm{kpc},~\log\,(\mathcal{M}_{*}/\mathcal{M}_{\odot}) < 10.4}) =
  0.96\log\,(\mathcal{M}_{*}/\mathcal{M}_{\odot}) - 6.12
\end{equation}
\begin{equation}\label{eq:mu1kpc_high_mstar}
  \log(\Sigma^{*}_{1\rm{kpc},~\log\,(\mathcal{M}_{*}/\mathcal{M}_{\odot}) > 10.4}) =
  0.48\log\,(\mathcal{M}_{*}/\mathcal{M}_{\odot}) -1.10
\end{equation}
Where $\Sigma^{*}_{\rm{kpc}}$ is in units of $\mathcal{M}_{\odot} /
\rm{kpc}^{2}$.  The low-mass-end fit is very close to that found by
\citet{saracco17}, though we find a slightly shallower high-mass-end
fit.  Recently, \citet{arora21} compared this relation between many different studies, and found all of them showed similar behavior: LTGs and low-mass ETGs show similar slopes, while high-mass ETGs show a much shallower slope.  They replicate this in their own study as well, though they provide a different location for the high-mass bend than either study cited above: $\log\,(\mathcal{M}_{*}/\mathcal{M}_{\odot}) \sim 10.7$.

The bend does appear to be real, and not just a result of
increased scatter at the high-mass end of the relation: if we do a
similar double-fit for LTGs, we find slopes in the
$\mu_{3.6}$--$\log\,(\mathcal{M}_{*}/\mathcal{M}_{\odot})$ relation of
$-2.1$ and $-1.9$ for the low- and high-mass end, respectively, both
very similar to the low-mass end fit for the ETGs despite the higher
scatter.  Likewise, a fit to the entire S$^{4}$G+ETG population, shown
in black, has the form:
\begin{equation}\label{eq:mu1kpc_noe}
  \langle \mu_{3.6, 1\rm{kpc}} \rangle =
  -2.30\log\,(\mathcal{M}_{*}/\mathcal{M}_{\odot}) + 41.74
\end{equation}
which, when converted to mass surface density, has a slope of
$\sim0.93$, very close still to the slope of $\sim1.02$ given by
\citet{saracco17} for the low-mass end of their ETG relation, and very similar to the ETG+LTG best-fit slope of 0.91 found by \citet{arora21}.  Both
S0s and ellipticals contribute to the bend in the relation at high stellar mass; the relation's high-mass-end slope is $-1.23$ including only ellipticals, and $-1.28$ including only S0s.

The range of values in $\langle\mu_{3.6, 1\rm{kpc}}\rangle$ for ETGs
above $10^{10.4} \mathcal{M}_{\odot}$ are the same as those for LTGs
in the same mass range.  The most evident morphological divide in this
relation lies between ETGs and LTGs below this stellar mass, where ETGs
occupy the high-surface-brightness end of the distribution at a given
stellar mass, which is interesting if ETGs indeed evolve directly from
LTGs.  In disk galaxies, 1~kpc is well within the typical sizes of
bars and possibly related structures like rings or lenses \citep[on
  average, $\sim 4$\,kpc;][and references therein]{laurikainen11,
  laurikainen13}, hence this metric $\langle\mu_{3.6,
  1\rm{kpc}}\rangle$ traces the regime of central structure, such as
bulges.

If ETGs do evolve directly from LTGs, the bend at high
stellar masses could imply that whatever mechanisms causing the
evolution increase central densities only in lower mass galaxies,
leaving high-mass ETG cores untouched relative to their LTG
ancestors.  \citet{kim12}, for example, found evidence for the immutability of high-mass ETGs by demonstrating that the cores of ETGs in their sample with clear tidal debris lie on the same \citet{kormendy77} relation as their undisturbed counterparts.  This could be related to bulge formation.  The bulge-to-total ratio for S0s increases with increasing galaxy mass \citep[see Fig.~7 of][]{laurikainen10}, and, as stated previously, large bulges tend to inhibit bar formation \citep[and references therein]{kataria20}.  This in turn could stabilize gas and dust against angular momentum loss \citep[e.g.,][]{athanassoula92,
  maciejewski02, carles16}, or inhibit the formation of inner rings \citep[assuming such structures are related to bar resonances; e.g.,][]{buta96, rautiainen00, athanassoula10}, which are often sites of intense star formation even among S0s \citep[e.g.,][]{comeron13}. Inner structures, given their deeper gravitational potentials, are less likely to be stripped of gas and dust than disks, meaning central star formation (and thus central mass growth) should outlast disk star formation (and thus disk mass growth) during many quenching scenarios, leading to the higher central surface brightnesses we see here.  Otherwise, some subsequent mechanism in high-mass ETGs (both S0s and ellipticals) must act to decrease their core densities
relative to their lower mass counterparts.

In summary, no matter the metric used, ETG stellar mass scales positively with ETG core surface brightnesses, with little dependence on galaxy morphology.  We reproduce well the low-scatter log-linear relationship between total stellar mass and the mean stellar mass surface density within 1~kpc, providing further evidence that the cores of ETGs have evolved little over cosmic time.  This is also true of LTGs, albeit with subtle differences from ETGs: we find that low-mass ETGs have typically higher core surface brightnesses than LTGs of the same stellar mass, suggesting that whatever mechanism might transform the latter to the former typically halts star formation from the outside-in: disks quench first, followed by bulges.

\subsection{m$_{3.6} - $m$_{4.5}$ Color}\label{sec:color}

\begin{figure*}
  \centering
  \includegraphics[scale=1.0]{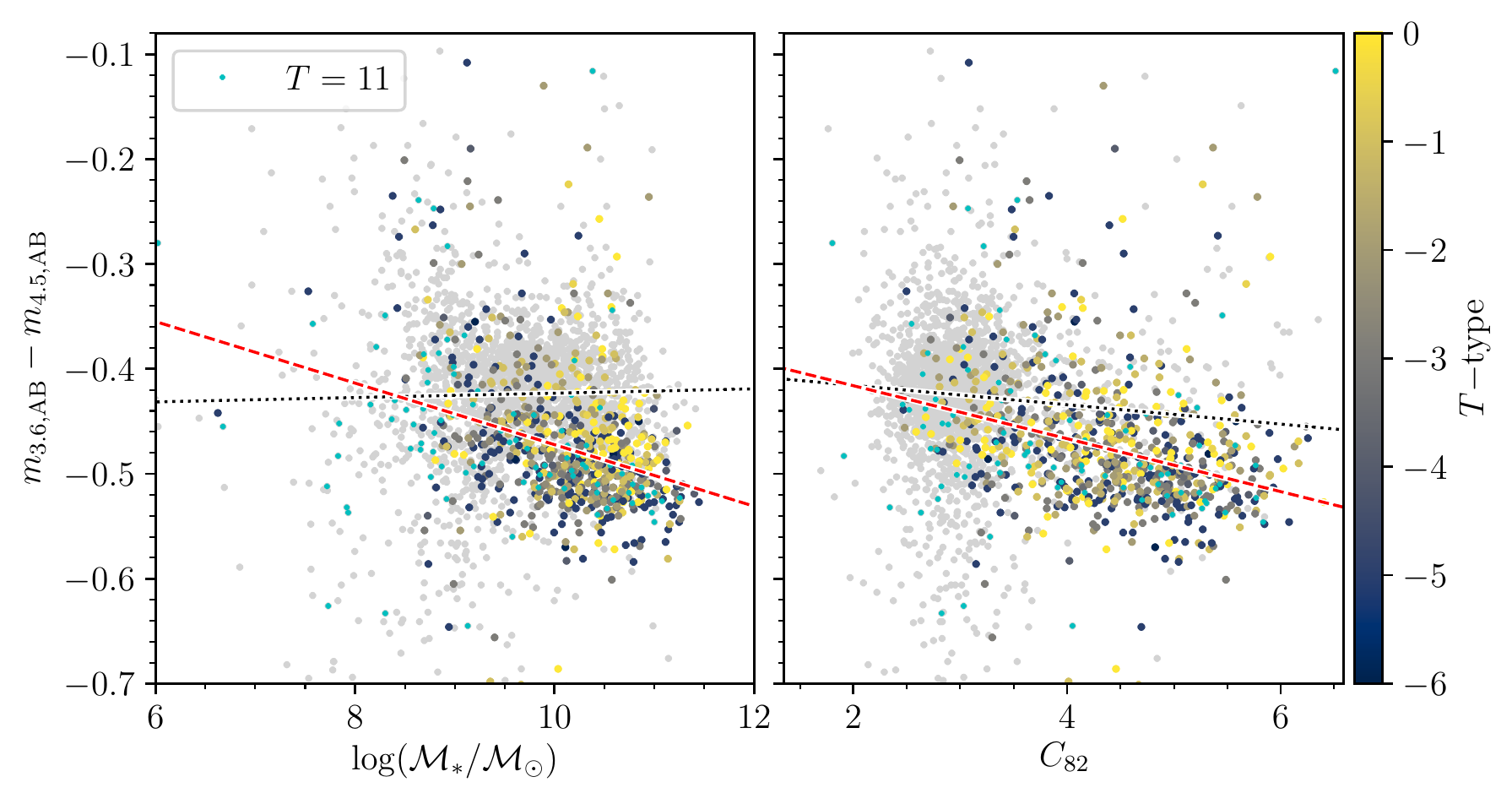}
  \caption{Relationships between $m_{3.6} - m_{4.5}$ color and
    stellar mass (\emph{left} panel) or concentration (\emph{right}
    panel).  The dashed red lines show linear fits to the ETG
    populations only, while dotted black lines shows fits to the
    LTGs.  ETG data points are multi-colored, while LTGs are shown in gray.
    \label{fig:color}}
\end{figure*}

The 3.6\,$\mu$m and 4.5\,$\mu$m bands are fairly insensitive to star
formation history, making the $m_{3.6} - m_{4.5}$ color nearly
constant across stellar populations \citep[e.g.,][]{meidt12, bonanos10, meidt14, querejeta15,
  williams16}.  Indeed, given how they trace only the Rayleigh-Jeans
tail in normal stellar populations, these colors are most often
employed in either very low-mass stellar or high-mass planetary
studies \citep[e.g.,][]{allers06, luhman08}, or else are used to trace
highly evolved stars \citep[e.g.,][]{mould08, bonanos09, bonanos10,
  williams16}.

Due to a CO absorption feature present in the 4.5\,$\mu$m band
\citep{cohen96}, $m_{3.6} - m_{4.5}$ can be a tracer of
metallicity in integrated stellar populations.  In a detailed study of
the NIR colors of elliptical galaxies, \citet{peletier12} demonstrated
that the mass-metallicity relation is traceable in these bands via a
slight blueward $m_{3.6} - m_{4.5}$ color trend with
increasing mass, as lower metallicities yield higher average stellar
temperatures, which weaken the absorption in the 4.5\,$\mu$m band \citep[see also][]{meidt14}.
Young stellar populations are also hotter.  Hence, this
relation should be clearer for galaxies with predominantly old stellar
populations than for mixed young and old populations.  For \citet{peletier12}, this manifested as a higher
scatter in the color--velocity dispersion relation for colors measured
within $R_{\rm eff}/8$ than for those measured within $R_{\rm eff}$,
as many galaxies in their sample had centrally concentrated star
formation and dust.  However, from synthesized stellar population models calibrated in the NIR using the $m_{3.6} - m_{4.5}$ colors of giant stars, \citet{meidt14} found that stellar age has an extremely limited impact on these colors (their Figs. 2 and 3), suggesting that dust and other non-stellar emission and absorption processes constitute the primary factor behind this increased dispersion.  Stellar age does influence this color when dust emission is taken into account, however, as dust heated by young stars contributes to both 3.6\,$\mu$m and 4.5\,$\mu$m luminosities \citep[e.g.,][]{zhu10, meidt12, jarrett13, querejeta15}, making the exact correlation between $m_{3.6} - m_{4.5}$ and stellar population age quite complicated.

We show this correlation for integrated colors in the left panel of
Fig.~\ref{fig:color}, using our 3.6\,$\mu$m and 4.5\,$\mu$m asymptotic
magnitudes \citep[in the AB system, hence they show an offset from the Vega magnitudes used by][]{peletier12}.  Despite using asymptotic rather than central colors, we
also find a blueward trend with increasing stellar mass for ETGs
(shown via the red dashed line, a linear fit to all ETG points; the corresponding Pearson's correlation coefficient is $\rho = -0.136$, with $p = 0.00015$).
While \citet{peletier12} studied only elliptical galaxies, we find that S0s
follow the same trend.  LTGs do not; the linear fit to
LTGs is shown as a black dotted line, and is nearly flat, with a
Pearson's $\rho$ correlation coefficient that is not statistically
significant ($p\textrm{-value}=0.86$, meaning the null hypothesis of no
correlation cannot be rejected).

In the right panel of Fig.~\ref{fig:color}, we show the correlation
between integrated $m_{3.6} - m_{4.5}$ color and the concentration
parameter $C_{82}$ (Eq.~\ref{eq:c82}).  This shows a similar blueward
trend (red dashed line; $\rho = -0.120$, $p = 2.67 \times 10^{-10}$) as the stellar mass correlation, as expected
given that $C_{82}$ correlates already with stellar mass
(Fig.~\ref{fig:c82mass}).  Again, LTGs show no
significant correlation.  They do, however, fall at the tail end of
the ETG correlation, appearing as a cloud of points near $C_{82} \sim
2.6$ (corresponding to $n=1$).  Indeed, a fit to the entire population
of S$^{4}$G+ETG galaxies gives a very similar slope to the ETG fit
alone ($-0.024$ and $-0.025$, respectively).  \citet{meidt14} also found that metallicity had the strongest impact on dust-free synthetic stellar population colors, due to CO absorption in the photospheres of giant stars \citep[echoing][and references therein]{peletier12}.  Given their lower masses
(Fig.~\ref{fig:hists}), LTGs are also on average lower in metallicity
\citep[e.g.,][]{tremonti04, gallazzi05}, so this may be primarily what is
driving this behavior.  A simple calculation provides evidence for
this.  From our linear fit:
\begin{equation}\label{eq:colc82}
  m_{3.6}-m_{4.5} = -0.025C_{82} - 0.365
\end{equation}
we can predict the expected color difference between the two
populations.  The median value of $C_{82}$ for ETGs is 4.40 (Section
\ref{sec:histograms}), while that of LTGs is 2.96, yielding a
difference in $C_{82}$ of 1.44.  The expected color difference from
Eq.~\ref{eq:colc82} is thus $-0.025 \times 1.44 \approx -0.04$ mag.  The
median color of the ETG sample is -0.48 mag, while that of the LTG
sample is -0.42 mag, yielding a median color difference of -0.06 mag,
slightly redward from the expectation from the linear relation.  If age has little impact on these colors in uncontaminated stellar photosphere emission, the remaining $\sim 0.02$ mag color difference could be attributable to non-stellar emission, which \citet{meidt12} showed has a fairly small impact on integrated quantities (of order 10\% on stellar mass estimates, for example).

In summary, we replicate the blue-ward $m_{3.6} - m_{4.5}$ color trend with stellar mass among ETGs using integrated magnitudes, in contrast to the LTG population, which shows no color trend with mass.  This trend leads to a color-concentration relation as well, with LTGs occupying the low-concentration, red color end of the entire S$^{4}$G+ETG population, when metallicity and dust effects are taken into consideration.

\section{Data online access}\label{sec:access}

As for the initial S$^{4}$G results \citep[MM2015;][]{salo15, querejeta15}, all of our data will be made available online at a dedicated webpage
at the NASA/IPAC Infrared Science Archive.  We will provide, for each galaxy, the P1 images and weight maps, the P2
masks, and the P3 radial profiles (as ASCII tables).  Once they become
available, we will also upload the P4 multi-component decompositions and any additional pipeline products to this same location.

Additionally, we will make all of our derived parameters available in
table format: asymptotic magnitudes in both photometric bands
(and associated uncertainties), both apparent and absolute; concentration parameters
in both photometric bands; stellar masses; distances; effective radii
in both photometric bands; local sky estimates in both photometric
bands (and associated uncertainties); isophotal radii ($R_{25.5}$ and
$R_{26.5}$) in both photometric bands; isophotal parameters
associated with these radii (position angles and ellipticities); and revised morphological classifications based on the 3.6\,$\mu$m images.  For
the sake of reproducibility, we will also include the limits within
which we made our linear fits for deriving our asymptotic
magnitudes (drawn from the radial profiles).  Also following MM2015,
we are including ancillary values from NED and HyperLEDA\footnote{These
  values are not dynamically linked to either catalogue, hence changes
  to either will not be reflected in these tables.}, including optical
colors, gas content, internal kinematics, etc., if the user wishes to
define specific subsamples selected according to specific criteria.

Finally, we will also make all of the above tables and FITS images available at the CDS via anonymous ftp to \url{cdsarc.u-strasbg.fr} (130.79.128.5) or via \url{http://cdsarc.u-strasbg.fr/viz-bin/cat/J/A+A}.

\section{Summary}\label{sec:summary}

We have presented the data release of the early-type galaxy extension
to the S$^{4}$G, an NIR survey of nearby galaxies in a
wavelength regime (3.6\,$\mu$m and 4.5\,$\mu$m) that is a nearly direct
tracer of stellar mass.  This extension includes newly calibrated
images of 465 nearby ETGs not observed during the original
S$^{4}$G.  We have now processed these images up through Pipeline 3,
which derives surface photometry and associated parameters for each
galaxy.  This includes radial surface brightness profiles, curves of
growth, asymptotic magnitudes (converted as well to stellar masses),
effective radii, isophotal radii, and concentration parameters.  We also
include updated CVHRS morphological classifications for these ETGs.

Using these parameters, we have conducted a preliminary investigation
into the following scaling relations:
\begin{itemize}
  \item Stellar mass--concentration, in which we show that S0 and
    elliptical galaxies follow roughly the same correlation, though
    disk galaxies tend to scatter to lower concentrations at a given
    stellar mass (which is also true of LTGs).  We also demonstrate that
    single-component fits to ETGs tend to underestimate concentration
    by overestimating the luminosity of the galaxies' outskirts.  This
    results in single-component magnitudes that are slightly too
    bright compared to their empirical counterparts;
    multi-component fits are necessary even for these otherwise visually
    simple elliptical galaxies.

  \item Stellar mass--size, including both effective and isophotal
    radius.  We find good agreement with past measures of the
    mass--effective radius relation for ETGs at $z=0$, taking into
    account methodological differences.  We find that the
    corresponding isophotal radius relation has surprisingly low
    scatter once galaxy inclination is corrected for, including among
    the more pressure-supported, triaxial ellipticals.

  \item Surface brightness profile shape, including relations between
    stellar mass and central surface brightness, median surface
    brightness within the effective radius, and median surface
    brightness within 1~kpc radius.  Surface brightness profiles of
    high-mass ETGs are brighter and, when normalized by effective
    radius, more extended than those of low-mass ETGs, but there is no
    noticeable trend with ETG morphological class.  We reproduce
    the log-linear mass--central surface brightness trend found by
    other studies, but we note that, as with concentration parameters,
    single-component fits used to estimate this tend to fall short in
    galaxies with complex structure, often leading to extreme fitted
    S\'{e}rsic indices.  The tightest of these surface brightness
    scaling relations is between stellar mass and surface brightness
    within 1~kpc, with little variation even between ETGs and LTGs,
    suggesting that the central 1~kpc of most galaxies is only mildly
    affected by whatever processes lead to morphological
    differentiation over cosmic time.

  \item Stellar mass--NIR color, in which we reproduce the tendency
    for $m_{3.6} - m_{4.5}$ to become bluer for higher mass
    galaxies, a result of a CO absorption feature in the 4.5\,$\mu$m
    band.  This trend is echoed in the concentration-color relation,
    where LTGs form roughly the low-concentration, reddest color tail
    of the ETG trend.
\end{itemize}

Altogether, our preliminary look into these scaling relations shows
good agreement with those of previous studies, but also includes some
details that are worthy of further investigations.  All of these
images, profiles, and derived parameters will be made available via the
NASA/IPAC Infrared Science Archive.

\begin{acknowledgements}
AW acknowledges support from the STFC [ST/S00615X/1].  SC acknowledges funding from the State Research Agency (AEI-MCINN) of the Spanish Ministry of Science and Innovation under the grants “The structure and evolution of galaxies and their central regions” with reference PID2019-105602GB-I00/10.13039/501100011033, and  “Thick discs, relics of the infancy of galaxies" with reference PID2020-113213GA-I00. This project
has received funding from the European Union’s Horizon 2020 research
and innovation programme under the Marie Sk{\l}odowska-Curie grant
agreement No.~893673 (SDG), as well as grant agreement No.~721463 to
the SUNDIAL ITN network (AHS, HS, JHK), from the State Research Agency
(AEI-MCINN) of the Spanish Ministry of Science and Innovation under
the grant “The structure and evolution of galaxies and their central
regions” with reference PID2019-105602GB-I00/10.13039/501100011033,
and from the IAC project P/300724 which is financed by the Ministry of
Science and Innovation, through the State Budget and by the Canary
Islands Department of Economy, Knowledge and Employment, through the
Regional Budget of the Autonomous Community.  EA and AB gratefully acknowledge financial support from CNES (Centre National d'\'{E}tudes Spatiales, France).  LCH is supported by the National Science Foundation of China (11721303 and 11991052) and the National Key R\&D Program of China (2016YFA0400702). TK acknowledges support from the Basic Science Research Program through the National Research Foundation of Korea (NRF) funded by the Ministry of Education (NRF-2019R1I1A3A02062242). KMD thanks the support of the Serrapilheira Institute (grant Serra-1709-17357) as well as that of the Brazilian National Council for Scientific and Technological Development (CNPq grant 312702/2017-5) and of the Carlos Chagas Filho Foundation for Supporting Research in the State of Rio de Janeiro (FAPERJ grant E-26/203.184/2017), Brazil. This work is
based on observations made with the Spitzer Space Telescope, which
was operated by the Jet Propulsion Laboratory, California Institute
of Technology under a contract with NASA. Support for this work was
provided by NASA through an award issued by JPL/Caltech.  This research has made
use of the NASA/IPAC Extragalactic Database (NED), which is funded by
the National Aeronautics and Space Administration and operated by the
California Institute of Technology.  This work made use of the Python
packages Numpy \citep{harris20}, Scipy \citep{virtanen20}, and
Matplotlib \citep{hunter07}.  We thank Jarkko Laine for providing us with some of the IRAF scripts used to produce the radial profiles discussed in MM2015, which were critical to us for keeping the methodology consistent across such a gap of time.  Finally, we thank the anonymous referee for their valuable feedback.
\end{acknowledgements}

\bibliographystyle{aa}
\bibliography{refs}

\begin{thebibliography}{259}
\expandafter\ifx\csname natexlab\endcsname\relax\def\natexlab#1{#1}\fi

\bibitem[{{Abraham} {et~al.}(1994){Abraham}, {Valdes}, {Yee}, \& {van den
  Bergh}}]{abraham94}
{Abraham}, R.~G., {Valdes}, F., {Yee}, H.~K.~C., \& {van den Bergh}, S. 1994,
  \apj, 432, 75

\bibitem[{{Aguerri}(2012)}]{aguerri12}
{Aguerri}, J. A.~L. 2012, Advances in Astronomy, 2012, 382674

\bibitem[{{Aguerri} {et~al.}(2001){Aguerri}, {Balcells}, \&
  {Peletier}}]{aguerri01}
{Aguerri}, J.~A.~L., {Balcells}, M., \& {Peletier}, R.~F. 2001, \aap, 367, 428

\bibitem[{{Aguerri} {et~al.}(2005){Aguerri}, {Elias-Rosa}, {Corsini}, \&
  {Mu{\~n}oz-Tu{\~n}{\'o}n}}]{aguerri05}
{Aguerri}, J.~A.~L., {Elias-Rosa}, N., {Corsini}, E.~M., \&
  {Mu{\~n}oz-Tu{\~n}{\'o}n}, C. 2005, \aap, 434, 109

\bibitem[{{Aihara} {et~al.}(2018){Aihara}, {Armstrong}, {Bickerton}, {Bosch},
  {Coupon}, {Furusawa}, {Hayashi}, {Ikeda}, {Kamata}, {Karoji}, {Kawanomoto},
  {Koike}, {Komiyama}, {Lang}, {Lupton}, {Mineo}, {Miyatake}, {Miyazaki},
  {Morokuma}, {Obuchi}, {Oishi}, {Okura}, {Price}, {Takata}, {Tanaka},
  {Tanaka}, {Tanaka}, {Uchida}, {Uraguchi}, {Utsumi}, {Wang}, {Yamada},
  {Yamanoi}, {Yasuda}, {Arimoto}, {Chiba}, {Finet}, {Fujimori}, {Fujimoto},
  {Furusawa}, {Goto}, {Goulding}, {Gunn}, {Harikane}, {Hattori}, {Hayashi},
  {He{\l}miniak}, {Higuchi}, {Hikage}, {Ho}, {Hsieh}, {Huang}, {Huang},
  {Imanishi}, {Iwata}, {Jaelani}, {Jian}, {Kashikawa}, {Katayama}, {Kojima},
  {Konno}, {Koshida}, {Kusakabe}, {Leauthaud}, {Lee}, {Lin}, {Lin},
  {Mandelbaum}, {Matsuoka}, {Medezinski}, {Miyama}, {Momose}, {More}, {More},
  {Mukae}, {Murata}, {Murayama}, {Nagao}, {Nakata}, {Niida}, {Niikura},
  {Nishizawa}, {Oguri}, {Okabe}, {Ono}, {Onodera}, {Onoue}, {Ouchi}, {Pyo},
  {Shibuya}, {Shimasaku}, {Simet}, {Speagle}, {Spergel}, {Strauss}, {Sugahara},
  {Sugiyama}, {Suto}, {Suzuki}, {Tait}, {Takada}, {Terai}, {Toba}, {Turner},
  {Uchiyama}, {Umetsu}, {Urata}, {Usuda}, {Yeh}, \& {Yuma}}]{aihara18}
{Aihara}, H., {Armstrong}, R., {Bickerton}, S., {et~al.} 2018, \pasj, 70, S8

\bibitem[{{Allers} {et~al.}(2006){Allers}, {Jaffe}, {van der Bliek}, {Allard},
  \& {Baraffe}}]{allers06}
{Allers}, K.~N., {Jaffe}, D.~T., {van der Bliek}, N.~S., {Allard}, F., \&
  {Baraffe}, I. 2006, in Astronomical Society of the Pacific Conference Series,
  Vol. 357, The Spitzer Space Telescope: New Views of the Cosmos, ed.
  L.~{Armus} \& W.~T. {Reach}, 77

\bibitem[{{Andreon} {et~al.}(2016){Andreon}, {Dong}, \& {Raichoor}}]{andreon16}
{Andreon}, S., {Dong}, H., \& {Raichoor}, A. 2016, \aap, 593, A2

\bibitem[{{Arora} {et~al.}(2021){Arora}, {Stone}, {Courteau}, \&
  {Jarrett}}]{arora21}
{Arora}, N., {Stone}, C., {Courteau}, S., \& {Jarrett}, T.~H. 2021, \mnras,
  505, 3135

\bibitem[{{Athanassoula}(1992)}]{athanassoula92}
{Athanassoula}, E. 1992, \mnras, 259, 345

\bibitem[{{Athanassoula}(2003)}]{athanassoula03}
{Athanassoula}, E. 2003, \mnras, 341, 1179

\bibitem[{{Athanassoula}(2005)}]{athanassoula05}
{Athanassoula}, E. 2005, \mnras, 358, 1477

\bibitem[{{Athanassoula} {et~al.}(2005){Athanassoula}, {Lambert}, \&
  {Dehnen}}]{athanassoula05b}
{Athanassoula}, E., {Lambert}, J.~C., \& {Dehnen}, W. 2005, \mnras, 363, 496

\bibitem[{{Athanassoula} {et~al.}(2016){Athanassoula}, {Rodionov}, {Peschken},
  \& {Lambert}}]{athanassoula16}
{Athanassoula}, E., {Rodionov}, S.~A., {Peschken}, N., \& {Lambert}, J.~C.
  2016, \apj, 821, 90

\bibitem[{{Athanassoula} {et~al.}(2010){Athanassoula}, {Romero-G{\'o}mez},
  {Bosma}, \& {Masdemont}}]{athanassoula10}
{Athanassoula}, E., {Romero-G{\'o}mez}, M., {Bosma}, A., \& {Masdemont}, J.~J.
  2010, \mnras, 407, 1433

\bibitem[{{Balcells} {et~al.}(2003){Balcells}, {Aguerri}, \&
  {Peletier}}]{balcells03}
{Balcells}, M., {Aguerri}, J. A.~L., \& {Peletier}, R.~F. 2003, in The Mass of
  Galaxies at Low and High Redshift, ed. R.~{Bender} \& A.~{Renzini}, 97

\bibitem[{{Baldry} {et~al.}(2004){Baldry}, {Glazebrook}, {Brinkmann},
  {Ivezi{\'c}}, {Lupton}, {Nichol}, \& {Szalay}}]{baldry04}
{Baldry}, I.~K., {Glazebrook}, K., {Brinkmann}, J., {et~al.} 2004, \apj, 600,
  681

\bibitem[{{Barnes}(1988)}]{barnes88}
{Barnes}, J.~E. 1988, \apj, 331, 699

\bibitem[{{Barnes}(1990)}]{barnes90}
{Barnes}, J.~E. 1990, \nat, 344, 379

\bibitem[{{Begelman} {et~al.}(1980){Begelman}, {Blandford}, \&
  {Rees}}]{begelman80}
{Begelman}, M.~C., {Blandford}, R.~D., \& {Rees}, M.~J. 1980, \nat, 287, 307

\bibitem[{{Bekki} {et~al.}(2002){Bekki}, {Couch}, \& {Shioya}}]{bekki02}
{Bekki}, K., {Couch}, W.~J., \& {Shioya}, Y. 2002, \apj, 577, 651

\bibitem[{{Bendo} \& {Barnes}(2000)}]{bendo00}
{Bendo}, G.~J. \& {Barnes}, J.~E. 2000, \mnras, 316, 315

\bibitem[{{Bershady} {et~al.}(2000){Bershady}, {Jangren}, \&
  {Conselice}}]{bershady00}
{Bershady}, M.~A., {Jangren}, A., \& {Conselice}, C.~J. 2000, \aj, 119, 2645

\bibitem[{{Bertin} \& {Arnouts}(1996)}]{bertin96}
{Bertin}, E. \& {Arnouts}, S. 1996, \aaps, 117, 393

\bibitem[{{Bertola} \& {Capaccioli}(1975)}]{bertola75}
{Bertola}, F. \& {Capaccioli}, M. 1975, \apj, 200, 439

\bibitem[{{Bezanson} {et~al.}(2009){Bezanson}, {van Dokkum}, {Tal},
  {Marchesini}, {Kriek}, {Franx}, \& {Coppi}}]{bezanson09}
{Bezanson}, R., {van Dokkum}, P.~G., {Tal}, T., {et~al.} 2009, \apj, 697, 1290

\bibitem[{{Binggeli} \& {Cameron}(1991)}]{binggeli91}
{Binggeli}, B. \& {Cameron}, L.~M. 1991, \aap, 252, 27

\bibitem[{{Binggeli} {et~al.}(1985){Binggeli}, {Sandage}, \&
  {Tammann}}]{binggeli85}
{Binggeli}, B., {Sandage}, A., \& {Tammann}, G.~A. 1985, \aj, 90, 1681

\bibitem[{{Binney}(1978)}]{binney78}
{Binney}, J. 1978, \mnras, 183, 501

\bibitem[{{Bonanos} {et~al.}(2010){Bonanos}, {Lennon}, {K{\"o}hlinger}, {van
  Loon}, {Massa}, {Sewilo}, {Evans}, {Panagia}, {Babler}, {Block}, {Bracker},
  {Engelbracht}, {Gordon}, {Hora}, {Indebetouw}, {Meade}, {Meixner}, {Misselt},
  {Robitaille}, {Shiao}, \& {Whitney}}]{bonanos10}
{Bonanos}, A.~Z., {Lennon}, D.~J., {K{\"o}hlinger}, F., {et~al.} 2010, \aj,
  140, 416

\bibitem[{{Bonanos} {et~al.}(2009){Bonanos}, {Massa}, {Sewilo}, {Lennon},
  {Panagia}, {Smith}, {Meixner}, {Babler}, {Bracker}, {Meade}, {Gordon},
  {Hora}, {Indebetouw}, \& {Whitney}}]{bonanos09}
{Bonanos}, A.~Z., {Massa}, D.~L., {Sewilo}, M., {et~al.} 2009, \aj, 138, 1003

\bibitem[{{Boselli} {et~al.}(2008){Boselli}, {Boissier}, {Cortese}, \&
  {Gavazzi}}]{boselli08}
{Boselli}, A., {Boissier}, S., {Cortese}, L., \& {Gavazzi}, G. 2008, \aap, 489,
  1015

\bibitem[{{Bouquin} {et~al.}(2018){Bouquin}, {Gil de Paz}, {Mu{\~n}oz-Mateos},
  {Boissier}, {Sheth}, {Zaritsky}, {Peletier}, {Knapen}, \&
  {Gallego}}]{bouquin18}
{Bouquin}, A. Y.~K., {Gil de Paz}, A., {Mu{\~n}oz-Mateos}, J.~C., {et~al.}
  2018, \apjs, 234, 18

\bibitem[{{Bournaud} {et~al.}(2007){Bournaud}, {Jog}, \& {Combes}}]{bournaud07}
{Bournaud}, F., {Jog}, C.~J., \& {Combes}, F. 2007, \aap, 476, 1179

\bibitem[{{Buitrago} {et~al.}(2008){Buitrago}, {Trujillo}, {Conselice},
  {Bouwens}, {Dickinson}, \& {Yan}}]{buitrago08}
{Buitrago}, F., {Trujillo}, I., {Conselice}, C.~J., {et~al.} 2008, \apjl, 687,
  L61

\bibitem[{{Burstein} {et~al.}(2005){Burstein}, {Ho}, {Huchra}, \&
  {Macri}}]{burstein05}
{Burstein}, D., {Ho}, L.~C., {Huchra}, J.~P., \& {Macri}, L.~M. 2005, \apj,
  621, 246

\bibitem[{{Busko}(1996)}]{busko96}
{Busko}, I.~C. 1996, in Astronomical Society of the Pacific Conference Series,
  Vol. 101, Astronomical Data Analysis Software and Systems V, ed. G.~H.
  {Jacoby} \& J.~{Barnes}, 139

\bibitem[{{Buta} \& {Combes}(1996)}]{buta96}
{Buta}, R. \& {Combes}, F. 1996, \fcp, 17, 95

\bibitem[{{Buta} {et~al.}(2007){Buta}, {Corwin}, \& {Odewahn}}]{buta07}
{Buta}, R.~J., {Corwin}, H.~G., \& {Odewahn}, S.~C., eds. 2007, {The de
  Vaucouleurs Atlas of Galaxies} (Cambridge University Press, Cambridge, UK)

\bibitem[{{Buta} {et~al.}(2015){Buta}, {Sheth}, {Athanassoula}, {Bosma},
  {Knapen}, {Laurikainen}, {Salo}, {Elmegreen}, {Ho}, {Zaritsky}, {Courtois},
  {Hinz}, {Mu{\~n}oz-Mateos}, {Kim}, {Regan}, {Gadotti}, {Gil de Paz}, {Laine},
  {Men{\'e}ndez-Delmestre}, {Comer{\'o}n}, {Erroz Ferrer}, {Seibert},
  {Mizusawa}, {Holwerda}, \& {Madore}}]{buta15}
{Buta}, R.~J., {Sheth}, K., {Athanassoula}, E., {et~al.} 2015, \apjs, 217, 32

\bibitem[{{Buta} {et~al.}(2010){Buta}, {Sheth}, {Regan}, {Hinz}, {Gil de Paz},
  {Men{\'e}ndez-Delmestre}, {Munoz-Mateos}, {Seibert}, {Laurikainen}, {Salo},
  {Gadotti}, {Athanassoula}, {Bosma}, {Knapen}, {Ho}, {Madore}, {Elmegreen},
  {Masters}, {Comer{\'o}n}, {Aravena}, \& {Kim}}]{buta10}
{Buta}, R.~J., {Sheth}, K., {Regan}, M., {et~al.} 2010, \apjs, 190, 147

\bibitem[{{Calvi} {et~al.}(2012){Calvi}, {Poggianti}, {Fasano}, \&
  {Vulcani}}]{calvi12}
{Calvi}, R., {Poggianti}, B.~M., {Fasano}, G., \& {Vulcani}, B. 2012, \mnras,
  419, L14

\bibitem[{{Calzetti}(2001)}]{calzetti01}
{Calzetti}, D. 2001, \pasp, 113, 1449

\bibitem[{{Caon} {et~al.}(1993){Caon}, {Capaccioli}, \& {D'Onofrio}}]{caon93}
{Caon}, N., {Capaccioli}, M., \& {D'Onofrio}, M. 1993, \mnras, 265, 1013

\bibitem[{{Cappellari}(2016)}]{cappellari16}
{Cappellari}, M. 2016, \araa, 54, 597

\bibitem[{{Cappellari} {et~al.}(2007){Cappellari}, {Emsellem}, {Bacon},
  {Bureau}, {Davies}, {de Zeeuw}, {Falc{\'o}n-Barroso}, {Krajnovi{\'c}},
  {Kuntschner}, {McDermid}, {Peletier}, {Sarzi}, {van den Bosch}, \& {van de
  Ven}}]{cappellari07}
{Cappellari}, M., {Emsellem}, E., {Bacon}, R., {et~al.} 2007, \mnras, 379, 418

\bibitem[{{Cappellari} {et~al.}(2011{\natexlab{a}}){Cappellari}, {Emsellem},
  {Krajnovi{\'c}}, {McDermid}, {Scott}, {Verdoes Kleijn}, {Young}, {Alatalo},
  {Bacon}, {Blitz}, {Bois}, {Bournaud}, {Bureau}, {Davies}, {Davis}, {de
  Zeeuw}, {Duc}, {Khochfar}, {Kuntschner}, {Lablanche}, {Morganti}, {Naab},
  {Oosterloo}, {Sarzi}, {Serra}, \& {Weijmans}}]{cappellari11}
{Cappellari}, M., {Emsellem}, E., {Krajnovi{\'c}}, D., {et~al.}
  2011{\natexlab{a}}, \mnras, 413, 813

\bibitem[{{Cappellari} {et~al.}(2011{\natexlab{b}}){Cappellari}, {Emsellem},
  {Krajnovi{\'c}}, {McDermid}, {Serra}, {Alatalo}, {Blitz}, {Bois}, {Bournaud},
  {Bureau}, {Davies}, {Davis}, {de Zeeuw}, {Khochfar}, {Kuntschner},
  {Lablanche}, {Morganti}, {Naab}, {Oosterloo}, {Sarzi}, {Scott}, {Weijmans},
  \& {Young}}]{cappellari11b}
{Cappellari}, M., {Emsellem}, E., {Krajnovi{\'c}}, D., {et~al.}
  2011{\natexlab{b}}, \mnras, 416, 1680

\bibitem[{{Carles} {et~al.}(2016){Carles}, {Martel}, {Ellison}, \&
  {Kawata}}]{carles16}
{Carles}, C., {Martel}, H., {Ellison}, S.~L., \& {Kawata}, D. 2016, \mnras,
  463, 1074

\bibitem[{{Chamba} {et~al.}(2020){Chamba}, {Trujillo}, \& {Knapen}}]{chamba20}
{Chamba}, N., {Trujillo}, I., \& {Knapen}, J.~H. 2020, \aap, 633, L3

\bibitem[{{Cohen} {et~al.}(1996){Cohen}, {Witteborn}, {Bregman}, {Wooden},
  {Salama}, \& {Metcalfe}}]{cohen96}
{Cohen}, M., {Witteborn}, F.~C., {Bregman}, J.~D., {et~al.} 1996, \aj, 112, 241

\bibitem[{{Combes} \& {Sanders}(1981)}]{combes81}
{Combes}, F. \& {Sanders}, R.~H. 1981, \aap, 96, 164

\bibitem[{{Comer{\'o}n}(2013)}]{comeron13}
{Comer{\'o}n}, S. 2013, \aap, 555, L4

\bibitem[{{Comer{\'o}n} {et~al.}(2011){Comer{\'o}n}, {Elmegreen}, {Knapen},
  {Salo}, {Laurikainen}, {Laine}, {Athanassoula}, {Bosma}, {Sheth}, {Regan},
  {Hinz}, {Gil de Paz}, {Men{\'e}ndez-Delmestre}, {Mizusawa},
  {Mu{\~n}oz-Mateos}, {Seibert}, {Kim}, {Elmegreen}, {Gadotti}, {Ho},
  {Holwerda}, {Lappalainen}, {Schinnerer}, \& {Skibba}}]{comeron11}
{Comer{\'o}n}, S., {Elmegreen}, B.~G., {Knapen}, J.~H., {et~al.} 2011, \apj,
  741, 28

\bibitem[{{Comer{\'o}n} {et~al.}(2018){Comer{\'o}n}, {Salo}, \&
  {Knapen}}]{comeron18}
{Comer{\'o}n}, S., {Salo}, H., \& {Knapen}, J.~H. 2018, \aap, 610, A5

\bibitem[{{Comer{\'o}n} {et~al.}(2016){Comer{\'o}n}, {Salo}, {Peletier}, \&
  {Mentz}}]{comeron16}
{Comer{\'o}n}, S., {Salo}, H., {Peletier}, R.~F., \& {Mentz}, J. 2016, \aap,
  593, L6

\bibitem[{{Conselice}(2003)}]{conselice03}
{Conselice}, C.~J. 2003, \apjs, 147, 1

\bibitem[{{Conselice}(2006)}]{conselice06}
{Conselice}, C.~J. 2006, \mnras, 373, 1389

\bibitem[{{Corsini} {et~al.}(2017){Corsini}, {Wegner}, {Thomas}, {Saglia}, \&
  {Bender}}]{corsini17}
{Corsini}, E.~M., {Wegner}, G.~A., {Thomas}, J., {Saglia}, R.~P., \& {Bender},
  R. 2017, \mnras, 466, 974

\bibitem[{{C{\^o}t{\'e}} {et~al.}(2006){C{\^o}t{\'e}}, {Piatek}, {Ferrarese},
  {Jord{\'a}n}, {Merritt}, {Peng}, {Ha{\c{s}}egan}, {Blakeslee}, {Mei}, {West},
  {Milosavljevi{\'c}}, \& {Tonry}}]{cote06}
{C{\^o}t{\'e}}, P., {Piatek}, S., {Ferrarese}, L., {et~al.} 2006, \apjs, 165,
  57

\bibitem[{{Daddi} {et~al.}(2005){Daddi}, {Renzini}, {Pirzkal}, {Cimatti},
  {Malhotra}, {Stiavelli}, {Xu}, {Pasquali}, {Rhoads}, {Brusa}, {di Serego
  Alighieri}, {Ferguson}, {Koekemoer}, {Moustakas}, {Panagia}, \&
  {Windhorst}}]{daddi05}
{Daddi}, E., {Renzini}, A., {Pirzkal}, N., {et~al.} 2005, \apj, 626, 680

\bibitem[{{Davari} {et~al.}(2017){Davari}, {Ho}, {Mobasher}, \&
  {Canalizo}}]{davari17}
{Davari}, R.~H., {Ho}, L.~C., {Mobasher}, B., \& {Canalizo}, G. 2017, \apj,
  836, 75

\bibitem[{{Davies} {et~al.}(1983){Davies}, {Efstathiou}, {Fall}, {Illingworth},
  \& {Schechter}}]{davies83b}
{Davies}, R.~L., {Efstathiou}, G., {Fall}, S.~M., {Illingworth}, G., \&
  {Schechter}, P.~L. 1983, \apj, 266, 41

\bibitem[{{Davies} \& {Illingworth}(1983)}]{davies83}
{Davies}, R.~L. \& {Illingworth}, G. 1983, \apj, 266, 516

\bibitem[{{de Vaucouleurs}(1948)}]{devau48}
{de Vaucouleurs}, G. 1948, Annales d'Astrophysique, 11, 247

\bibitem[{{de Vaucouleurs}(1959)}]{devau59}
{de Vaucouleurs}, G. 1959, Handbuch der Physik, 53, 275

\bibitem[{{de Vaucouleurs}(1977)}]{devau77}
{de Vaucouleurs}, G. 1977, in Evolution of Galaxies and Stellar Populations,
  ed. B.~M. {Tinsley} \& D.~C. {Larson}, Richard B.~Gehret, 43

\bibitem[{{de Vaucouleurs} \& {Ag{\"u}ero}(1973)}]{devau73}
{de Vaucouleurs}, G. \& {Ag{\"u}ero}, E. 1973, \pasp, 85, 150

\bibitem[{{de Vaucouleurs} {et~al.}(1991){de Vaucouleurs}, {de Vaucouleurs},
  {Corwin}, {Buta}, {Paturel}, \& {Fouque}}]{devau91}
{de Vaucouleurs}, G., {de Vaucouleurs}, A., {Corwin}, Herold~G., J., {et~al.}
  1991, {Third Reference Catalogue of Bright Galaxies} (Springer, New York, NY
  {USA})

\bibitem[{{de Zeeuw} \& {Franx}(1991)}]{dezeeuw91}
{de Zeeuw}, T. \& {Franx}, M. 1991, \araa, 29, 239

\bibitem[{{Debattista} {et~al.}(2006){Debattista}, {Mayer}, {Carollo}, {Moore},
  {Wadsley}, \& {Quinn}}]{debattista06}
{Debattista}, V.~P., {Mayer}, L., {Carollo}, C.~M., {et~al.} 2006, \apj, 645,
  209

\bibitem[{{den Brok} {et~al.}(2011){den Brok}, {Peletier}, {Valentijn},
  {Balcells}, {Carter}, {Erwin}, {Ferguson}, {Goudfrooij}, {Graham}, {Hammer},
  {Lucey}, {Trentham}, {Guzm{\'a}n}, {Hoyos}, {Verdoes Kleijn}, {Jogee},
  {Karick}, {Marinova}, {Mouhcine}, \& {Weinzirl}}]{denbrok11}
{den Brok}, M., {Peletier}, R.~F., {Valentijn}, E.~A., {et~al.} 2011, \mnras,
  414, 3052

\bibitem[{{D{\'\i}az-Garc{\'\i}a} {et~al.}(2019){D{\'\i}az-Garc{\'\i}a},
  {Salo}, {Knapen}, \& {Herrera-Endoqui}}]{diazgarcia19}
{D{\'\i}az-Garc{\'\i}a}, S., {Salo}, H., {Knapen}, J.~H., \& {Herrera-Endoqui},
  M. 2019, \aap, 631, A94

\bibitem[{{D{\'\i}az-Garc{\'\i}a} {et~al.}(2016){D{\'\i}az-Garc{\'\i}a},
  {Salo}, \& {Laurikainen}}]{diazgarcia16}
{D{\'\i}az-Garc{\'\i}a}, S., {Salo}, H., \& {Laurikainen}, E. 2016, \aap, 596,
  A84

\bibitem[{{Djorgovski} \& {Davis}(1987)}]{djorgovski87}
{Djorgovski}, S. \& {Davis}, M. 1987, \apj, 313, 59

\bibitem[{{Draine} \& {Lee}(1984)}]{draine84}
{Draine}, B.~T. \& {Lee}, H.~M. 1984, \apj, 285, 89

\bibitem[{{Dressler}(1980)}]{dressler80}
{Dressler}, A. 1980, \apj, 236, 351

\bibitem[{{Dressler} {et~al.}(1987){Dressler}, {Lynden-Bell}, {Burstein},
  {Davies}, {Faber}, {Terlevich}, \& {Wegner}}]{dressler87}
{Dressler}, A., {Lynden-Bell}, D., {Burstein}, D., {et~al.} 1987, \apj, 313, 42

\bibitem[{{Driver} {et~al.}(2016){Driver}, {Wright}, {Andrews}, {Davies},
  {Kafle}, {Lange}, {Moffett}, {Mannering}, {Robotham}, {Vinsen}, {Alpaslan},
  {Andrae}, {Baldry}, {Bauer}, {Bamford}, {Bland-Hawthorn}, {Bourne}, {Brough},
  {Brown}, {Cluver}, {Croom}, {Colless}, {Conselice}, {da Cunha}, {De Propris},
  {Drinkwater}, {Dunne}, {Eales}, {Edge}, {Frenk}, {Graham}, {Grootes},
  {Holwerda}, {Hopkins}, {Ibar}, {van Kampen}, {Kelvin}, {Jarrett}, {Jones},
  {Lara-Lopez}, {Liske}, {Lopez-Sanchez}, {Loveday}, {Maddox}, {Madore},
  {Mahajan}, {Meyer}, {Norberg}, {Penny}, {Phillipps}, {Popescu}, {Tuffs},
  {Peacock}, {Pimbblet}, {Prescott}, {Rowlands}, {Sansom}, {Seibert}, {Smith},
  {Sutherland}, {Taylor}, {Valiante}, {Vazquez-Mata}, {Wang}, {Wilkins}, \&
  {Williams}}]{driver16}
{Driver}, S.~P., {Wright}, A.~H., {Andrews}, S.~K., {et~al.} 2016, \mnras, 455,
  3911

\bibitem[{{Dullo} \& {Graham}(2012)}]{dullo12}
{Dullo}, B.~T. \& {Graham}, A.~W. 2012, \apj, 755, 163

\bibitem[{{Ebisuzaki} {et~al.}(1991){Ebisuzaki}, {Makino}, \&
  {Okumura}}]{ebisuzaki91}
{Ebisuzaki}, T., {Makino}, J., \& {Okumura}, S.~K. 1991, \nat, 354, 212

\bibitem[{{Eggen} {et~al.}(1962){Eggen}, {Lynden-Bell}, \& {Sandage}}]{eggen62}
{Eggen}, O.~J., {Lynden-Bell}, D., \& {Sandage}, A.~R. 1962, \apj, 136, 748

\bibitem[{{Eliche-Moral} {et~al.}(2015){Eliche-Moral}, {Borlaff}, {Beckman}, \&
  {Guti{\'e}rrez}}]{eliche15}
{Eliche-Moral}, M.~C., {Borlaff}, A., {Beckman}, J.~E., \& {Guti{\'e}rrez}, L.
  2015, \aap, 580, A33

\bibitem[{{Emsellem} {et~al.}(2011{\natexlab{a}}){Emsellem}, {Cappellari},
  {Krajnovi{\'c}}, {Alatalo}, {Blitz}, {Bois}, {Bournaud}, {Bureau}, {Davies},
  {Davis}, {de Zeeuw}, {Khochfar}, {Kuntschner}, {Lablanche}, {McDermid},
  {Morganti}, {Naab}, {Oosterloo}, {Sarzi}, {Scott}, {Serra}, {van de Ven},
  {Weijmans}, \& {Young}}]{cappellari11c}
{Emsellem}, E., {Cappellari}, M., {Krajnovi{\'c}}, D., {et~al.}
  2011{\natexlab{a}}, \mnras, 414, 888

\bibitem[{{Emsellem} {et~al.}(2011{\natexlab{b}}){Emsellem}, {Cappellari},
  {Krajnovi{\'c}}, {Alatalo}, {Blitz}, {Bois}, {Bournaud}, {Bureau}, {Davies},
  {Davis}, {de Zeeuw}, {Khochfar}, {Kuntschner}, {Lablanche}, {McDermid},
  {Morganti}, {Naab}, {Oosterloo}, {Sarzi}, {Scott}, {Serra}, {van de Ven},
  {Weijmans}, \& {Young}}]{emsellem11}
{Emsellem}, E., {Cappellari}, M., {Krajnovi{\'c}}, D., {et~al.}
  2011{\natexlab{b}}, \mnras, 414, 888

\bibitem[{{Emsellem} {et~al.}(2007){Emsellem}, {Cappellari}, {Krajnovi{\'c}},
  {van de Ven}, {Bacon}, {Bureau}, {Davies}, {de Zeeuw}, {Falc{\'o}n-Barroso},
  {Kuntschner}, {McDermid}, {Peletier}, \& {Sarzi}}]{emsellem07}
{Emsellem}, E., {Cappellari}, M., {Krajnovi{\'c}}, D., {et~al.} 2007, \mnras,
  379, 401

\bibitem[{{Erwin} {et~al.}(2008){Erwin}, {Pohlen}, \& {Beckman}}]{erwin08}
{Erwin}, P., {Pohlen}, M., \& {Beckman}, J.~E. 2008, \aj, 135, 20

\bibitem[{{Eskew} {et~al.}(2012){Eskew}, {Zaritsky}, \& {Meidt}}]{eskew12}
{Eskew}, M., {Zaritsky}, D., \& {Meidt}, S. 2012, \aj, 143, 139

\bibitem[{{Faber} {et~al.}(1997){Faber}, {Tremaine}, {Ajhar}, {Byun},
  {Dressler}, {Gebhardt}, {Grillmair}, {Kormendy}, {Lauer}, \&
  {Richstone}}]{faber97}
{Faber}, S.~M., {Tremaine}, S., {Ajhar}, E.~A., {et~al.} 1997, \aj, 114, 1771

\bibitem[{{Falc{\'o}n-Barroso} {et~al.}(2019){Falc{\'o}n-Barroso}, {van de
  Ven}, {Lyubenova}, {Mendez-Abreu}, {Aguerri}, {Garc{\'\i}a-Lorenzo},
  {Bekerait{\'e}}, {S{\'a}nchez}, {Husemann}, {Garc{\'\i}a-Benito},
  {Gonz{\'a}lez Delgado}, {Mast}, {Walcher}, {Zibetti}, {Zhu},
  {Barrera-Ballesteros}, {Galbany}, {S{\'a}nchez-Bl{\'a}zquez}, {Singh}, {van
  den Bosch}, {Wild}, {Bland-Hawthorn}, {Cid Fernandes}, {de
  Lorenzo-C{\'a}ceres}, {Gallazzi}, {Marino}, {M{\'a}rquez}, {Peletier},
  {P{\'e}rez}, {P{\'e}rez}, {Roth}, {Rosales-Ortega}, {Ruiz-Lara}, {Wisotzki},
  \& {Ziegler}}]{falcon19}
{Falc{\'o}n-Barroso}, J., {van de Ven}, G., {Lyubenova}, M., {et~al.} 2019,
  \aap, 632, A59

\bibitem[{{Fathi} \& {Peletier}(2003)}]{fathi03}
{Fathi}, K. \& {Peletier}, R.~F. 2003, \aap, 407, 61

\bibitem[{{Fazio} {et~al.}(2004){Fazio}, {Hora}, {Allen}, {Ashby}, {Barmby},
  {Deutsch}, {Huang}, {Kleiner}, {Marengo}, {Megeath}, {Melnick}, {Pahre},
  {Patten}, {Polizotti}, {Smith}, {Taylor}, {Wang}, {Willner}, {Hoffmann},
  {Pipher}, {Forrest}, {McMurty}, {McCreight}, {McKelvey}, {McMurray}, {Koch},
  {Moseley}, {Arendt}, {Mentzell}, {Marx}, {Losch}, {Mayman}, {Eichhorn},
  {Krebs}, {Jhabvala}, {Gezari}, {Fixsen}, {Flores}, {Shakoorzadeh}, {Jungo},
  {Hakun}, {Workman}, {Karpati}, {Kichak}, {Whitley}, {Mann}, {Tollestrup},
  {Eisenhardt}, {Stern}, {Gorjian}, {Bhattacharya}, {Carey}, {Nelson},
  {Glaccum}, {Lacy}, {Lowrance}, {Laine}, {Reach}, {Stauffer}, {Surace},
  {Wilson}, {Wright}, {Hoffman}, {Domingo}, \& {Cohen}}]{fazio04}
{Fazio}, G.~G., {Hora}, J.~L., {Allen}, L.~E., {et~al.} 2004, \apjs, 154, 10

\bibitem[{{Ferrarese} {et~al.}(2006){Ferrarese}, {C{\^o}t{\'e}}, {Jord{\'a}n},
  {Peng}, {Blakeslee}, {Piatek}, {Mei}, {Merritt}, {Milosavljevi{\'c}},
  {Tonry}, \& {West}}]{ferrarese06}
{Ferrarese}, L., {C{\^o}t{\'e}}, P., {Jord{\'a}n}, A., {et~al.} 2006, \apjs,
  164, 334

\bibitem[{{Fliri} \& {Trujillo}(2016)}]{fliri16}
{Fliri}, J. \& {Trujillo}, I. 2016, \mnras, 456, 1359

\bibitem[{{Fraser}(1972)}]{fraser72}
{Fraser}, C.~W. 1972, The Observatory, 92, 51

\bibitem[{{Frenk} {et~al.}(1985){Frenk}, {White}, {Efstathiou}, \&
  {Davis}}]{frenk85}
{Frenk}, C.~S., {White}, S.~D.~M., {Efstathiou}, G., \& {Davis}, M. 1985, \nat,
  317, 595

\bibitem[{{Fruchter} \& {Hook}(2002)}]{fruchter02}
{Fruchter}, A.~S. \& {Hook}, R.~N. 2002, \pasp, 114, 144

\bibitem[{{Gallazzi} {et~al.}(2005){Gallazzi}, {Charlot}, {Brinchmann},
  {White}, \& {Tremonti}}]{gallazzi05}
{Gallazzi}, A., {Charlot}, S., {Brinchmann}, J., {White}, S. D.~M., \&
  {Tremonti}, C.~A. 2005, \mnras, 362, 41

\bibitem[{{Gott}(1977)}]{gott77}
{Gott}, J.~R., I. 1977, \araa, 15, 235

\bibitem[{{Gott}(1973)}]{gott73}
{Gott}, Richard~J., I. 1973, \apj, 186, 481

\bibitem[{{Graham}(2001)}]{graham01a}
{Graham}, A.~W. 2001, \aj, 121, 820

\bibitem[{{Graham}(2019)}]{graham19}
{Graham}, A.~W. 2019, \pasa, 36, e035

\bibitem[{{Graham} \& {Driver}(2005)}]{graham05}
{Graham}, A.~W. \& {Driver}, S.~P. 2005, \pasa, 22, 118

\bibitem[{{Graham} \& {Guzm{\'a}n}(2003)}]{graham03}
{Graham}, A.~W. \& {Guzm{\'a}n}, R. 2003, \aj, 125, 2936

\bibitem[{{Graham} {et~al.}(2001){Graham}, {Trujillo}, \& {Caon}}]{graham01b}
{Graham}, A.~W., {Trujillo}, I., \& {Caon}, N. 2001, \aj, 122, 1707

\bibitem[{{Gudehus}(1973)}]{gudehus73}
{Gudehus}, D.~H. 1973, \aj, 78, 583

\bibitem[{{Gunn} \& {Gott}(1972)}]{gunn72}
{Gunn}, J.~E. \& {Gott}, J.~Richard, I. 1972, \apj, 176, 1

\bibitem[{{Hall} {et~al.}(2012){Hall}, {Courteau}, {Dutton}, {McDonald}, \&
  {Zhu}}]{hall12}
{Hall}, M., {Courteau}, S., {Dutton}, A.~A., {McDonald}, M., \& {Zhu}, Y. 2012,
  \mnras, 425, 2741

\bibitem[{Harris {et~al.}(2020)Harris, Millman, van~der Walt, Gommers,
  Virtanen, Cournapeau, Wieser, Taylor, Berg, Smith, Kern, Picus, Hoyer, van
  Kerkwijk, Brett, Haldane, del R{\'{i}}o, Wiebe, Peterson,
  G{\'{e}}rard-Marchant, Sheppard, Reddy, Weckesser, Abbasi, Gohlke, \&
  Oliphant}]{harris20}
Harris, C.~R., Millman, K.~J., van~der Walt, S.~J., {et~al.} 2020, Nature, 585,
  357

\bibitem[{{Hernquist}(1992)}]{hernquist92}
{Hernquist}, L. 1992, \apj, 400, 460

\bibitem[{{Herrera-Endoqui} {et~al.}(2015){Herrera-Endoqui},
  {D{\'\i}az-Garc{\'\i}a}, {Laurikainen}, \& {Salo}}]{herrera15}
{Herrera-Endoqui}, M., {D{\'\i}az-Garc{\'\i}a}, S., {Laurikainen}, E., \&
  {Salo}, H. 2015, \aap, 582, A86

\bibitem[{{Holwerda} {et~al.}(2014){Holwerda}, {Mu{\~n}oz-Mateos},
  {Comer{\'o}n}, {Meidt}, {Sheth}, {Laine}, {Hinz}, {Regan}, {Gil de Paz},
  {Men{\'e}ndez-Delmestre}, {Seibert}, {Kim}, {Mizusawa}, {Laurikainen},
  {Salo}, {Laine}, {Gadotti}, {Zaritsky}, {Erroz-Ferrer}, {Ho}, {Knapen},
  {Athanassoula}, {Bosma}, \& {Pirzkal}}]{holwerda14}
{Holwerda}, B.~W., {Mu{\~n}oz-Mateos}, J.~C., {Comer{\'o}n}, S., {et~al.} 2014,
  \apj, 781, 12

\bibitem[{{Hopkins} {et~al.}(2009){Hopkins}, {Bundy}, {Murray}, {Quataert},
  {Lauer}, \& {Ma}}]{hopkins09}
{Hopkins}, P.~F., {Bundy}, K., {Murray}, N., {et~al.} 2009, \mnras, 398, 898

\bibitem[{{Hopkins} {et~al.}(2008){Hopkins}, {Cox}, \& {Hernquist}}]{hopkins08}
{Hopkins}, P.~F., {Cox}, T.~J., \& {Hernquist}, L. 2008, \apj, 689, 17

\bibitem[{{Hora} {et~al.}(2012){Hora}, {Marengo}, {Park}, {Wood}, {Hoffmann},
  {Lowrance}, {Carey}, {Surace}, {Krick}, {Glaccum}, {Ingalls}, {Laine},
  {Fazio}, {Ashby}, \& {Wang}}]{hora12}
{Hora}, J.~L., {Marengo}, M., {Park}, R., {et~al.} 2012, in Society of
  Photo-Optical Instrumentation Engineers (SPIE) Conference Series, Vol. 8442,
  Space Telescopes and Instrumentation 2012: Optical, Infrared, and Millimeter
  Wave, ed. M.~C. {Clampin}, G.~G. {Fazio}, H.~A. {MacEwen}, \& J.~{Oschmann},
  Jacobus~M., 844239

\bibitem[{{Hubble}(1926)}]{hubble26}
{Hubble}, E.~P. 1926, \apj, 64, 321

\bibitem[{Hunter(2007)}]{hunter07}
Hunter, J.~D. 2007, Computing in Science \& Engineering, 9, 90

\bibitem[{{{\.I}kiz} {et~al.}(2020){{\.I}kiz}, {Peletier}, {Barthel}, \&
  {Ye{\c{s}}ilyaprak}}]{ikiz20}
{{\.I}kiz}, T., {Peletier}, R.~F., {Barthel}, P.~D., \& {Ye{\c{s}}ilyaprak}, C.
  2020, \aap, 640, A68

\bibitem[{{Illingworth}(1977)}]{illingworth77}
{Illingworth}, G. 1977, \apjl, 218, L43

\bibitem[{{Impey} {et~al.}(1996){Impey}, {Sprayberry}, {Irwin}, \&
  {Bothun}}]{impey96}
{Impey}, C.~D., {Sprayberry}, D., {Irwin}, M.~J., \& {Bothun}, G.~D. 1996,
  \apjs, 105, 209

\bibitem[{{Iodice} {et~al.}(2020){Iodice}, {Cantiello}, {Hilker}, {Rejkuba},
  {Arnaboldi}, {Spavone}, {Greggio}, {Forbes}, {D'Ago}, {Mieske}, {Spiniello},
  {La Marca}, {Rampazzo}, {Paolillo}, {Capaccioli}, \& {Schipani}}]{iodice20}
{Iodice}, E., {Cantiello}, M., {Hilker}, M., {et~al.} 2020, \aap, 642, A48

\bibitem[{{Ivezi{\'c}} {et~al.}(2019){Ivezi{\'c}}, {Kahn}, {Tyson}, {Abel},
  {Acosta}, {Allsman}, {Alonso}, {AlSayyad}, {Anderson}, {Andrew}, \&
  et~al.}]{ivezic19}
{Ivezi{\'c}}, {\v Z}., {Kahn}, S.~M., {Tyson}, J.~A., {et~al.} 2019, \apj, 873,
  111

\bibitem[{{Janz} {et~al.}(2014){Janz}, {Laurikainen}, {Lisker}, {Salo},
  {Peletier}, {Niemi}, {Toloba}, {Hensler}, {Falc{\'o}n-Barroso}, {Boselli},
  {den Brok}, {Hansson}, {Meyer}, {Ry{\'s}}, \& {Paudel}}]{janz14}
{Janz}, J., {Laurikainen}, E., {Lisker}, T., {et~al.} 2014, \apj, 786, 105

\bibitem[{{Janz} \& {Lisker}(2008)}]{janz08}
{Janz}, J. \& {Lisker}, T. 2008, \apjl, 689, L25

\bibitem[{{Jarrett} {et~al.}(2003){Jarrett}, {Chester}, {Cutri}, {Schneider},
  \& {Huchra}}]{jarrett03}
{Jarrett}, T.~H., {Chester}, T., {Cutri}, R., {Schneider}, S.~E., \& {Huchra},
  J.~P. 2003, \aj, 125, 525

\bibitem[{{Jarrett} {et~al.}(2013){Jarrett}, {Masci}, {Tsai}, {Petty},
  {Cluver}, {Assef}, {Benford}, {Blain}, {Bridge}, {Donoso}, {Eisenhardt},
  {Koribalski}, {Lake}, {Neill}, {Seibert}, {Sheth}, {Stanford}, \&
  {Wright}}]{jarrett13}
{Jarrett}, T.~H., {Masci}, F., {Tsai}, C.~W., {et~al.} 2013, \aj, 145, 6

\bibitem[{{Jedrzejewski}(1987)}]{jedrzejewski87}
{Jedrzejewski}, R.~I. 1987, \mnras, 226, 747

\bibitem[{{Jun} \& {Im}(2008)}]{jun08}
{Jun}, H.~D. \& {Im}, M. 2008, \apjl, 678, L97

\bibitem[{{Kaneda} {et~al.}(2007){Kaneda}, {Onaka}, \& {Sakon}}]{kaneda07}
{Kaneda}, H., {Onaka}, T., \& {Sakon}, I. 2007, \apjl, 666, L21

\bibitem[{{Karachentsev} {et~al.}(2010){Karachentsev}, {Makarov},
  {Karachentseva}, \& {Melnyk}}]{karachentsev10}
{Karachentsev}, I.~D., {Makarov}, D.~I., {Karachentseva}, V.~E., \& {Melnyk},
  O.~V. 2010, in Astronomical Society of the Pacific Conference Series, Vol.
  421, Galaxies in Isolation: Exploring Nature Versus Nurture, ed.
  L.~{Verdes-Montenegro}, A.~{Del Olmo}, \& J.~{Sulentic}, 69

\bibitem[{{Kataria} {et~al.}(2020){Kataria}, {Das}, \& {Barway}}]{kataria20}
{Kataria}, S.~K., {Das}, M., \& {Barway}, S. 2020, \aap, 640, A14

\bibitem[{{Katkov} {et~al.}(2019){Katkov}, {Kniazev}, {Kasparova}, \&
  {Sil'chenko}}]{katkov19}
{Katkov}, I.~Y., {Kniazev}, A.~Y., {Kasparova}, A.~V., \& {Sil'chenko}, O.~K.
  2019, \mnras, 483, 2413

\bibitem[{{Kauffmann} {et~al.}(1993){Kauffmann}, {White}, \&
  {Guiderdoni}}]{kauffmann93}
{Kauffmann}, G., {White}, S.~D.~M., \& {Guiderdoni}, B. 1993, \mnras, 264, 201

\bibitem[{{Kauffmann} {et~al.}(2004){Kauffmann}, {White}, {Heckman},
  {M{\'e}nard}, {Brinchmann}, {Charlot}, {Tremonti}, \&
  {Brinkmann}}]{kauffmann04}
{Kauffmann}, G., {White}, S. D.~M., {Heckman}, T.~M., {et~al.} 2004, \mnras,
  353, 713

\bibitem[{{Kent}(1985)}]{kent85}
{Kent}, S.~M. 1985, \apjs, 59, 115

\bibitem[{{Kim} {et~al.}(2016){Kim}, {Gadotti}, {Athanassoula}, {Bosma},
  {Sheth}, \& {Lee}}]{kim16}
{Kim}, T., {Gadotti}, D.~A., {Athanassoula}, E., {et~al.} 2016, \mnras, 462,
  3430

\bibitem[{{Kim} {et~al.}(2012){Kim}, {Sheth}, {Hinz}, {Lee}, {Zaritsky},
  {Gadotti}, {Knapen}, {Schinnerer}, {Ho}, {Laurikainen}, {Salo},
  {Athanassoula}, {Bosma}, {de Swardt}, {Mu{\~n}oz-Mateos}, {Madore},
  {Comer{\'o}n}, {Regan}, {Men{\'e}ndez-Delmestre}, {Gil de Paz}, {Seibert},
  {Laine}, {Erroz-Ferrer}, \& {Mizusawa}}]{kim12}
{Kim}, T., {Sheth}, K., {Hinz}, J.~L., {et~al.} 2012, \apj, 753, 43

\bibitem[{{King} \& {Minkowski}(1966)}]{king66}
{King}, I.~R. \& {Minkowski}, R. 1966, \apj, 143, 1002

\bibitem[{{Koda} {et~al.}(2015){Koda}, {Yagi}, {Yamanoi}, \&
  {Komiyama}}]{koda15}
{Koda}, J., {Yagi}, M., {Yamanoi}, H., \& {Komiyama}, Y. 2015, \apjl, 807, L2

\bibitem[{{Kormendy}(1977)}]{kormendy77}
{Kormendy}, J. 1977, \apj, 218, 333

\bibitem[{{Kormendy}(1982)}]{kormendy82}
{Kormendy}, J. 1982, \apj, 257, 75

\bibitem[{{Kormendy}(1999)}]{kormendy99}
{Kormendy}, J. 1999, in Astronomical Society of the Pacific Conference Series,
  Vol. 182, Galaxy Dynamics - A Rutgers Symposium, ed. D.~R. {Merritt},
  M.~{Valluri}, \& J.~A. {Sellwood}, 124

\bibitem[{{Kormendy} \& {Bender}(1996)}]{kormendy96}
{Kormendy}, J. \& {Bender}, R. 1996, \apjl, 464, L119

\bibitem[{{Kormendy} \& {Bender}(2012)}]{kormendy12}
{Kormendy}, J. \& {Bender}, R. 2012, \apjs, 198, 2

\bibitem[{{Kormendy} {et~al.}(2009){Kormendy}, {Fisher}, {Cornell}, \&
  {Bender}}]{kormendy09}
{Kormendy}, J., {Fisher}, D.~B., {Cornell}, M.~E., \& {Bender}, R. 2009, \apjs,
  182, 216

\bibitem[{{Kormendy} \& {Kennicutt}(2004)}]{kormendy04}
{Kormendy}, J. \& {Kennicutt}, Robert~C., J. 2004, \araa, 42, 603

\bibitem[{{Krajnovi{\'c}}(2015)}]{krajnovic15}
{Krajnovi{\'c}}, D. 2015, in Galaxy Masses as Constraints of Formation Models,
  ed. M.~{Cappellari} \& S.~{Courteau}, Vol. 311, 45--48

\bibitem[{{Krajnovi{\'c}} {et~al.}(2020){Krajnovi{\'c}}, {Ural}, {Kuntschner},
  {Goudfrooij}, {Wolfe}, {Cappellari}, {Davies}, {de Zeeuw}, {Duc}, {Emsellem},
  {Karick}, {McDermid}, {Mei}, \& {Naab}}]{krajnovic20}
{Krajnovi{\'c}}, D., {Ural}, U., {Kuntschner}, H., {et~al.} 2020, \aap, 635,
  A129

\bibitem[{{La Barbera} {et~al.}(2004){La Barbera}, {Merluzzi}, {Busarello},
  {Massarotti}, \& {Mercurio}}]{labarbera04}
{La Barbera}, F., {Merluzzi}, P., {Busarello}, G., {Massarotti}, M., \&
  {Mercurio}, A. 2004, \aap, 425, 797

\bibitem[{{Laine} {et~al.}(2016){Laine}, {Laurikainen}, \& {Salo}}]{laine16}
{Laine}, J., {Laurikainen}, E., \& {Salo}, H. 2016, \aap, 596, A25

\bibitem[{{Laine} {et~al.}(2014){Laine}, {Knapen}, {Mu{\~n}oz-Mateos}, {Kim},
  {Comer{\'o}n}, {Martig}, {Holwerda}, {Athanassoula}, {Bosma}, {Johansson},
  {Erroz-Ferrer}, {Gadotti}, {de Paz}, {Hinz}, {Laine}, {Laurikainen},
  {Men{\'e}ndez-Delmestre}, {Mizusawa}, {Regan}, {Salo}, {Sheth}, {Seibert},
  {Buta}, {Cisternas}, {Elmegreen}, {Elmegreen}, {Ho}, {Madore}, \&
  {Zaritsky}}]{laine14}
{Laine}, S., {Knapen}, J.~H., {Mu{\~n}oz-Mateos}, J.-C., {et~al.} 2014, \mnras,
  444, 3015

\bibitem[{{Larson}(1974{\natexlab{a}})}]{larson74a}
{Larson}, R.~B. 1974{\natexlab{a}}, \mnras, 166, 585

\bibitem[{{Larson}(1974{\natexlab{b}})}]{larson74b}
{Larson}, R.~B. 1974{\natexlab{b}}, \mnras, 169, 229

\bibitem[{{Larson}(1975)}]{larson75}
{Larson}, R.~B. 1975, \mnras, 173, 671

\bibitem[{{Lauer} {et~al.}(2007){Lauer}, {Gebhardt}, {Faber}, {Richstone},
  {Tremaine}, {Kormendy}, {Aller}, {Bender}, {Dressler}, {Filippenko}, {Green},
  \& {Ho}}]{lauer07}
{Lauer}, T.~R., {Gebhardt}, K., {Faber}, S.~M., {et~al.} 2007, \apj, 664, 226

\bibitem[{{Laureijs} {et~al.}(2010){Laureijs}, {Duvet}, {Escudero Sanz},
  {Gondoin}, {Lumb}, {Oosterbroek}, \& {Saavedra Criado}}]{laureijs10}
{Laureijs}, R.~J., {Duvet}, L., {Escudero Sanz}, I., {et~al.} 2010, in Society
  of Photo-Optical Instrumentation Engineers (SPIE) Conference Series, Vol.
  7731, \procspie, 77311H

\bibitem[{{Laurikainen} \& {Salo}(2017)}]{laurikainen17}
{Laurikainen}, E. \& {Salo}, H. 2017, \aap, 598, A10

\bibitem[{{Laurikainen} {et~al.}(2013){Laurikainen}, {Salo}, {Athanassoula},
  {Bosma}, {Buta}, \& {Janz}}]{laurikainen13}
{Laurikainen}, E., {Salo}, H., {Athanassoula}, E., {et~al.} 2013, \mnras, 430,
  3489

\bibitem[{{Laurikainen} {et~al.}(2007){Laurikainen}, {Salo}, {Buta}, \&
  {Knapen}}]{laurikainen07}
{Laurikainen}, E., {Salo}, H., {Buta}, R., \& {Knapen}, J.~H. 2007, \mnras,
  381, 401

\bibitem[{{Laurikainen} {et~al.}(2011){Laurikainen}, {Salo}, {Buta}, \&
  {Knapen}}]{laurikainen11}
{Laurikainen}, E., {Salo}, H., {Buta}, R., \& {Knapen}, J.~H. 2011, Advances in
  Astronomy, 2011, 516739

\bibitem[{{Laurikainen} {et~al.}(2010){Laurikainen}, {Salo}, {Buta}, {Knapen},
  \& {Comer{\'o}n}}]{laurikainen10}
{Laurikainen}, E., {Salo}, H., {Buta}, R., {Knapen}, J.~H., \& {Comer{\'o}n},
  S. 2010, \mnras, 405, 1089

\bibitem[{{Longhetti} {et~al.}(2007){Longhetti}, {Saracco}, {Severgnini},
  {Della Ceca}, {Mannucci}, {Bender}, {Drory}, {Feulner}, \&
  {Hopp}}]{longhetti07}
{Longhetti}, M., {Saracco}, P., {Severgnini}, P., {et~al.} 2007, \mnras, 374,
  614

\bibitem[{{Luhman} {et~al.}(2008){Luhman}, {Hern{\'a}ndez}, {Downes},
  {Hartmann}, \& {Brice{\~n}o}}]{luhman08}
{Luhman}, K.~L., {Hern{\'a}ndez}, J., {Downes}, J.~J., {Hartmann}, L., \&
  {Brice{\~n}o}, C. 2008, \apj, 688, 362

\bibitem[{{Maciejewski} {et~al.}(2002){Maciejewski}, {Teuben}, {Sparke}, \&
  {Stone}}]{maciejewski02}
{Maciejewski}, W., {Teuben}, P.~J., {Sparke}, L.~S., \& {Stone}, J.~M. 2002,
  \mnras, 329, 502

\bibitem[{{Mahajan} {et~al.}(2015){Mahajan}, {Drinkwater}, {Driver}, {Kelvin},
  {Hopkins}, {Baldry}, {Phillipps}, {Bland-Hawthorn}, {Brough}, {Loveday},
  {Penny}, \& {Robotham}}]{mahajan15}
{Mahajan}, S., {Drinkwater}, M.~J., {Driver}, S., {et~al.} 2015, \mnras, 446,
  2967

\bibitem[{{Mart{\'\i}n-Navarro} {et~al.}(2012){Mart{\'\i}n-Navarro}, {Bakos},
  {Trujillo}, {Knapen}, {Athanassoula}, {Bosma}, {Comer{\'o}n}, {Elmegreen},
  {Erroz-Ferrer}, {Gadotti}, {Gil de Paz}, {Hinz}, {Ho}, {Holwerda}, {Kim},
  {Laine}, {Laurikainen}, {Men{\'e}ndez-Delmestre}, {Mizusawa},
  {Mu{\~n}oz-Mateos}, {Regan}, {Salo}, {Seibert}, \& {Sheth}}]{martinnavarro12}
{Mart{\'\i}n-Navarro}, I., {Bakos}, J., {Trujillo}, I., {et~al.} 2012, \mnras,
  427, 1102

\bibitem[{{Meidt} {et~al.}(2012){Meidt}, {Schinnerer}, {Knapen}, {Bosma},
  {Athanassoula}, {Sheth}, {Buta}, {Zaritsky}, {Laurikainen}, {Elmegreen},
  {Elmegreen}, {Gadotti}, {Salo}, {Regan}, {Ho}, {Madore}, {Hinz}, {Skibba},
  {Gil de Paz}, {Mu{\~n}oz-Mateos}, {Men{\'e}ndez-Delmestre}, {Seibert}, {Kim},
  {Mizusawa}, {Laine}, \& {Comer{\'o}n}}]{meidt12}
{Meidt}, S.~E., {Schinnerer}, E., {Knapen}, J.~H., {et~al.} 2012, \apj, 744, 17

\bibitem[{{Meidt} {et~al.}(2014){Meidt}, {Schinnerer}, {van de Ven},
  {Zaritsky}, {Peletier}, {Knapen}, {Sheth}, {Regan}, {Querejeta},
  {Mu{\~n}oz-Mateos}, {Kim}, {Hinz}, {Gil de Paz}, {Athanassoula}, {Bosma},
  {Buta}, {Cisternas}, {Ho}, {Holwerda}, {Skibba}, {Laurikainen}, {Salo},
  {Gadotti}, {Laine}, {Erroz-Ferrer}, {Comer{\'o}n}, {Men{\'e}ndez-Delmestre},
  {Seibert}, \& {Mizusawa}}]{meidt14}
{Meidt}, S.~E., {Schinnerer}, E., {van de Ven}, G., {et~al.} 2014, \apj, 788,
  144

\bibitem[{{M{\'e}ndez-Abreu} {et~al.}(2021){M{\'e}ndez-Abreu}, {de
  Lorenzo-C{\'a}ceres}, \& {S{\'a}nchez}}]{mendezabreu21}
{M{\'e}ndez-Abreu}, J., {de Lorenzo-C{\'a}ceres}, A., \& {S{\'a}nchez}, S.~F.
  2021, \mnras, 504, 3058

\bibitem[{{Mihos} {et~al.}(2015){Mihos}, {Durrell}, {Ferrarese}, {Feldmeier},
  {C{\^o}t{\'e}}, {Peng}, {Harding}, {Liu}, {Gwyn}, \& {Cuillandre}}]{mihos15}
{Mihos}, J.~C., {Durrell}, P.~R., {Ferrarese}, L., {et~al.} 2015, \apjl, 809,
  L21

\bibitem[{{Mihos} \& {Hernquist}(1994)}]{mihos94}
{Mihos}, J.~C. \& {Hernquist}, L. 1994, \apjl, 437, L47

\bibitem[{{Milosavljevi{\'c}} \& {Merritt}(2001)}]{milosavljevic01}
{Milosavljevi{\'c}}, M. \& {Merritt}, D. 2001, \apj, 563, 34

\bibitem[{{Moore} {et~al.}(1996){Moore}, {Katz}, {Lake}, {Dressler}, \&
  {Oemler}}]{moore96}
{Moore}, B., {Katz}, N., {Lake}, G., {Dressler}, A., \& {Oemler}, A. 1996,
  \nat, 379, 613

\bibitem[{{Morgan}(1958)}]{morgan58}
{Morgan}, W.~W. 1958, \pasp, 70, 364

\bibitem[{{Morgan} \& {Mayall}(1957)}]{morgan57}
{Morgan}, W.~W. \& {Mayall}, N.~U. 1957, \pasp, 69, 291

\bibitem[{{Mould} {et~al.}(2008){Mould}, {Barmby}, {Gordon}, {Willner},
  {Ashby}, {Gehrz}, {Humphreys}, \& {Woodward}}]{mould08}
{Mould}, J., {Barmby}, P., {Gordon}, K., {et~al.} 2008, \apj, 687, 230

\bibitem[{{Mu{\~n}oz-Mateos} {et~al.}(2013){Mu{\~n}oz-Mateos}, {Sheth}, {Gil de
  Paz}, {Meidt}, {Athanassoula}, {Bosma}, {Comer{\'o}n}, {Elmegreen},
  {Elmegreen}, {Erroz-Ferrer}, {Gadotti}, {Hinz}, {Ho}, {Holwerda}, {Jarrett},
  {Kim}, {Knapen}, {Laine}, {Laurikainen}, {Madore}, {Menendez-Delmestre},
  {Mizusawa}, {Regan}, {Salo}, {Schinnerer}, {Seibert}, {Skibba}, \&
  {Zaritsky}}]{munoz13}
{Mu{\~n}oz-Mateos}, J.~C., {Sheth}, K., {Gil de Paz}, A., {et~al.} 2013, \apj,
  771, 59

\bibitem[{{Mu{\~n}oz-Mateos} {et~al.}(2015){Mu{\~n}oz-Mateos}, {Sheth},
  {Regan}, {Kim}, {Laine}, {Erroz-Ferrer}, {Gil de Paz}, {Comeron}, {Hinz},
  {Laurikainen}, {Salo}, {Athanassoula}, {Bosma}, {Bouquin}, {Schinnerer},
  {Ho}, {Zaritsky}, {Gadotti}, {Madore}, {Holwerda}, {Men{\'e}ndez-Delmestre},
  {Knapen}, {Meidt}, {Querejeta}, {Mizusawa}, {Seibert}, {Laine}, \&
  {Courtois}}]{munoz15}
{Mu{\~n}oz-Mateos}, J.~C., {Sheth}, K., {Regan}, M., {et~al.} 2015, \apjs, 219,
  3

\bibitem[{{Naab} {et~al.}(2009){Naab}, {Johansson}, \& {Ostriker}}]{naab09}
{Naab}, T., {Johansson}, P.~H., \& {Ostriker}, J.~P. 2009, \apjl, 699, L178

\bibitem[{{Newman} {et~al.}(2012){Newman}, {Ellis}, {Bundy}, \&
  {Treu}}]{newman12}
{Newman}, A.~B., {Ellis}, R.~S., {Bundy}, K., \& {Treu}, T. 2012, \apj, 746,
  162

\bibitem[{{Nipoti} {et~al.}(2003){Nipoti}, {Londrillo}, \& {Ciotti}}]{nipoti03}
{Nipoti}, C., {Londrillo}, P., \& {Ciotti}, L. 2003, \mnras, 342, 501

\bibitem[{{Norman} {et~al.}(1996){Norman}, {Sellwood}, \& {Hasan}}]{norman96}
{Norman}, C.~A., {Sellwood}, J.~A., \& {Hasan}, H. 1996, \apj, 462, 114

\bibitem[{{Oemler}(1974)}]{oemler74}
{Oemler}, Augustus, J. 1974, \apj, 194, 1

\bibitem[{{Okamura} {et~al.}(1984){Okamura}, {Kodaira}, \&
  {Watanabe}}]{okamura84}
{Okamura}, S., {Kodaira}, K., \& {Watanabe}, M. 1984, \apj, 280, 7

\bibitem[{{Oke}(1974)}]{oke74}
{Oke}, J.~B. 1974, \apjs, 27, 21

\bibitem[{{Ouellette} {et~al.}(2017){Ouellette}, {Courteau}, {Holtzman},
  {Dutton}, {Cappellari}, {Dalcanton}, {McDonald}, {Roediger}, {Taylor},
  {Tully}, {C{\^o}t{\'e}}, {Ferrarese}, \& {Peng}}]{ouellette17}
{Ouellette}, N. N.~Q., {Courteau}, S., {Holtzman}, J.~A., {et~al.} 2017, \apj,
  843, 74

\bibitem[{{Pahre} {et~al.}(2004){Pahre}, {Ashby}, {Fazio}, \&
  {Willner}}]{pahre04}
{Pahre}, M.~A., {Ashby}, M.~L.~N., {Fazio}, G.~G., \& {Willner}, S.~P. 2004,
  \apjs, 154, 235

\bibitem[{{Paturel} {et~al.}(2003){Paturel}, {Petit}, {Prugniel}, {Theureau},
  {Rousseau}, {Brouty}, {Dubois}, \& {Cambr{\'e}sy}}]{paturel03}
{Paturel}, G., {Petit}, C., {Prugniel}, P., {et~al.} 2003, \aap, 412, 45

\bibitem[{{Peletier} {et~al.}(2020){Peletier}, {Iodice}, {Venhola},
  {Capaccioli}, {Cantiello}, {D'Abrusco}, {Falc{\'o}n-Barroso}, {Grado},
  {Hilker}, {Limatola}, {Mieske}, {Napolitano}, {Paolillo}, {Spavone},
  {Valentijn}, {van de Ven}, \& {Verdoes Kleijn}}]{peletier20}
{Peletier}, R., {Iodice}, E., {Venhola}, A., {et~al.} 2020, arXiv e-prints,
  arXiv:2008.12633

\bibitem[{{Peletier} {et~al.}(1990){Peletier}, {Davies}, {Illingworth},
  {Davis}, \& {Cawson}}]{peletier90}
{Peletier}, R.~F., {Davies}, R.~L., {Illingworth}, G.~D., {Davis}, L.~E., \&
  {Cawson}, M. 1990, \aj, 100, 1091

\bibitem[{{Peletier} {et~al.}(2012){Peletier}, {Kutdemir}, {van der Wolk},
  {Falc{\'o}n-Barroso}, {Bacon}, {Bureau}, {Cappellari}, {Davies}, {de Zeeuw},
  {Emsellem}, {Krajnovi{\'c}}, {Kuntschner}, {McDermid}, {Sarzi}, {Scott},
  {Shapiro}, {van den Bosch}, \& {van de Ven}}]{peletier12}
{Peletier}, R.~F., {Kutdemir}, E., {van der Wolk}, G., {et~al.} 2012, \mnras,
  419, 2031

\bibitem[{{Peng} {et~al.}(2002){Peng}, {Ho}, {Impey}, \& {Rix}}]{peng02}
{Peng}, C.~Y., {Ho}, L.~C., {Impey}, C.~D., \& {Rix}, H.-W. 2002, \aj, 124, 266

\bibitem[{{Peng} {et~al.}(2010){Peng}, {Ho}, {Impey}, \& {Rix}}]{peng10}
{Peng}, C.~Y., {Ho}, L.~C., {Impey}, C.~D., \& {Rix}, H.-W. 2010, \aj, 139,
  2097

\bibitem[{{Petrosian}(1976)}]{petrosian76}
{Petrosian}, V. 1976, \apjl, 210, L53

\bibitem[{{Pohlen} \& {Trujillo}(2006)}]{pohlen06}
{Pohlen}, M. \& {Trujillo}, I. 2006, \aap, 454, 759

\bibitem[{{Pozzetti} {et~al.}(2010){Pozzetti}, {Bolzonella}, {Zucca},
  {Zamorani}, {Lilly}, {Renzini}, {Moresco}, {Mignoli}, {Cassata}, {Tasca},
  {Lamareille}, {Maier}, {Meneux}, {Halliday}, {Oesch}, {Vergani}, {Caputi},
  {Kova{\v{c}}}, {Cimatti}, {Cucciati}, {Iovino}, {Peng}, {Carollo}, {Contini},
  {Kneib}, {Le F{\'e}vre}, {Mainieri}, {Scodeggio}, {Bardelli}, {Bongiorno},
  {Coppa}, {de la Torre}, {de Ravel}, {Franzetti}, {Garilli}, {Kampczyk},
  {Knobel}, {Le Borgne}, {Le Brun}, {Pell{\`o}}, {Perez Montero},
  {Ricciardelli}, {Silverman}, {Tanaka}, {Tresse}, {Abbas}, {Bottini}, {Cappi},
  {Guzzo}, {Koekemoer}, {Leauthaud}, {Maccagni}, {Marinoni}, {McCracken},
  {Memeo}, {Porciani}, {Scaramella}, {Scarlata}, \& {Scoville}}]{pozzetti10}
{Pozzetti}, L., {Bolzonella}, M., {Zucca}, E., {et~al.} 2010, \aap, 523, A13

\bibitem[{{Querejeta} {et~al.}(2015){Querejeta}, {Eliche-Moral}, {Tapia},
  {Borlaff}, {van de Ven}, {Lyubenova}, {Martig}, {Falc{\'o}n-Barroso}, \&
  {M{\'e}ndez-Abreu}}]{querejeta15}
{Querejeta}, M., {Eliche-Moral}, M.~C., {Tapia}, T., {et~al.} 2015, \aap, 579,
  L2

\bibitem[{{Ramos} {et~al.}(2015){Ramos}, {Men{\'e}ndez-Delmestre}, {Kim}, \&
  {Sheth}}]{ramos15}
{Ramos}, B. H.~F., {Men{\'e}ndez-Delmestre}, K., {Kim}, T., \& {Sheth}, K.
  2015, Highlights of Astronomy, 16, 333

\bibitem[{{Rautiainen} \& {Salo}(2000)}]{rautiainen00}
{Rautiainen}, P. \& {Salo}, H. 2000, \aap, 362, 465

\bibitem[{{Ravindranath} {et~al.}(2001){Ravindranath}, {Ho}, {Peng},
  {Filippenko}, \& {Sargent}}]{ravindranath01}
{Ravindranath}, S., {Ho}, L.~C., {Peng}, C.~Y., {Filippenko}, A.~V., \&
  {Sargent}, W. L.~W. 2001, \aj, 122, 653

\bibitem[{{Reach} {et~al.}(2005){Reach}, {Megeath}, {Cohen}, {Hora}, {Carey},
  {Surace}, {Willner}, {Barmby}, {Wilson}, {Glaccum}, {Lowrance}, {Marengo}, \&
  {Fazio}}]{reach05}
{Reach}, W.~T., {Megeath}, S.~T., {Cohen}, M., {et~al.} 2005, \pasp, 117, 978

\bibitem[{{Rizzo} {et~al.}(2018){Rizzo}, {Fraternali}, \& {Iorio}}]{rizzo18}
{Rizzo}, F., {Fraternali}, F., \& {Iorio}, G. 2018, \mnras, 476, 2137

\bibitem[{{Rom{\'a}n} \& {Trujillo}(2018)}]{roman18}
{Rom{\'a}n}, J. \& {Trujillo}, I. 2018, Research Notes of the American
  Astronomical Society, 2, 144

\bibitem[{{Saha} \& {Cortesi}(2018)}]{saha18}
{Saha}, K. \& {Cortesi}, A. 2018, \apjl, 862, L12

\bibitem[{{Saintonge} \& {Spekkens}(2011)}]{saintonge11}
{Saintonge}, A. \& {Spekkens}, K. 2011, \apj, 726, 77

\bibitem[{{Salo} {et~al.}(2015){Salo}, {Laurikainen}, {Laine}, {Comer{\'o}n},
  {Gadotti}, {Buta}, {Sheth}, {Zaritsky}, {Ho}, {Knapen}, {Athanassoula},
  {Bosma}, {Laine}, {Cisternas}, {Kim}, {Mu{\~n}oz-Mateos}, {Regan}, {Hinz},
  {Gil de Paz}, {Menendez-Delmestre}, {Mizusawa}, {Erroz-Ferrer}, {Meidt}, \&
  {Querejeta}}]{salo15}
{Salo}, H., {Laurikainen}, E., {Laine}, J., {et~al.} 2015, \apjs, 219, 4

\bibitem[{{S{\'a}nchez Almeida}(2020)}]{sanchezalmeida20}
{S{\'a}nchez Almeida}, J. 2020, \mnras, 495, 78

\bibitem[{{Sandage}(1961)}]{sandage61}
{Sandage}, A. 1961, {The Hubble Atlas of Galaxies}

\bibitem[{{Saracco} {et~al.}(2017){Saracco}, {Gargiulo}, {Ciocca}, \&
  {Marchesini}}]{saracco17}
{Saracco}, P., {Gargiulo}, A., {Ciocca}, F., \& {Marchesini}, D. 2017, \aap,
  597, A122

\bibitem[{{Saracco} {et~al.}(2012){Saracco}, {Gargiulo}, \&
  {Longhetti}}]{saracco12}
{Saracco}, P., {Gargiulo}, A., \& {Longhetti}, M. 2012, \mnras, 422, 3107

\bibitem[{{Savorgnan} {et~al.}(2013){Savorgnan}, {Graham}, {Marconi}, {Sani},
  {Hunt}, {Vika}, \& {Driver}}]{savorgnan13}
{Savorgnan}, G., {Graham}, A.~W., {Marconi}, A., {et~al.} 2013, \mnras, 434,
  387

\bibitem[{{Schaye}(2004)}]{schaye04}
{Schaye}, J. 2004, \apj, 609, 667

\bibitem[{{Schlafly} \& {Finkbeiner}(2011)}]{schlafly11}
{Schlafly}, E.~F. \& {Finkbeiner}, D.~P. 2011, \apj, 737, 103

\bibitem[{{Schombert}(1986)}]{schombert86}
{Schombert}, J.~M. 1986, \apjs, 60, 603

\bibitem[{{Schombert}(2015)}]{schombert15}
{Schombert}, J.~M. 2015, \aj, 150, 162

\bibitem[{{Sedgwick} {et~al.}(2019){Sedgwick}, {Baldry}, {James}, \&
  {Kelvin}}]{sedgwick19}
{Sedgwick}, T.~M., {Baldry}, I.~K., {James}, P.~A., \& {Kelvin}, L.~S. 2019,
  arXiv e-prints, arXiv:1909.04535

\bibitem[{{Sellwood}(1980)}]{sellwood80}
{Sellwood}, J.~A. 1980, \aap, 89, 296

\bibitem[{{Sellwood} \& {Binney}(2002)}]{sellwood02}
{Sellwood}, J.~A. \& {Binney}, J.~J. 2002, \mnras, 336, 785

\bibitem[{{Shen} {et~al.}(2003){Shen}, {Mo}, {White}, {Blanton}, {Kauffmann},
  {Voges}, {Brinkmann}, \& {Csabai}}]{shen03}
{Shen}, S., {Mo}, H.~J., {White}, S. D.~M., {et~al.} 2003, \mnras, 343, 978

\bibitem[{{Sheth} {et~al.}(2013){Sheth}, {Armus}, {Athanassoula}, {Bosma},
  {Gadotti}, {Munoz-Mateos}, {Hinz}, {Regan}, {Laurikainen}, {Jarrett},
  {Zaritsky}, {Menendez-Delmestre}, {Madore}, {Elmegreen}, {Knapen}, {Salo},
  {Schinnerer}, {Kim}, {Ho}, {Elmegreen}, {Buta}, {Cisternas}, {Laine},
  {Comeron}, {Donovan Meyer}, {D'Onghia}, \& {Salim}}]{sheth13}
{Sheth}, K., {Armus}, L., {Athanassoula}, E., {et~al.} 2013, {Not Dead Yet!
  Completing Spitzer's Legacy with Early Type Galaxies}, Spitzer Proposal

\bibitem[{{Sheth} {et~al.}(2008){Sheth}, {Elmegreen}, {Elmegreen}, {Capak},
  {Abraham}, {Athanassoula}, {Ellis}, {Mobasher}, {Salvato}, {Schinnerer},
  {Scoville}, {Spalsbury}, {Strubbe}, {Carollo}, {Rich}, \& {West}}]{sheth08}
{Sheth}, K., {Elmegreen}, D.~M., {Elmegreen}, B.~G., {et~al.} 2008, \apj, 675,
  1141

\bibitem[{{Sheth} {et~al.}(2010){Sheth}, {Regan}, {Hinz}, {Gil de Paz},
  {Men{\'e}ndez-Delmestre}, {Mu{\~n}oz-Mateos}, {Seibert}, {Kim},
  {Laurikainen}, {Salo}, {Gadotti}, {Laine}, {Mizusawa}, {Armus},
  {Athanassoula}, {Bosma}, {Buta}, {Capak}, {Jarrett}, {Elmegreen},
  {Elmegreen}, {Knapen}, {Koda}, {Helou}, {Ho}, {Madore}, {Masters},
  {Mobasher}, {Ogle}, {Peng}, {Schinnerer}, {Surace}, {Zaritsky},
  {Comer{\'o}n}, {de Swardt}, {Meidt}, {Kasliwal}, \& {Aravena}}]{sheth10}
{Sheth}, K., {Regan}, M., {Hinz}, J.~L., {et~al.} 2010, \pasp, 122, 1397

\bibitem[{{Sil'chenko}(2013)}]{silchenko13}
{Sil'chenko}, O. 2013, Memorie della Societa Astronomica Italiana Supplementi,
  25, 93

\bibitem[{{Skibba} {et~al.}(2011){Skibba}, {Engelbracht}, {Dale}, {Hinz},
  {Zibetti}, {Crocker}, {Groves}, {Hunt}, {Johnson}, {Meidt}, {Murphy},
  {Appleton}, {Armus}, {Bolatto}, {Brandl}, {Calzetti}, {Croxall}, {Galametz},
  {Gordon}, {Kennicutt}, {Koda}, {Krause}, {Montiel}, {Rix}, {Roussel},
  {Sandstrom}, {Sauvage}, {Schinnerer}, {Smith}, {Walter}, {Wilson}, \&
  {Wolfire}}]{skibba11}
{Skibba}, R.~A., {Engelbracht}, C.~W., {Dale}, D., {et~al.} 2011, \apj, 738, 89

\bibitem[{{Spergel} {et~al.}(2015){Spergel}, {Gehrels}, {Baltay}, {Bennett},
  {Breckinridge}, {Donahue}, {Dressler}, {Gaudi}, {Greene}, {Guyon}, {Hirata},
  {Kalirai}, {Kasdin}, {Macintosh}, {Moos}, {Perlmutter}, {Postman},
  {Rauscher}, {Rhodes}, {Wang}, {Weinberg}, {Benford}, {Hudson}, {Jeong},
  {Mellier}, {Traub}, {Yamada}, {Capak}, {Colbert}, {Masters}, {Penny},
  {Savransky}, {Stern}, {Zimmerman}, {Barry}, {Bartusek}, {Carpenter}, {Cheng},
  {Content}, {Dekens}, {Demers}, {Grady}, {Jackson}, {Kuan}, {Kruk}, {Melton},
  {Nemati}, {Parvin}, {Poberezhskiy}, {Peddie}, {Ruffa}, {Wallace}, {Whipple},
  {Wollack}, \& {Zhao}}]{spergel15}
{Spergel}, D., {Gehrels}, N., {Baltay}, C., {et~al.} 2015, arXiv e-prints,
  arXiv:1503.03757

\bibitem[{{Spitzer} \& {Baade}(1951)}]{spitzer51}
{Spitzer}, Lyman, J. \& {Baade}, W. 1951, \apj, 113, 413

\bibitem[{{Szomoru} {et~al.}(2012){Szomoru}, {Franx}, \& {van
  Dokkum}}]{szomoru12}
{Szomoru}, D., {Franx}, M., \& {van Dokkum}, P.~G. 2012, \apj, 749, 121

\bibitem[{{Tacchella} {et~al.}(2016){Tacchella}, {Dekel}, {Carollo},
  {Ceverino}, {DeGraf}, {Lapiner}, {Mandelker}, \& {Primack}}]{tacchella16}
{Tacchella}, S., {Dekel}, A., {Carollo}, C.~M., {et~al.} 2016, \mnras, 458, 242

\bibitem[{{Taylor-Mager} {et~al.}(2007){Taylor-Mager}, {Conselice},
  {Windhorst}, \& {Jansen}}]{taylor07}
{Taylor-Mager}, V.~A., {Conselice}, C.~J., {Windhorst}, R.~A., \& {Jansen},
  R.~A. 2007, \apj, 659, 162

\bibitem[{{Thomas} {et~al.}(2009){Thomas}, {Jesseit}, {Saglia}, {Bender},
  {Burkert}, {Corsini}, {Gebhardt}, {Magorrian}, {Naab}, {Thomas}, \&
  {Wegner}}]{thomas09}
{Thomas}, J., {Jesseit}, R., {Saglia}, R.~P., {et~al.} 2009, \mnras, 393, 641

\bibitem[{{Thomas} {et~al.}(2014){Thomas}, {Saglia}, {Bender}, {Erwin}, \&
  {Fabricius}}]{thomas14}
{Thomas}, J., {Saglia}, R.~P., {Bender}, R., {Erwin}, P., \& {Fabricius}, M.
  2014, \apj, 782, 39

\bibitem[{{Thuan}(1975)}]{thuan75}
{Thuan}, T.~X. 1975, \nat, 257, 774

\bibitem[{{Toft} {et~al.}(2007){Toft}, {van Dokkum}, {Franx}, {Labbe},
  {F{\"o}rster Schreiber}, {Wuyts}, {Webb}, {Rudnick}, {Zirm}, {Kriek}, {van
  der Werf}, {Blakeslee}, {Illingworth}, {Rix}, {Papovich}, \&
  {Moorwood}}]{toft07}
{Toft}, S., {van Dokkum}, P., {Franx}, M., {et~al.} 2007, \apj, 671, 285

\bibitem[{{Tokunaga} {et~al.}(1991){Tokunaga}, {Sellgren}, {Smith}, {Nagata},
  {Sakata}, \& {Nakada}}]{tokunaga91}
{Tokunaga}, A.~T., {Sellgren}, K., {Smith}, R.~G., {et~al.} 1991, \apj, 380,
  452

\bibitem[{{Toomre}(1977)}]{toomre77}
{Toomre}, A. 1977, in Evolution of Galaxies and Stellar Populations, ed. B.~M.
  {Tinsley} \& D.~C. {Larson}, Richard B.~Gehret, 401

\bibitem[{{Tremonti} {et~al.}(2004){Tremonti}, {Heckman}, {Kauffmann},
  {Brinchmann}, {Charlot}, {White}, {Seibert}, {Peng}, {Schlegel}, {Uomoto},
  {Fukugita}, \& {Brinkmann}}]{tremonti04}
{Tremonti}, C.~A., {Heckman}, T.~M., {Kauffmann}, G., {et~al.} 2004, \apj, 613,
  898

\bibitem[{{Trujillo} {et~al.}(2020){Trujillo}, {Chamba}, \&
  {Knapen}}]{trujillo20}
{Trujillo}, I., {Chamba}, N., \& {Knapen}, J.~H. 2020, \mnras, 493, 87

\bibitem[{{Trujillo} {et~al.}(2007){Trujillo}, {Conselice}, {Bundy}, {Cooper},
  {Eisenhardt}, \& {Ellis}}]{trujillo07}
{Trujillo}, I., {Conselice}, C.~J., {Bundy}, K., {et~al.} 2007, \mnras, 382,
  109

\bibitem[{{Trujillo} {et~al.}(2001){Trujillo}, {Graham}, \&
  {Caon}}]{trujillo01}
{Trujillo}, I., {Graham}, A.~W., \& {Caon}, N. 2001, \mnras, 326, 869

\bibitem[{{van den Bergh}(1976)}]{vandenbergh76}
{van den Bergh}, S. 1976, \apj, 206, 883

\bibitem[{{van den Bergh}(1998)}]{vandenbergh98}
{van den Bergh}, S. 1998, {Galaxy Morphology and Classification} (Cambridge
  University Press, Cambridge, NY {USA})

\bibitem[{{van der Wel}(2008)}]{vanderwel08}
{van der Wel}, A. 2008, \apjl, 675, L13

\bibitem[{{van der Wel} {et~al.}(2009){van der Wel}, {Bell}, {van den Bosch},
  {Gallazzi}, \& {Rix}}]{vanderwel09}
{van der Wel}, A., {Bell}, E.~F., {van den Bosch}, F.~C., {Gallazzi}, A., \&
  {Rix}, H.-W. 2009, \apj, 698, 1232

\bibitem[{{van der Wel} {et~al.}(2014){van der Wel}, {Franx}, {van Dokkum},
  {Skelton}, {Momcheva}, {Whitaker}, {Brammer}, {Bell}, {Rix}, {Wuyts},
  {Ferguson}, {Holden}, {Barro}, {Koekemoer}, {Chang}, {McGrath},
  {H{\"a}ussler}, {Dekel}, {Behroozi}, {Fumagalli}, {Leja}, {Lundgren},
  {Maseda}, {Nelson}, {Wake}, {Patel}, {Labb{\'e}}, {Faber}, {Grogin}, \&
  {Kocevski}}]{vanderwel14}
{van der Wel}, A., {Franx}, M., {van Dokkum}, P.~G., {et~al.} 2014, \apj, 788,
  28

\bibitem[{{van Dokkum} {et~al.}(2015){van Dokkum}, {Abraham}, {Merritt},
  {Zhang}, {Geha}, \& {Conroy}}]{vandokkum15}
{van Dokkum}, P.~G., {Abraham}, R., {Merritt}, A., {et~al.} 2015, \apjl, 798,
  L45

\bibitem[{{van Dokkum} {et~al.}(2008){van Dokkum}, {Franx}, {Kriek}, {Holden},
  {Illingworth}, {Magee}, {Bouwens}, {Marchesini}, {Quadri}, {Rudnick},
  {Taylor}, \& {Toft}}]{vandokkum08}
{van Dokkum}, P.~G., {Franx}, M., {Kriek}, M., {et~al.} 2008, \apjl, 677, L5

\bibitem[{{Venhola} {et~al.}(2018){Venhola}, {Peletier}, {Laurikainen}, {Salo},
  {Iodice}, {Mieske}, {Hilker}, {Wittmann}, {Lisker}, {Paolillo}, {Cantiello},
  {Janz}, {Spavone}, {D'Abrusco}, {Ven}, {Napolitano}, {Kleijn}, {Maddox},
  {Capaccioli}, {Grado}, {Valentijn}, {Falc{\'o}n-Barroso}, \&
  {Limatola}}]{venhola18}
{Venhola}, A., {Peletier}, R., {Laurikainen}, E., {et~al.} 2018, \aap, 620,
  A165

\bibitem[{Virtanen {et~al.}(2020)Virtanen, Gommers, Oliphant, Haberland, Reddy,
  Cournapeau, Burovski, Peterson, Weckesser, Bright, {van der Walt}, Brett,
  Wilson, Millman, Mayorov, Nelson, Jones, Kern, Larson, Carey, Polat, Feng,
  Moore, {VanderPlas}, Laxalde, Perktold, Cimrman, Henriksen, Quintero, Harris,
  Archibald, Ribeiro, Pedregosa, {van Mulbregt}, \& {SciPy 1.0
  Contributors}}]{virtanen20}
Virtanen, P., Gommers, R., Oliphant, T.~E., {et~al.} 2020, Nature Methods, 17,
  261

\bibitem[{{Watkins} {et~al.}(2019){Watkins}, {Laine}, {Comer{\'o}n}, {Janz}, \&
  {Salo}}]{watkins19}
{Watkins}, A.~E., {Laine}, J., {Comer{\'o}n}, S., {Janz}, J., \& {Salo}, H.
  2019, \aap, 625, A36

\bibitem[{{Wegner} {et~al.}(2012){Wegner}, {Corsini}, {Thomas}, {Saglia},
  {Bender}, \& {Pu}}]{wegner12}
{Wegner}, G.~A., {Corsini}, E.~M., {Thomas}, J., {et~al.} 2012, \aj, 144, 78

\bibitem[{{Werner} {et~al.}(2004){Werner}, {Roellig}, {Low}, {Rieke}, {Rieke},
  {Hoffmann}, {Young}, {Houck}, {Brandl}, {Fazio}, {Hora}, {Gehrz}, {Helou},
  {Soifer}, {Stauffer}, {Keene}, {Eisenhardt}, {Gallagher}, {Gautier}, {Irace},
  {Lawrence}, {Simmons}, {Van Cleve}, {Jura}, {Wright}, \&
  {Cruikshank}}]{werner04}
{Werner}, M.~W., {Roellig}, T.~L., {Low}, F.~J., {et~al.} 2004, \apjs, 154, 1

\bibitem[{{White}(1983)}]{white83}
{White}, S.~D.~M. 1983, in IAU Symposium, Vol. 100, Internal Kinematics and
  Dynamics of Galaxies, ed. E.~{Athanassoula}, 337--344

\bibitem[{{White} \& {Rees}(1978)}]{white78}
{White}, S.~D.~M. \& {Rees}, M.~J. 1978, \mnras, 183, 341

\bibitem[{{Williams} \& {Bonanos}(2016)}]{williams16}
{Williams}, S.~J. \& {Bonanos}, A.~Z. 2016, \aap, 587, A121

\bibitem[{{Willner} {et~al.}(1977){Willner}, {Soifer}, {Russell}, {Joyce}, \&
  {Gillett}}]{willner77}
{Willner}, S.~P., {Soifer}, B.~T., {Russell}, R.~W., {Joyce}, R.~R., \&
  {Gillett}, F.~C. 1977, \apjl, 217, L121

\bibitem[{{Wilman} \& {Erwin}(2012)}]{wilman12}
{Wilman}, D.~J. \& {Erwin}, P. 2012, \apj, 746, 160

\bibitem[{{Wyse} \& {Jones}(1984)}]{wyse84}
{Wyse}, R.~F.~G. \& {Jones}, B.~J.~T. 1984, \apj, 286, 88

\bibitem[{{Young} \& {Currie}(1994)}]{young94}
{Young}, C.~K. \& {Currie}, M.~J. 1994, \mnras, 268, L11

\bibitem[{{Zaritsky} {et~al.}(2013){Zaritsky}, {Salo}, {Laurikainen},
  {Elmegreen}, {Athanassoula}, {Bosma}, {Comer{\'o}n}, {Erroz-Ferrer},
  {Elmegreen}, {Gadotti}, {Gil de Paz}, {Hinz}, {Ho}, {Holwerda}, {Kim},
  {Knapen}, {Laine}, {Laine}, {Madore}, {Meidt}, {Menendez-Delmestre},
  {Mizusawa}, {Mu{\~n}oz-Mateos}, {Regan}, {Seibert}, \& {Sheth}}]{zaritsky13}
{Zaritsky}, D., {Salo}, H., {Laurikainen}, E., {et~al.} 2013, \apj, 772, 135

\bibitem[{{Zhu} {et~al.}(2010){Zhu}, {Wu}, {Li}, \& {Cao}}]{zhu10}
{Zhu}, Y.-N., {Wu}, H., {Li}, H.-N., \& {Cao}, C. 2010, Research in Astronomy
  and Astrophysics, 10, 329

\end{thebibliography}

\begin{appendix}

\section{Updated ETG morphological classifications}

We will make our updated Comprehensive de Vaucouleurs revised
Hubble-Sandage \citep[CVHRS;][]{buta07} morphological classifications (Sec. \ref{sec:sample}) for all of the newly observed ETGs in the S$^{4}$G+ETG sample available at IPAC, alongside all of the other quantities we have discussed in this paper.  Table \ref{tab:app_ttypes} shows a sub-sample of the full table we will provide, which will contain morphological classifications for all 465 newly observed galaxies.  This includes the full CVRHS classification, outer and inner structure classifications (rings, lenses, barlenses, etc.), the numerical family index corresponding to the presence or absence of a bar, and the numerical $T-$type (which we have used throughout this paper).

\begin{table}
  \caption{CVRHS Classifications
 for newly observed S$^4$G Early-Type Galaxies}
  \label{tab:app_ttypes}
  \centering
  \begin{tabular}{lllclr}
    \hline\hline
    Name & CVRHS & Outer & Family & Inner & Stage \\
      & Type & Variety & Index & Variety & Index \\
    1 & 2 & 3 & 4 & 5 & 6 \\
    \hline \\
    NGC   216      & S0d spw / E(d)6 pec     &          &    .. &            &    $-$2: \\
    NGC   357      & (RL)SB(rs,bl)0/a        & (RL)     &     4 & (rs,bl)    &     0\\                                      
    NGC   448      & S0$^{-}$ sp / E(d)4         &          &    .. &            &    $-$3\\                                  
    NGC   502      & SA(l)0$^{+}$                &          &     0 & (l)        &    $-$1\\                                   
    NGC   509      & S0$^{+}$ sp / E(d)5         &          &    .. &            &    $-$1\\ 
    ... & ... & ... & ... & ... & ... \\
    \hline
  \end{tabular}
  \tablefoot{Col. 1 $-$ Galaxy name; col. 2:
 classification in system of R. Buta et al. 
 ApJS, 217, 32, 2015; col. 3: outer feature type;
 col. 4: numerical family index (0=SA, 1=S$\underline{\rm A}$B, 2=SAB, 3=SA$\underline{\rm B}$, 4=SB); 
col. 5: inner feature type; col. 6: $T$-type (RC3 scale)}
\end{table}
\end{appendix}

\end{document}